\documentclass[longauth]{aaEC}

\usepackage{graphicx}
\usepackage{natbib}
\usepackage{scalerel}
\usepackage{diagbox}
\usepackage[table]{xcolor}

\usepackage{longtable}
\usepackage{caption}
\usepackage{txfonts}
\usepackage[pdfencoding=auto,psdextra]{hyperref}
\hypersetup{
    colorlinks=true,
    linkcolor=blue,
    filecolor=magenta,      
    urlcolor=blue,
    citecolor=blue
}
\usepackage[nameinlink,capitalise]{cleveref}
\crefname{section}{Sect.}{Sects.}
\Crefname{section}{Section}{Sections}
\crefname{figure}{Fig.}{Figs.}
\Crefname{figure}{Figure}{Figures}
\crefname{equation}{Eq.}{Eqs.}
\Crefname{equation}{Equation}{Equations}
\crefname{table}{Table}{Tables}
\crefname{appendix}{Appendix}{Appendices}

\makeatletter
\renewcommand*\aa@pageof{, page \thepage{} of \pageref*{LastPage}}
\makeatother

\usepackage[utf8]{inputenc}

\usepackage[switch, modulo]{lineno}
\usepackage{orcidlink} %this has to come after hyperref
\newcommand{\orcid}[1]{\orcidlink{#1}}

\usepackage{euclid}

\newcommand{\pz}{\phantom{0}}

\begin{document}
% Put the title and authors of your (Standard Project) paper here
%

\title{Euclid Quick Data Release (Q1)}
\subtitle{The average far-infrared properties of \Euclid-selected star-forming galaxies}

%%%% Version Friday 26th of September 2025 03:47:06 PM UT												
%%%% Please do not edit the author list -- contact ECEB Bureau for changes
%\newcommand{\orcid}[1]{} %% if already defined in aa.cls: comment, or use renewcommand			   
\author{Euclid Collaboration: R.~Hill\orcid{0009-0008-8718-0644}\thanks{\email{ryleyhill@phas.ubc.ca}}\inst{\ref{aff1}}
\and A.~Abghari\orcid{0009-0003-2250-3880}\inst{\ref{aff1}}
\and D.~Scott\orcid{0000-0002-6878-9840}\inst{\ref{aff1}}
\and M.~Bethermin\orcid{0000-0002-3915-2015}\inst{\ref{aff2}}
\and S.~C.~Chapman\orcid{0000-0002-8487-3153}\inst{\ref{aff1},\ref{aff3},\ref{aff4}}
\and D.~L.~Clements\orcid{0000-0002-9548-5033}\inst{\ref{aff5}}
\and S.~Eales\inst{\ref{aff6}}
\and A.~Enia\orcid{0000-0002-0200-2857}\inst{\ref{aff7},\ref{aff8}}
\and B.~Jego\orcid{0009-0006-6399-7858}\inst{\ref{aff2}}
\and A.~Parmar\orcid{0009-0009-9390-9232}\inst{\ref{aff5}}
\and P.~Tanouri\orcid{0009-0006-3261-6397}\inst{\ref{aff1}}
\and L.~Wang\orcid{0000-0002-6736-9158}\inst{\ref{aff9},\ref{aff10}}
\and S.~Andreon\orcid{0000-0002-2041-8784}\inst{\ref{aff11}}
\and N.~Auricchio\orcid{0000-0003-4444-8651}\inst{\ref{aff7}}
\and C.~Baccigalupi\orcid{0000-0002-8211-1630}\inst{\ref{aff12},\ref{aff13},\ref{aff14},\ref{aff15}}
\and M.~Baldi\orcid{0000-0003-4145-1943}\inst{\ref{aff8},\ref{aff7},\ref{aff16}}
\and A.~Balestra\orcid{0000-0002-6967-261X}\inst{\ref{aff17}}
\and S.~Bardelli\orcid{0000-0002-8900-0298}\inst{\ref{aff7}}
\and P.~Battaglia\orcid{0000-0002-7337-5909}\inst{\ref{aff7}}
\and A.~Biviano\orcid{0000-0002-0857-0732}\inst{\ref{aff13},\ref{aff12}}
\and E.~Branchini\orcid{0000-0002-0808-6908}\inst{\ref{aff18},\ref{aff19},\ref{aff11}}
\and M.~Brescia\orcid{0000-0001-9506-5680}\inst{\ref{aff20},\ref{aff21}}
\and S.~Camera\orcid{0000-0003-3399-3574}\inst{\ref{aff22},\ref{aff23},\ref{aff24}}
\and G.~Ca\~nas-Herrera\orcid{0000-0003-2796-2149}\inst{\ref{aff25},\ref{aff26}}
\and V.~Capobianco\orcid{0000-0002-3309-7692}\inst{\ref{aff24}}
\and C.~Carbone\orcid{0000-0003-0125-3563}\inst{\ref{aff27}}
\and J.~Carretero\orcid{0000-0002-3130-0204}\inst{\ref{aff28},\ref{aff29}}
\and M.~Castellano\orcid{0000-0001-9875-8263}\inst{\ref{aff30}}
\and G.~Castignani\orcid{0000-0001-6831-0687}\inst{\ref{aff7}}
\and S.~Cavuoti\orcid{0000-0002-3787-4196}\inst{\ref{aff21},\ref{aff31}}
\and K.~C.~Chambers\orcid{0000-0001-6965-7789}\inst{\ref{aff32}}
\and A.~Cimatti\inst{\ref{aff33}}
\and C.~Colodro-Conde\inst{\ref{aff34}}
\and G.~Congedo\orcid{0000-0003-2508-0046}\inst{\ref{aff35}}
\and C.~J.~Conselice\orcid{0000-0003-1949-7638}\inst{\ref{aff36}}
\and L.~Conversi\orcid{0000-0002-6710-8476}\inst{\ref{aff37},\ref{aff38}}
\and Y.~Copin\orcid{0000-0002-5317-7518}\inst{\ref{aff39}}
\and A.~Costille\inst{\ref{aff40}}
\and F.~Courbin\orcid{0000-0003-0758-6510}\inst{\ref{aff41},\ref{aff42},\ref{aff43}}
\and H.~M.~Courtois\orcid{0000-0003-0509-1776}\inst{\ref{aff44}}
\and M.~Cropper\orcid{0000-0003-4571-9468}\inst{\ref{aff45}}
\and A.~Da~Silva\orcid{0000-0002-6385-1609}\inst{\ref{aff46},\ref{aff47}}
\and H.~Degaudenzi\orcid{0000-0002-5887-6799}\inst{\ref{aff48}}
\and G.~De~Lucia\orcid{0000-0002-6220-9104}\inst{\ref{aff13}}
\and H.~Dole\orcid{0000-0002-9767-3839}\inst{\ref{aff49}}
\and F.~Dubath\orcid{0000-0002-6533-2810}\inst{\ref{aff48}}
\and X.~Dupac\inst{\ref{aff38}}
\and S.~Dusini\orcid{0000-0002-1128-0664}\inst{\ref{aff50}}
\and S.~Escoffier\orcid{0000-0002-2847-7498}\inst{\ref{aff51}}
\and M.~Farina\orcid{0000-0002-3089-7846}\inst{\ref{aff52}}
\and F.~Faustini\orcid{0000-0001-6274-5145}\inst{\ref{aff30},\ref{aff53}}
\and S.~Ferriol\inst{\ref{aff39}}
\and F.~Finelli\orcid{0000-0002-6694-3269}\inst{\ref{aff7},\ref{aff54}}
\and N.~Fourmanoit\orcid{0009-0005-6816-6925}\inst{\ref{aff51}}
\and M.~Frailis\orcid{0000-0002-7400-2135}\inst{\ref{aff13}}
\and E.~Franceschi\orcid{0000-0002-0585-6591}\inst{\ref{aff7}}
\and M.~Fumana\orcid{0000-0001-6787-5950}\inst{\ref{aff27}}
\and S.~Galeotta\orcid{0000-0002-3748-5115}\inst{\ref{aff13}}
\and K.~George\orcid{0000-0002-1734-8455}\inst{\ref{aff55}}
\and B.~Gillis\orcid{0000-0002-4478-1270}\inst{\ref{aff35}}
\and C.~Giocoli\orcid{0000-0002-9590-7961}\inst{\ref{aff7},\ref{aff16}}
\and J.~Gracia-Carpio\inst{\ref{aff56}}
\and A.~Grazian\orcid{0000-0002-5688-0663}\inst{\ref{aff17}}
\and F.~Grupp\inst{\ref{aff56},\ref{aff57}}
\and S.~V.~H.~Haugan\orcid{0000-0001-9648-7260}\inst{\ref{aff58}}
\and W.~Holmes\inst{\ref{aff59}}
\and I.~M.~Hook\orcid{0000-0002-2960-978X}\inst{\ref{aff60}}
\and F.~Hormuth\inst{\ref{aff61}}
\and A.~Hornstrup\orcid{0000-0002-3363-0936}\inst{\ref{aff62},\ref{aff63}}
\and K.~Jahnke\orcid{0000-0003-3804-2137}\inst{\ref{aff64}}
\and M.~Jhabvala\inst{\ref{aff65}}
\and B.~Joachimi\orcid{0000-0001-7494-1303}\inst{\ref{aff66}}
\and E.~Keih\"anen\orcid{0000-0003-1804-7715}\inst{\ref{aff67}}
\and S.~Kermiche\orcid{0000-0002-0302-5735}\inst{\ref{aff51}}
\and A.~Kiessling\orcid{0000-0002-2590-1273}\inst{\ref{aff59}}
\and B.~Kubik\orcid{0009-0006-5823-4880}\inst{\ref{aff39}}
\and M.~K\"ummel\orcid{0000-0003-2791-2117}\inst{\ref{aff57}}
\and M.~Kunz\orcid{0000-0002-3052-7394}\inst{\ref{aff68}}
\and H.~Kurki-Suonio\orcid{0000-0002-4618-3063}\inst{\ref{aff69},\ref{aff70}}
\and A.~M.~C.~Le~Brun\orcid{0000-0002-0936-4594}\inst{\ref{aff71}}
\and D.~Le~Mignant\orcid{0000-0002-5339-5515}\inst{\ref{aff40}}
\and S.~Ligori\orcid{0000-0003-4172-4606}\inst{\ref{aff24}}
\and P.~B.~Lilje\orcid{0000-0003-4324-7794}\inst{\ref{aff58}}
\and V.~Lindholm\orcid{0000-0003-2317-5471}\inst{\ref{aff69},\ref{aff70}}
\and I.~Lloro\orcid{0000-0001-5966-1434}\inst{\ref{aff72}}
\and G.~Mainetti\orcid{0000-0003-2384-2377}\inst{\ref{aff73}}
\and D.~Maino\inst{\ref{aff74},\ref{aff27},\ref{aff75}}
\and E.~Maiorano\orcid{0000-0003-2593-4355}\inst{\ref{aff7}}
\and O.~Mansutti\orcid{0000-0001-5758-4658}\inst{\ref{aff13}}
\and S.~Marcin\inst{\ref{aff76}}
\and O.~Marggraf\orcid{0000-0001-7242-3852}\inst{\ref{aff77}}
\and M.~Martinelli\orcid{0000-0002-6943-7732}\inst{\ref{aff30},\ref{aff78}}
\and N.~Martinet\orcid{0000-0003-2786-7790}\inst{\ref{aff40}}
\and F.~Marulli\orcid{0000-0002-8850-0303}\inst{\ref{aff79},\ref{aff7},\ref{aff16}}
\and R.~J.~Massey\orcid{0000-0002-6085-3780}\inst{\ref{aff80}}
\and E.~Medinaceli\orcid{0000-0002-4040-7783}\inst{\ref{aff7}}
\and S.~Mei\orcid{0000-0002-2849-559X}\inst{\ref{aff81},\ref{aff82}}
\and M.~Melchior\inst{\ref{aff83}}
\and Y.~Mellier\inst{\ref{aff84},\ref{aff85}}
\and M.~Meneghetti\orcid{0000-0003-1225-7084}\inst{\ref{aff7},\ref{aff16}}
\and E.~Merlin\orcid{0000-0001-6870-8900}\inst{\ref{aff30}}
\and G.~Meylan\inst{\ref{aff86}}
\and A.~Mora\orcid{0000-0002-1922-8529}\inst{\ref{aff87}}
\and M.~Moresco\orcid{0000-0002-7616-7136}\inst{\ref{aff79},\ref{aff7}}
\and L.~Moscardini\orcid{0000-0002-3473-6716}\inst{\ref{aff79},\ref{aff7},\ref{aff16}}
\and R.~Nakajima\orcid{0009-0009-1213-7040}\inst{\ref{aff77}}
\and C.~Neissner\orcid{0000-0001-8524-4968}\inst{\ref{aff88},\ref{aff29}}
\and S.-M.~Niemi\orcid{0009-0005-0247-0086}\inst{\ref{aff25}}
\and C.~Padilla\orcid{0000-0001-7951-0166}\inst{\ref{aff88}}
\and S.~Paltani\orcid{0000-0002-8108-9179}\inst{\ref{aff48}}
\and F.~Pasian\orcid{0000-0002-4869-3227}\inst{\ref{aff13}}
\and K.~Pedersen\inst{\ref{aff89}}
\and W.~J.~Percival\orcid{0000-0002-0644-5727}\inst{\ref{aff90},\ref{aff91},\ref{aff92}}
\and V.~Pettorino\orcid{0000-0002-4203-9320}\inst{\ref{aff25}}
\and S.~Pires\orcid{0000-0002-0249-2104}\inst{\ref{aff93}}
\and G.~Polenta\orcid{0000-0003-4067-9196}\inst{\ref{aff53}}
\and M.~Poncet\inst{\ref{aff94}}
\and L.~A.~Popa\inst{\ref{aff95}}
\and L.~Pozzetti\orcid{0000-0001-7085-0412}\inst{\ref{aff7}}
\and F.~Raison\orcid{0000-0002-7819-6918}\inst{\ref{aff56}}
\and R.~Rebolo\orcid{0000-0003-3767-7085}\inst{\ref{aff34},\ref{aff96},\ref{aff97}}
\and A.~Renzi\orcid{0000-0001-9856-1970}\inst{\ref{aff98},\ref{aff50}}
\and J.~Rhodes\orcid{0000-0002-4485-8549}\inst{\ref{aff59}}
\and G.~Riccio\inst{\ref{aff21}}
\and E.~Romelli\orcid{0000-0003-3069-9222}\inst{\ref{aff13}}
\and M.~Roncarelli\orcid{0000-0001-9587-7822}\inst{\ref{aff7}}
\and R.~Saglia\orcid{0000-0003-0378-7032}\inst{\ref{aff57},\ref{aff56}}
\and Z.~Sakr\orcid{0000-0002-4823-3757}\inst{\ref{aff99},\ref{aff100},\ref{aff101}}
\and D.~Sapone\orcid{0000-0001-7089-4503}\inst{\ref{aff102}}
\and B.~Sartoris\orcid{0000-0003-1337-5269}\inst{\ref{aff57},\ref{aff13}}
\and M.~Sauvage\orcid{0000-0002-0809-2574}\inst{\ref{aff93}}
\and M.~Schirmer\orcid{0000-0003-2568-9994}\inst{\ref{aff64}}
\and P.~Schneider\orcid{0000-0001-8561-2679}\inst{\ref{aff77}}
\and T.~Schrabback\orcid{0000-0002-6987-7834}\inst{\ref{aff103}}
\and A.~Secroun\orcid{0000-0003-0505-3710}\inst{\ref{aff51}}
\and G.~Seidel\orcid{0000-0003-2907-353X}\inst{\ref{aff64}}
\and S.~Serrano\orcid{0000-0002-0211-2861}\inst{\ref{aff104},\ref{aff105},\ref{aff106}}
\and C.~Sirignano\orcid{0000-0002-0995-7146}\inst{\ref{aff98},\ref{aff50}}
\and G.~Sirri\orcid{0000-0003-2626-2853}\inst{\ref{aff16}}
\and L.~Stanco\orcid{0000-0002-9706-5104}\inst{\ref{aff50}}
\and J.-L.~Starck\orcid{0000-0003-2177-7794}\inst{\ref{aff93}}
\and J.~Steinwagner\orcid{0000-0001-7443-1047}\inst{\ref{aff56}}
\and P.~Tallada-Cresp\'{i}\orcid{0000-0002-1336-8328}\inst{\ref{aff28},\ref{aff29}}
\and A.~N.~Taylor\inst{\ref{aff35}}
\and H.~I.~Teplitz\orcid{0000-0002-7064-5424}\inst{\ref{aff107}}
\and I.~Tereno\orcid{0000-0002-4537-6218}\inst{\ref{aff46},\ref{aff108}}
\and N.~Tessore\orcid{0000-0002-9696-7931}\inst{\ref{aff66},\ref{aff45}}
\and S.~Toft\orcid{0000-0003-3631-7176}\inst{\ref{aff109},\ref{aff110}}
\and R.~Toledo-Moreo\orcid{0000-0002-2997-4859}\inst{\ref{aff111}}
\and F.~Torradeflot\orcid{0000-0003-1160-1517}\inst{\ref{aff29},\ref{aff28}}
\and I.~Tutusaus\orcid{0000-0002-3199-0399}\inst{\ref{aff106},\ref{aff104},\ref{aff100}}
\and L.~Valenziano\orcid{0000-0002-1170-0104}\inst{\ref{aff7},\ref{aff54}}
\and J.~Valiviita\orcid{0000-0001-6225-3693}\inst{\ref{aff69},\ref{aff70}}
\and T.~Vassallo\orcid{0000-0001-6512-6358}\inst{\ref{aff13},\ref{aff55}}
\and G.~Verdoes~Kleijn\orcid{0000-0001-5803-2580}\inst{\ref{aff10}}
\and A.~Veropalumbo\orcid{0000-0003-2387-1194}\inst{\ref{aff11},\ref{aff19},\ref{aff18}}
\and Y.~Wang\orcid{0000-0002-4749-2984}\inst{\ref{aff107}}
\and J.~Weller\orcid{0000-0002-8282-2010}\inst{\ref{aff57},\ref{aff56}}
\and A.~Zacchei\orcid{0000-0003-0396-1192}\inst{\ref{aff13},\ref{aff12}}
\and G.~Zamorani\orcid{0000-0002-2318-301X}\inst{\ref{aff7}}
\and F.~M.~Zerbi\inst{\ref{aff11}}
\and I.~A.~Zinchenko\orcid{0000-0002-2944-2449}\inst{\ref{aff112}}
\and E.~Zucca\orcid{0000-0002-5845-8132}\inst{\ref{aff7}}
\and V.~Allevato\orcid{0000-0001-7232-5152}\inst{\ref{aff21}}
\and M.~Ballardini\orcid{0000-0003-4481-3559}\inst{\ref{aff113},\ref{aff114},\ref{aff7}}
\and M.~Bolzonella\orcid{0000-0003-3278-4607}\inst{\ref{aff7}}
\and E.~Bozzo\orcid{0000-0002-8201-1525}\inst{\ref{aff48}}
\and C.~Burigana\orcid{0000-0002-3005-5796}\inst{\ref{aff115},\ref{aff54}}
\and R.~Cabanac\orcid{0000-0001-6679-2600}\inst{\ref{aff100}}
\and M.~Calabrese\orcid{0000-0002-2637-2422}\inst{\ref{aff116},\ref{aff27}}
\and A.~Cappi\inst{\ref{aff7},\ref{aff117}}
\and J.~A.~Escartin~Vigo\inst{\ref{aff56}}
\and L.~Gabarra\orcid{0000-0002-8486-8856}\inst{\ref{aff118}}
\and W.~G.~Hartley\inst{\ref{aff48}}
\and M.~Huertas-Company\orcid{0000-0002-1416-8483}\inst{\ref{aff34},\ref{aff119},\ref{aff120},\ref{aff121}}
\and R.~Maoli\orcid{0000-0002-6065-3025}\inst{\ref{aff122},\ref{aff30}}
\and J.~Mart\'{i}n-Fleitas\orcid{0000-0002-8594-569X}\inst{\ref{aff123}}
\and S.~Matthew\orcid{0000-0001-8448-1697}\inst{\ref{aff35}}
\and N.~Mauri\orcid{0000-0001-8196-1548}\inst{\ref{aff33},\ref{aff16}}
\and R.~B.~Metcalf\orcid{0000-0003-3167-2574}\inst{\ref{aff79},\ref{aff7}}
\and A.~Pezzotta\orcid{0000-0003-0726-2268}\inst{\ref{aff11}}
\and M.~P\"ontinen\orcid{0000-0001-5442-2530}\inst{\ref{aff69}}
\and I.~Risso\orcid{0000-0003-2525-7761}\inst{\ref{aff11},\ref{aff19}}
\and V.~Scottez\orcid{0009-0008-3864-940X}\inst{\ref{aff84},\ref{aff124}}
\and M.~Sereno\orcid{0000-0003-0302-0325}\inst{\ref{aff7},\ref{aff16}}
\and M.~Tenti\orcid{0000-0002-4254-5901}\inst{\ref{aff16}}
\and M.~Viel\orcid{0000-0002-2642-5707}\inst{\ref{aff12},\ref{aff13},\ref{aff15},\ref{aff14},\ref{aff125}}
\and M.~Wiesmann\orcid{0009-0000-8199-5860}\inst{\ref{aff58}}
\and Y.~Akrami\orcid{0000-0002-2407-7956}\inst{\ref{aff126},\ref{aff127}}
\and I.~T.~Andika\orcid{0000-0001-6102-9526}\inst{\ref{aff128},\ref{aff129}}
\and S.~Anselmi\orcid{0000-0002-3579-9583}\inst{\ref{aff50},\ref{aff98},\ref{aff130}}
\and M.~Archidiacono\orcid{0000-0003-4952-9012}\inst{\ref{aff74},\ref{aff75}}
\and F.~Atrio-Barandela\orcid{0000-0002-2130-2513}\inst{\ref{aff131}}
\and D.~Bertacca\orcid{0000-0002-2490-7139}\inst{\ref{aff98},\ref{aff17},\ref{aff50}}
\and L.~Bisigello\orcid{0000-0003-0492-4924}\inst{\ref{aff17}}
\and A.~Blanchard\orcid{0000-0001-8555-9003}\inst{\ref{aff100}}
\and L.~Blot\orcid{0000-0002-9622-7167}\inst{\ref{aff132},\ref{aff71}}
\and H.~B\"ohringer\orcid{0000-0001-8241-4204}\inst{\ref{aff56},\ref{aff133},\ref{aff134}}
\and M.~Bonici\orcid{0000-0002-8430-126X}\inst{\ref{aff90},\ref{aff27}}
\and S.~Borgani\orcid{0000-0001-6151-6439}\inst{\ref{aff135},\ref{aff12},\ref{aff13},\ref{aff14},\ref{aff125}}
\and M.~L.~Brown\orcid{0000-0002-0370-8077}\inst{\ref{aff36}}
\and S.~Bruton\orcid{0000-0002-6503-5218}\inst{\ref{aff136}}
\and A.~Calabro\orcid{0000-0003-2536-1614}\inst{\ref{aff30}}
\and B.~Camacho~Quevedo\orcid{0000-0002-8789-4232}\inst{\ref{aff12},\ref{aff15},\ref{aff13}}
\and F.~Caro\inst{\ref{aff30}}
\and C.~S.~Carvalho\inst{\ref{aff108}}
\and T.~Castro\orcid{0000-0002-6292-3228}\inst{\ref{aff13},\ref{aff14},\ref{aff12},\ref{aff125}}
\and Y.~Charles\inst{\ref{aff40}}
\and F.~Cogato\orcid{0000-0003-4632-6113}\inst{\ref{aff79},\ref{aff7}}
\and S.~Conseil\orcid{0000-0002-3657-4191}\inst{\ref{aff39}}
\and A.~R.~Cooray\orcid{0000-0002-3892-0190}\inst{\ref{aff137}}
\and O.~Cucciati\orcid{0000-0002-9336-7551}\inst{\ref{aff7}}
\and S.~Davini\orcid{0000-0003-3269-1718}\inst{\ref{aff19}}
\and F.~De~Paolis\orcid{0000-0001-6460-7563}\inst{\ref{aff138},\ref{aff139},\ref{aff140}}
\and G.~Desprez\orcid{0000-0001-8325-1742}\inst{\ref{aff10}}
\and A.~D\'iaz-S\'anchez\orcid{0000-0003-0748-4768}\inst{\ref{aff141}}
\and J.~J.~Diaz\orcid{0000-0003-2101-1078}\inst{\ref{aff34}}
\and S.~Di~Domizio\orcid{0000-0003-2863-5895}\inst{\ref{aff18},\ref{aff19}}
\and J.~M.~Diego\orcid{0000-0001-9065-3926}\inst{\ref{aff142}}
\and P.-A.~Duc\orcid{0000-0003-3343-6284}\inst{\ref{aff2}}
\and M.~Y.~Elkhashab\orcid{0000-0001-9306-2603}\inst{\ref{aff13},\ref{aff14},\ref{aff135},\ref{aff12}}
\and A.~Finoguenov\orcid{0000-0002-4606-5403}\inst{\ref{aff69}}
\and A.~Fontana\orcid{0000-0003-3820-2823}\inst{\ref{aff30}}
\and F.~Fontanot\orcid{0000-0003-4744-0188}\inst{\ref{aff13},\ref{aff12}}
\and A.~Franco\orcid{0000-0002-4761-366X}\inst{\ref{aff139},\ref{aff138},\ref{aff140}}
\and K.~Ganga\orcid{0000-0001-8159-8208}\inst{\ref{aff81}}
\and J.~Garc\'ia-Bellido\orcid{0000-0002-9370-8360}\inst{\ref{aff126}}
\and T.~Gasparetto\orcid{0000-0002-7913-4866}\inst{\ref{aff30}}
\and V.~Gautard\inst{\ref{aff143}}
\and E.~Gaztanaga\orcid{0000-0001-9632-0815}\inst{\ref{aff106},\ref{aff104},\ref{aff144}}
\and F.~Giacomini\orcid{0000-0002-3129-2814}\inst{\ref{aff16}}
\and F.~Gianotti\orcid{0000-0003-4666-119X}\inst{\ref{aff7}}
\and A.~H.~Gonzalez\orcid{0000-0002-0933-8601}\inst{\ref{aff145}}
\and G.~Gozaliasl\orcid{0000-0002-0236-919X}\inst{\ref{aff146},\ref{aff69}}
\and M.~Guidi\orcid{0000-0001-9408-1101}\inst{\ref{aff8},\ref{aff7}}
\and C.~M.~Gutierrez\orcid{0000-0001-7854-783X}\inst{\ref{aff147}}
\and A.~Hall\orcid{0000-0002-3139-8651}\inst{\ref{aff35}}
\and S.~Hemmati\orcid{0000-0003-2226-5395}\inst{\ref{aff148}}
\and C.~Hern\'andez-Monteagudo\orcid{0000-0001-5471-9166}\inst{\ref{aff97},\ref{aff34}}
\and H.~Hildebrandt\orcid{0000-0002-9814-3338}\inst{\ref{aff149}}
\and J.~Hjorth\orcid{0000-0002-4571-2306}\inst{\ref{aff89}}
\and J.~J.~E.~Kajava\orcid{0000-0002-3010-8333}\inst{\ref{aff150},\ref{aff151}}
\and Y.~Kang\orcid{0009-0000-8588-7250}\inst{\ref{aff48}}
\and V.~Kansal\orcid{0000-0002-4008-6078}\inst{\ref{aff152},\ref{aff153}}
\and D.~Karagiannis\orcid{0000-0002-4927-0816}\inst{\ref{aff113},\ref{aff154}}
\and K.~Kiiveri\inst{\ref{aff67}}
\and J.~Kim\orcid{0000-0003-2776-2761}\inst{\ref{aff118}}
\and C.~C.~Kirkpatrick\inst{\ref{aff67}}
\and S.~Kruk\orcid{0000-0001-8010-8879}\inst{\ref{aff38}}
\and J.~Le~Graet\orcid{0000-0001-6523-7971}\inst{\ref{aff51}}
\and L.~Legrand\orcid{0000-0003-0610-5252}\inst{\ref{aff155},\ref{aff156}}
\and M.~Lembo\orcid{0000-0002-5271-5070}\inst{\ref{aff85},\ref{aff113},\ref{aff114}}
\and F.~Lepori\orcid{0009-0000-5061-7138}\inst{\ref{aff157}}
\and G.~Leroy\orcid{0009-0004-2523-4425}\inst{\ref{aff158},\ref{aff80}}
\and G.~F.~Lesci\orcid{0000-0002-4607-2830}\inst{\ref{aff79},\ref{aff7}}
\and J.~Lesgourgues\orcid{0000-0001-7627-353X}\inst{\ref{aff159}}
\and T.~I.~Liaudat\orcid{0000-0002-9104-314X}\inst{\ref{aff160}}
\and A.~Loureiro\orcid{0000-0002-4371-0876}\inst{\ref{aff161},\ref{aff5}}
\and J.~Macias-Perez\orcid{0000-0002-5385-2763}\inst{\ref{aff162}}
\and M.~Magliocchetti\orcid{0000-0001-9158-4838}\inst{\ref{aff52}}
\and E.~A.~Magnier\orcid{0000-0002-7965-2815}\inst{\ref{aff32}}
\and F.~Mannucci\orcid{0000-0002-4803-2381}\inst{\ref{aff163}}
\and C.~J.~A.~P.~Martins\orcid{0000-0002-4886-9261}\inst{\ref{aff164},\ref{aff165}}
\and L.~Maurin\orcid{0000-0002-8406-0857}\inst{\ref{aff49}}
\and C.~J.~R.~McPartland\orcid{0000-0003-0639-025X}\inst{\ref{aff63},\ref{aff110}}
\and M.~Miluzio\inst{\ref{aff38},\ref{aff166}}
\and P.~Monaco\orcid{0000-0003-2083-7564}\inst{\ref{aff135},\ref{aff13},\ref{aff14},\ref{aff12}}
\and C.~Moretti\orcid{0000-0003-3314-8936}\inst{\ref{aff13},\ref{aff12},\ref{aff14},\ref{aff15}}
\and G.~Morgante\inst{\ref{aff7}}
\and K.~Naidoo\orcid{0000-0002-9182-1802}\inst{\ref{aff144},\ref{aff66}}
\and A.~Navarro-Alsina\orcid{0000-0002-3173-2592}\inst{\ref{aff77}}
\and S.~Nesseris\orcid{0000-0002-0567-0324}\inst{\ref{aff126}}
\and D.~Paoletti\orcid{0000-0003-4761-6147}\inst{\ref{aff7},\ref{aff54}}
\and F.~Passalacqua\orcid{0000-0002-8606-4093}\inst{\ref{aff98},\ref{aff50}}
\and K.~Paterson\orcid{0000-0001-8340-3486}\inst{\ref{aff64}}
\and L.~Patrizii\inst{\ref{aff16}}
\and A.~Pisani\orcid{0000-0002-6146-4437}\inst{\ref{aff51}}
\and D.~Potter\orcid{0000-0002-0757-5195}\inst{\ref{aff157}}
\and S.~Quai\orcid{0000-0002-0449-8163}\inst{\ref{aff79},\ref{aff7}}
\and M.~Radovich\orcid{0000-0002-3585-866X}\inst{\ref{aff17}}
\and G.~Rodighiero\orcid{0000-0002-9415-2296}\inst{\ref{aff98},\ref{aff17}}
\and S.~Sacquegna\orcid{0000-0002-8433-6630}\inst{\ref{aff167}}
\and M.~Sahl\'en\orcid{0000-0003-0973-4804}\inst{\ref{aff168}}
\and D.~B.~Sanders\orcid{0000-0002-1233-9998}\inst{\ref{aff32}}
\and E.~Sarpa\orcid{0000-0002-1256-655X}\inst{\ref{aff15},\ref{aff125},\ref{aff14}}
\and A.~Schneider\orcid{0000-0001-7055-8104}\inst{\ref{aff157}}
\and D.~Sciotti\orcid{0009-0008-4519-2620}\inst{\ref{aff30},\ref{aff78}}
\and E.~Sellentin\inst{\ref{aff169},\ref{aff26}}
\and L.~C.~Smith\orcid{0000-0002-3259-2771}\inst{\ref{aff170}}
\and J.~G.~Sorce\orcid{0000-0002-2307-2432}\inst{\ref{aff171},\ref{aff49}}
\and S.~A.~Stanford\orcid{0000-0003-0122-0841}\inst{\ref{aff172}}
\and K.~Tanidis\orcid{0000-0001-9843-5130}\inst{\ref{aff118}}
\and C.~Tao\orcid{0000-0001-7961-8177}\inst{\ref{aff51}}
\and G.~Testera\inst{\ref{aff19}}
\and R.~Teyssier\orcid{0000-0001-7689-0933}\inst{\ref{aff173}}
\and S.~Tosi\orcid{0000-0002-7275-9193}\inst{\ref{aff18},\ref{aff19},\ref{aff11}}
\and A.~Troja\orcid{0000-0003-0239-4595}\inst{\ref{aff98},\ref{aff50}}
\and M.~Tucci\inst{\ref{aff48}}
\and C.~Valieri\inst{\ref{aff16}}
\and A.~Venhola\orcid{0000-0001-6071-4564}\inst{\ref{aff174}}
\and D.~Vergani\orcid{0000-0003-0898-2216}\inst{\ref{aff7}}
\and G.~Verza\orcid{0000-0002-1886-8348}\inst{\ref{aff175}}
\and P.~Vielzeuf\orcid{0000-0003-2035-9339}\inst{\ref{aff51}}
\and N.~A.~Walton\orcid{0000-0003-3983-8778}\inst{\ref{aff170}}}
										   
%%%% please do not edit the affiliation list -- contact ECEB Bureau for changes
\institute{Department of Physics and Astronomy, University of British Columbia, Vancouver, BC V6T 1Z1, Canada\label{aff1}
\and
Universit\'e de Strasbourg, CNRS, Observatoire astronomique de Strasbourg, UMR 7550, 67000 Strasbourg, France\label{aff2}
\and
Herzberg Astronomy and Astrophysics Research Centre, 5071 W. Saanich Rd. Victoria, BC, V9E 2E7, Canada\label{aff3}
\and
Department of Physics and Atmospheric Science, Dalhousie University, Halifax, NS, B3H 4R2, Canada\label{aff4}
\and
Astrophysics Group, Blackett Laboratory, Imperial College London, London SW7 2AZ, UK\label{aff5}
\and
School of Physics and Astronomy, Cardiff University, The Parade, Cardiff, CF24 3AA, UK\label{aff6}
\and
INAF-Osservatorio di Astrofisica e Scienza dello Spazio di Bologna, Via Piero Gobetti 93/3, 40129 Bologna, Italy\label{aff7}
\and
Dipartimento di Fisica e Astronomia, Universit\`a di Bologna, Via Gobetti 93/2, 40129 Bologna, Italy\label{aff8}
\and
SRON Netherlands Institute for Space Research, Landleven 12, 9747 AD, Groningen, The Netherlands\label{aff9}
\and
Kapteyn Astronomical Institute, University of Groningen, PO Box 800, 9700 AV Groningen, The Netherlands\label{aff10}
\and
INAF-Osservatorio Astronomico di Brera, Via Brera 28, 20122 Milano, Italy\label{aff11}
\and
IFPU, Institute for Fundamental Physics of the Universe, via Beirut 2, 34151 Trieste, Italy\label{aff12}
\and
INAF-Osservatorio Astronomico di Trieste, Via G. B. Tiepolo 11, 34143 Trieste, Italy\label{aff13}
\and
INFN, Sezione di Trieste, Via Valerio 2, 34127 Trieste TS, Italy\label{aff14}
\and
SISSA, International School for Advanced Studies, Via Bonomea 265, 34136 Trieste TS, Italy\label{aff15}
\and
INFN-Sezione di Bologna, Viale Berti Pichat 6/2, 40127 Bologna, Italy\label{aff16}
\and
INAF-Osservatorio Astronomico di Padova, Via dell'Osservatorio 5, 35122 Padova, Italy\label{aff17}
\and
Dipartimento di Fisica, Universit\`a di Genova, Via Dodecaneso 33, 16146, Genova, Italy\label{aff18}
\and
INFN-Sezione di Genova, Via Dodecaneso 33, 16146, Genova, Italy\label{aff19}
\and
Department of Physics "E. Pancini", University Federico II, Via Cinthia 6, 80126, Napoli, Italy\label{aff20}
\and
INAF-Osservatorio Astronomico di Capodimonte, Via Moiariello 16, 80131 Napoli, Italy\label{aff21}
\and
Dipartimento di Fisica, Universit\`a degli Studi di Torino, Via P. Giuria 1, 10125 Torino, Italy\label{aff22}
\and
INFN-Sezione di Torino, Via P. Giuria 1, 10125 Torino, Italy\label{aff23}
\and
INAF-Osservatorio Astrofisico di Torino, Via Osservatorio 20, 10025 Pino Torinese (TO), Italy\label{aff24}
\and
European Space Agency/ESTEC, Keplerlaan 1, 2201 AZ Noordwijk, The Netherlands\label{aff25}
\and
Leiden Observatory, Leiden University, Einsteinweg 55, 2333 CC Leiden, The Netherlands\label{aff26}
\and
INAF-IASF Milano, Via Alfonso Corti 12, 20133 Milano, Italy\label{aff27}
\and
Centro de Investigaciones Energ\'eticas, Medioambientales y Tecnol\'ogicas (CIEMAT), Avenida Complutense 40, 28040 Madrid, Spain\label{aff28}
\and
Port d'Informaci\'{o} Cient\'{i}fica, Campus UAB, C. Albareda s/n, 08193 Bellaterra (Barcelona), Spain\label{aff29}
\and
INAF-Osservatorio Astronomico di Roma, Via Frascati 33, 00078 Monteporzio Catone, Italy\label{aff30}
\and
INFN section of Naples, Via Cinthia 6, 80126, Napoli, Italy\label{aff31}
\and
Institute for Astronomy, University of Hawaii, 2680 Woodlawn Drive, Honolulu, HI 96822, USA\label{aff32}
\and
Dipartimento di Fisica e Astronomia "Augusto Righi" - Alma Mater Studiorum Universit\`a di Bologna, Viale Berti Pichat 6/2, 40127 Bologna, Italy\label{aff33}
\and
Instituto de Astrof\'{\i}sica de Canarias, V\'{\i}a L\'actea, 38205 La Laguna, Tenerife, Spain\label{aff34}
\and
Institute for Astronomy, University of Edinburgh, Royal Observatory, Blackford Hill, Edinburgh EH9 3HJ, UK\label{aff35}
\and
Jodrell Bank Centre for Astrophysics, Department of Physics and Astronomy, University of Manchester, Oxford Road, Manchester M13 9PL, UK\label{aff36}
\and
European Space Agency/ESRIN, Largo Galileo Galilei 1, 00044 Frascati, Roma, Italy\label{aff37}
\and
ESAC/ESA, Camino Bajo del Castillo, s/n., Urb. Villafranca del Castillo, 28692 Villanueva de la Ca\~nada, Madrid, Spain\label{aff38}
\and
Universit\'e Claude Bernard Lyon 1, CNRS/IN2P3, IP2I Lyon, UMR 5822, Villeurbanne, F-69100, France\label{aff39}
\and
Aix-Marseille Universit\'e, CNRS, CNES, LAM, Marseille, France\label{aff40}
\and
Institut de Ci\`{e}ncies del Cosmos (ICCUB), Universitat de Barcelona (IEEC-UB), Mart\'{i} i Franqu\`{e}s 1, 08028 Barcelona, Spain\label{aff41}
\and
Instituci\'o Catalana de Recerca i Estudis Avan\c{c}ats (ICREA), Passeig de Llu\'{\i}s Companys 23, 08010 Barcelona, Spain\label{aff42}
\and
Institut de Ciencies de l'Espai (IEEC-CSIC), Campus UAB, Carrer de Can Magrans, s/n Cerdanyola del Vall\'es, 08193 Barcelona, Spain\label{aff43}
\and
UCB Lyon 1, CNRS/IN2P3, IUF, IP2I Lyon, 4 rue Enrico Fermi, 69622 Villeurbanne, France\label{aff44}
\and
Mullard Space Science Laboratory, University College London, Holmbury St Mary, Dorking, Surrey RH5 6NT, UK\label{aff45}
\and
Departamento de F\'isica, Faculdade de Ci\^encias, Universidade de Lisboa, Edif\'icio C8, Campo Grande, PT1749-016 Lisboa, Portugal\label{aff46}
\and
Instituto de Astrof\'isica e Ci\^encias do Espa\c{c}o, Faculdade de Ci\^encias, Universidade de Lisboa, Campo Grande, 1749-016 Lisboa, Portugal\label{aff47}
\and
Department of Astronomy, University of Geneva, ch. d'Ecogia 16, 1290 Versoix, Switzerland\label{aff48}
\and
Universit\'e Paris-Saclay, CNRS, Institut d'astrophysique spatiale, 91405, Orsay, France\label{aff49}
\and
INFN-Padova, Via Marzolo 8, 35131 Padova, Italy\label{aff50}
\and
Aix-Marseille Universit\'e, CNRS/IN2P3, CPPM, Marseille, France\label{aff51}
\and
INAF-Istituto di Astrofisica e Planetologia Spaziali, via del Fosso del Cavaliere, 100, 00100 Roma, Italy\label{aff52}
\and
Space Science Data Center, Italian Space Agency, via del Politecnico snc, 00133 Roma, Italy\label{aff53}
\and
INFN-Bologna, Via Irnerio 46, 40126 Bologna, Italy\label{aff54}
\and
University Observatory, LMU Faculty of Physics, Scheinerstrasse 1, 81679 Munich, Germany\label{aff55}
\and
Max Planck Institute for Extraterrestrial Physics, Giessenbachstr. 1, 85748 Garching, Germany\label{aff56}
\and
Universit\"ats-Sternwarte M\"unchen, Fakult\"at f\"ur Physik, Ludwig-Maximilians-Universit\"at M\"unchen, Scheinerstrasse 1, 81679 M\"unchen, Germany\label{aff57}
\and
Institute of Theoretical Astrophysics, University of Oslo, P.O. Box 1029 Blindern, 0315 Oslo, Norway\label{aff58}
\and
Jet Propulsion Laboratory, California Institute of Technology, 4800 Oak Grove Drive, Pasadena, CA, 91109, USA\label{aff59}
\and
Department of Physics, Lancaster University, Lancaster, LA1 4YB, UK\label{aff60}
\and
Felix Hormuth Engineering, Goethestr. 17, 69181 Leimen, Germany\label{aff61}
\and
Technical University of Denmark, Elektrovej 327, 2800 Kgs. Lyngby, Denmark\label{aff62}
\and
Cosmic Dawn Center (DAWN), Denmark\label{aff63}
\and
Max-Planck-Institut f\"ur Astronomie, K\"onigstuhl 17, 69117 Heidelberg, Germany\label{aff64}
\and
NASA Goddard Space Flight Center, Greenbelt, MD 20771, USA\label{aff65}
\and
Department of Physics and Astronomy, University College London, Gower Street, London WC1E 6BT, UK\label{aff66}
\and
Department of Physics and Helsinki Institute of Physics, Gustaf H\"allstr\"omin katu 2, University of Helsinki, 00014 Helsinki, Finland\label{aff67}
\and
Universit\'e de Gen\`eve, D\'epartement de Physique Th\'eorique and Centre for Astroparticle Physics, 24 quai Ernest-Ansermet, CH-1211 Gen\`eve 4, Switzerland\label{aff68}
\and
Department of Physics, P.O. Box 64, University of Helsinki, 00014 Helsinki, Finland\label{aff69}
\and
Helsinki Institute of Physics, Gustaf H{\"a}llstr{\"o}min katu 2, University of Helsinki, 00014 Helsinki, Finland\label{aff70}
\and
Laboratoire d'etude de l'Univers et des phenomenes eXtremes, Observatoire de Paris, Universit\'e PSL, Sorbonne Universit\'e, CNRS, 92190 Meudon, France\label{aff71}
\and
SKAO, Jodrell Bank, Lower Withington, Macclesfield SK11 9FT, United Kingdom\label{aff72}
\and
Centre de Calcul de l'IN2P3/CNRS, 21 avenue Pierre de Coubertin 69627 Villeurbanne Cedex, France\label{aff73}
\and
Dipartimento di Fisica "Aldo Pontremoli", Universit\`a degli Studi di Milano, Via Celoria 16, 20133 Milano, Italy\label{aff74}
\and
INFN-Sezione di Milano, Via Celoria 16, 20133 Milano, Italy\label{aff75}
\and
University of Applied Sciences and Arts of Northwestern Switzerland, School of Computer Science, 5210 Windisch, Switzerland\label{aff76}
\and
Universit\"at Bonn, Argelander-Institut f\"ur Astronomie, Auf dem H\"ugel 71, 53121 Bonn, Germany\label{aff77}
\and
INFN-Sezione di Roma, Piazzale Aldo Moro, 2 - c/o Dipartimento di Fisica, Edificio G. Marconi, 00185 Roma, Italy\label{aff78}
\and
Dipartimento di Fisica e Astronomia "Augusto Righi" - Alma Mater Studiorum Universit\`a di Bologna, via Piero Gobetti 93/2, 40129 Bologna, Italy\label{aff79}
\and
Department of Physics, Institute for Computational Cosmology, Durham University, South Road, Durham, DH1 3LE, UK\label{aff80}
\and
Universit\'e Paris Cit\'e, CNRS, Astroparticule et Cosmologie, 75013 Paris, France\label{aff81}
\and
CNRS-UCB International Research Laboratory, Centre Pierre Bin\'etruy, IRL2007, CPB-IN2P3, Berkeley, USA\label{aff82}
\and
University of Applied Sciences and Arts of Northwestern Switzerland, School of Engineering, 5210 Windisch, Switzerland\label{aff83}
\and
Institut d'Astrophysique de Paris, 98bis Boulevard Arago, 75014, Paris, France\label{aff84}
\and
Institut d'Astrophysique de Paris, UMR 7095, CNRS, and Sorbonne Universit\'e, 98 bis boulevard Arago, 75014 Paris, France\label{aff85}
\and
Institute of Physics, Laboratory of Astrophysics, Ecole Polytechnique F\'ed\'erale de Lausanne (EPFL), Observatoire de Sauverny, 1290 Versoix, Switzerland\label{aff86}
\and
Telespazio UK S.L. for European Space Agency (ESA), Camino bajo del Castillo, s/n, Urbanizacion Villafranca del Castillo, Villanueva de la Ca\~nada, 28692 Madrid, Spain\label{aff87}
\and
Institut de F\'{i}sica d'Altes Energies (IFAE), The Barcelona Institute of Science and Technology, Campus UAB, 08193 Bellaterra (Barcelona), Spain\label{aff88}
\and
DARK, Niels Bohr Institute, University of Copenhagen, Jagtvej 155, 2200 Copenhagen, Denmark\label{aff89}
\and
Waterloo Centre for Astrophysics, University of Waterloo, Waterloo, Ontario N2L 3G1, Canada\label{aff90}
\and
Department of Physics and Astronomy, University of Waterloo, Waterloo, Ontario N2L 3G1, Canada\label{aff91}
\and
Perimeter Institute for Theoretical Physics, Waterloo, Ontario N2L 2Y5, Canada\label{aff92}
\and
Universit\'e Paris-Saclay, Universit\'e Paris Cit\'e, CEA, CNRS, AIM, 91191, Gif-sur-Yvette, France\label{aff93}
\and
Centre National d'Etudes Spatiales -- Centre spatial de Toulouse, 18 avenue Edouard Belin, 31401 Toulouse Cedex 9, France\label{aff94}
\and
Institute of Space Science, Str. Atomistilor, nr. 409 M\u{a}gurele, Ilfov, 077125, Romania\label{aff95}
\and
Consejo Superior de Investigaciones Cientificas, Calle Serrano 117, 28006 Madrid, Spain\label{aff96}
\and
Universidad de La Laguna, Departamento de Astrof\'{\i}sica, 38206 La Laguna, Tenerife, Spain\label{aff97}
\and
Dipartimento di Fisica e Astronomia "G. Galilei", Universit\`a di Padova, Via Marzolo 8, 35131 Padova, Italy\label{aff98}
\and
Institut f\"ur Theoretische Physik, University of Heidelberg, Philosophenweg 16, 69120 Heidelberg, Germany\label{aff99}
\and
Institut de Recherche en Astrophysique et Plan\'etologie (IRAP), Universit\'e de Toulouse, CNRS, UPS, CNES, 14 Av. Edouard Belin, 31400 Toulouse, France\label{aff100}
\and
Universit\'e St Joseph; Faculty of Sciences, Beirut, Lebanon\label{aff101}
\and
Departamento de F\'isica, FCFM, Universidad de Chile, Blanco Encalada 2008, Santiago, Chile\label{aff102}
\and
Universit\"at Innsbruck, Institut f\"ur Astro- und Teilchenphysik, Technikerstr. 25/8, 6020 Innsbruck, Austria\label{aff103}
\and
Institut d'Estudis Espacials de Catalunya (IEEC),  Edifici RDIT, Campus UPC, 08860 Castelldefels, Barcelona, Spain\label{aff104}
\and
Satlantis, University Science Park, Sede Bld 48940, Leioa-Bilbao, Spain\label{aff105}
\and
Institute of Space Sciences (ICE, CSIC), Campus UAB, Carrer de Can Magrans, s/n, 08193 Barcelona, Spain\label{aff106}
\and
Infrared Processing and Analysis Center, California Institute of Technology, Pasadena, CA 91125, USA\label{aff107}
\and
Instituto de Astrof\'isica e Ci\^encias do Espa\c{c}o, Faculdade de Ci\^encias, Universidade de Lisboa, Tapada da Ajuda, 1349-018 Lisboa, Portugal\label{aff108}
\and
Cosmic Dawn Center (DAWN)\label{aff109}
\and
Niels Bohr Institute, University of Copenhagen, Jagtvej 128, 2200 Copenhagen, Denmark\label{aff110}
\and
Universidad Polit\'ecnica de Cartagena, Departamento de Electr\'onica y Tecnolog\'ia de Computadoras,  Plaza del Hospital 1, 30202 Cartagena, Spain\label{aff111}
\and
Astronomisches Rechen-Institut, Zentrum f\"ur Astronomie der Universit\"at Heidelberg, M\"onchhofstr. 12-14, 69120 Heidelberg, Germany\label{aff112}
\and
Dipartimento di Fisica e Scienze della Terra, Universit\`a degli Studi di Ferrara, Via Giuseppe Saragat 1, 44122 Ferrara, Italy\label{aff113}
\and
Istituto Nazionale di Fisica Nucleare, Sezione di Ferrara, Via Giuseppe Saragat 1, 44122 Ferrara, Italy\label{aff114}
\and
INAF, Istituto di Radioastronomia, Via Piero Gobetti 101, 40129 Bologna, Italy\label{aff115}
\and
Astronomical Observatory of the Autonomous Region of the Aosta Valley (OAVdA), Loc. Lignan 39, I-11020, Nus (Aosta Valley), Italy\label{aff116}
\and
Universit\'e C\^{o}te d'Azur, Observatoire de la C\^{o}te d'Azur, CNRS, Laboratoire Lagrange, Bd de l'Observatoire, CS 34229, 06304 Nice cedex 4, France\label{aff117}
\and
Department of Physics, Oxford University, Keble Road, Oxford OX1 3RH, UK\label{aff118}
\and
Instituto de Astrof\'isica de Canarias (IAC); Departamento de Astrof\'isica, Universidad de La Laguna (ULL), 38200, La Laguna, Tenerife, Spain\label{aff119}
\and
Universit\'e PSL, Observatoire de Paris, Sorbonne Universit\'e, CNRS, LERMA, 75014, Paris, France\label{aff120}
\and
Universit\'e Paris-Cit\'e, 5 Rue Thomas Mann, 75013, Paris, France\label{aff121}
\and
Dipartimento di Fisica, Sapienza Universit\`a di Roma, Piazzale Aldo Moro 2, 00185 Roma, Italy\label{aff122}
\and
Aurora Technology for European Space Agency (ESA), Camino bajo del Castillo, s/n, Urbanizacion Villafranca del Castillo, Villanueva de la Ca\~nada, 28692 Madrid, Spain\label{aff123}
\and
ICL, Junia, Universit\'e Catholique de Lille, LITL, 59000 Lille, France\label{aff124}
\and
ICSC - Centro Nazionale di Ricerca in High Performance Computing, Big Data e Quantum Computing, Via Magnanelli 2, Bologna, Italy\label{aff125}
\and
Instituto de F\'isica Te\'orica UAM-CSIC, Campus de Cantoblanco, 28049 Madrid, Spain\label{aff126}
\and
CERCA/ISO, Department of Physics, Case Western Reserve University, 10900 Euclid Avenue, Cleveland, OH 44106, USA\label{aff127}
\and
Technical University of Munich, TUM School of Natural Sciences, Physics Department, James-Franck-Str.~1, 85748 Garching, Germany\label{aff128}
\and
Max-Planck-Institut f\"ur Astrophysik, Karl-Schwarzschild-Str.~1, 85748 Garching, Germany\label{aff129}
\and
Laboratoire Univers et Th\'eorie, Observatoire de Paris, Universit\'e PSL, Universit\'e Paris Cit\'e, CNRS, 92190 Meudon, France\label{aff130}
\and
Departamento de F{\'\i}sica Fundamental. Universidad de Salamanca. Plaza de la Merced s/n. 37008 Salamanca, Spain\label{aff131}
\and
Center for Data-Driven Discovery, Kavli IPMU (WPI), UTIAS, The University of Tokyo, Kashiwa, Chiba 277-8583, Japan\label{aff132}
\and
Ludwig-Maximilians-University, Schellingstrasse 4, 80799 Munich, Germany\label{aff133}
\and
Max-Planck-Institut f\"ur Physik, Boltzmannstr. 8, 85748 Garching, Germany\label{aff134}
\and
Dipartimento di Fisica - Sezione di Astronomia, Universit\`a di Trieste, Via Tiepolo 11, 34131 Trieste, Italy\label{aff135}
\and
California Institute of Technology, 1200 E California Blvd, Pasadena, CA 91125, USA\label{aff136}
\and
Department of Physics \& Astronomy, University of California Irvine, Irvine CA 92697, USA\label{aff137}
\and
Department of Mathematics and Physics E. De Giorgi, University of Salento, Via per Arnesano, CP-I93, 73100, Lecce, Italy\label{aff138}
\and
INFN, Sezione di Lecce, Via per Arnesano, CP-193, 73100, Lecce, Italy\label{aff139}
\and
INAF-Sezione di Lecce, c/o Dipartimento Matematica e Fisica, Via per Arnesano, 73100, Lecce, Italy\label{aff140}
\and
Departamento F\'isica Aplicada, Universidad Polit\'ecnica de Cartagena, Campus Muralla del Mar, 30202 Cartagena, Murcia, Spain\label{aff141}
\and
Instituto de F\'isica de Cantabria, Edificio Juan Jord\'a, Avenida de los Castros, 39005 Santander, Spain\label{aff142}
\and
CEA Saclay, DFR/IRFU, Service d'Astrophysique, Bat. 709, 91191 Gif-sur-Yvette, France\label{aff143}
\and
Institute of Cosmology and Gravitation, University of Portsmouth, Portsmouth PO1 3FX, UK\label{aff144}
\and
Department of Astronomy, University of Florida, Bryant Space Science Center, Gainesville, FL 32611, USA\label{aff145}
\and
Department of Computer Science, Aalto University, PO Box 15400, Espoo, FI-00 076, Finland\label{aff146}
\and
Instituto de Astrof\'\i sica de Canarias, c/ Via Lactea s/n, La Laguna 38200, Spain. Departamento de Astrof\'\i sica de la Universidad de La Laguna, Avda. Francisco Sanchez, La Laguna, 38200, Spain\label{aff147}
\and
Caltech/IPAC, 1200 E. California Blvd., Pasadena, CA 91125, USA\label{aff148}
\and
Ruhr University Bochum, Faculty of Physics and Astronomy, Astronomical Institute (AIRUB), German Centre for Cosmological Lensing (GCCL), 44780 Bochum, Germany\label{aff149}
\and
Department of Physics and Astronomy, Vesilinnantie 5, University of Turku, 20014 Turku, Finland\label{aff150}
\and
Serco for European Space Agency (ESA), Camino bajo del Castillo, s/n, Urbanizacion Villafranca del Castillo, Villanueva de la Ca\~nada, 28692 Madrid, Spain\label{aff151}
\and
ARC Centre of Excellence for Dark Matter Particle Physics, Melbourne, Australia\label{aff152}
\and
Centre for Astrophysics \& Supercomputing, Swinburne University of Technology,  Hawthorn, Victoria 3122, Australia\label{aff153}
\and
Department of Physics and Astronomy, University of the Western Cape, Bellville, Cape Town, 7535, South Africa\label{aff154}
\and
DAMTP, Centre for Mathematical Sciences, Wilberforce Road, Cambridge CB3 0WA, UK\label{aff155}
\and
Kavli Institute for Cosmology Cambridge, Madingley Road, Cambridge, CB3 0HA, UK\label{aff156}
\and
Department of Astrophysics, University of Zurich, Winterthurerstrasse 190, 8057 Zurich, Switzerland\label{aff157}
\and
Department of Physics, Centre for Extragalactic Astronomy, Durham University, South Road, Durham, DH1 3LE, UK\label{aff158}
\and
Institute for Theoretical Particle Physics and Cosmology (TTK), RWTH Aachen University, 52056 Aachen, Germany\label{aff159}
\and
IRFU, CEA, Universit\'e Paris-Saclay 91191 Gif-sur-Yvette Cedex, France\label{aff160}
\and
Oskar Klein Centre for Cosmoparticle Physics, Department of Physics, Stockholm University, Stockholm, SE-106 91, Sweden\label{aff161}
\and
Univ. Grenoble Alpes, CNRS, Grenoble INP, LPSC-IN2P3, 53, Avenue des Martyrs, 38000, Grenoble, France\label{aff162}
\and
INAF-Osservatorio Astrofisico di Arcetri, Largo E. Fermi 5, 50125, Firenze, Italy\label{aff163}
\and
Centro de Astrof\'{\i}sica da Universidade do Porto, Rua das Estrelas, 4150-762 Porto, Portugal\label{aff164}
\and
Instituto de Astrof\'isica e Ci\^encias do Espa\c{c}o, Universidade do Porto, CAUP, Rua das Estrelas, PT4150-762 Porto, Portugal\label{aff165}
\and
HE Space for European Space Agency (ESA), Camino bajo del Castillo, s/n, Urbanizacion Villafranca del Castillo, Villanueva de la Ca\~nada, 28692 Madrid, Spain\label{aff166}
\and
INAF - Osservatorio Astronomico d'Abruzzo, Via Maggini, 64100, Teramo, Italy\label{aff167}
\and
Theoretical astrophysics, Department of Physics and Astronomy, Uppsala University, Box 516, 751 37 Uppsala, Sweden\label{aff168}
\and
Mathematical Institute, University of Leiden, Einsteinweg 55, 2333 CA Leiden, The Netherlands\label{aff169}
\and
Institute of Astronomy, University of Cambridge, Madingley Road, Cambridge CB3 0HA, UK\label{aff170}
\and
Univ. Lille, CNRS, Centrale Lille, UMR 9189 CRIStAL, 59000 Lille, France\label{aff171}
\and
Department of Physics and Astronomy, University of California, Davis, CA 95616, USA\label{aff172}
\and
Department of Astrophysical Sciences, Peyton Hall, Princeton University, Princeton, NJ 08544, USA\label{aff173}
\and
Space physics and astronomy research unit, University of Oulu, Pentti Kaiteran katu 1, FI-90014 Oulu, Finland\label{aff174}
\and
Center for Computational Astrophysics, Flatiron Institute, 162 5th Avenue, 10010, New York, NY, USA\label{aff175}}

\abstract{
%Context
The first Euclid Quick Data Release (Q1) contains millions of galaxies with excellent optical and near-infrared (IR) coverage.
%Aims
To complement this dataset, we investigate the average far-IR properties of \Euclid-selected main sequence (MS) galaxies using existing \Herschel and SCUBA-2 data. 
%Methods
We used 17.6\,deg$^2$ (2.4\,deg$^2$) of overlapping \Herschel (SCUBA-2) data, containing 2.6 million (240\,000) MS galaxies. We binned the \Euclid catalogue by stellar mass and photometric redshift and performed a stacking analysis following {\tt SimStack}, which accounts for galaxy clustering and bin-to-bin correlations.
%Results
We detected stacked far-IR flux densities across a significant fraction of the bins. We fitted modified blackbody spectral energy distributions in each bin and derived mean dust temperatures ($T_{\rm d}$), dust masses ($M_{\rm d}$), and star-formation rates (SFRs). We find similar mean SFRs compared to the \Euclid catalogue, and we show that the average dust-to-stellar mass ratios decreased from $z\,{\simeq}\,1$ to the present day. Average dust temperatures are largely independent of stellar mass and are well-described by the function $T_2\,{+}\,(T_1\,{-}\,T_2)\,{\rm e}^{-t/\tau}$, where $t$ is the age of the Universe, $T_1\,{=}\,(79.7\pm7.4)\,$K, $T_2\,{=}\,(23.2\pm0.1)\,$K, and $\tau\,{=}\,(1.6\pm0.1)\,$Gyr. We argue that since the dust temperatures converge to a non-zero value below $z\,{=}\,1$, the dust is now primarily heated by the existing cooler and older stellar population, as opposed to hot young stars in star-forming regions at higher redshifts. We show that since the dust temperatures are independent of stellar mass, the correlation between dust temperature and SFR depends on stellar mass. Lastly, we estimate the contribution of the \Euclid catalogue to the cosmic IR background (CIB), finding that it accounts for ${>}\,60\%$ of the CIB at 250, 350, and 500\,$\mu$m.
%Conclusions
As the \Euclid mission progresses, larger catalogues will allow us to probe the far-IR properties of MS galaxies out to higher redshifts and lower stellar masses, potentially recovering the complete CIB.
}
%   
% Provide up to five key words:
%
% from the list in
% https://www.aanda.org/for-authors/latex-issues/information-files#pop}
%
% Add short versions of title and author list for page headings
%
\titlerunning{Euclid Quick Data Release (Q1): Average far-IR properties}
\authorrunning{Euclid Collaboration: R.~Hill et al.}

\maketitle

\section{Introduction}
\label{sec:introduction}

The \Euclid mission \citep{EuclidSkyOverview} will observe 14\,000\,deg$^2$ of extragalactic sky, detecting billions of galaxies at optical wavelengths \citep[with the VIS instrument at 550--900\,nm;][]{EuclidSkyVIS} and near-infrared wavelengths \citep[with the Near-Infrared Spectrometer and Photometer, NISP at 1--2\,$\mu$m;][]{EuclidSkyNISP}. The first Quick Data Release \citep[Q1;][]{Q1cite} provided single-exposure observations covering three deep fields: Euclid Deep Field Fornax (EDF-F); Euclid Deep Field North (EDF-N); and Euclid Deep Field South (EDF-S). Even at the current depths (about magnitude 24.7 in VIS and 23.2 in NISP) the catalogues generated from these observations contain over 10 million galaxies detected in the \Euclid filters \citep{Q1-TP004}.

Spectral energy distributions (SEDs) have been fitted to the \Euclid-selected galaxies, providing robust photometric redshifts, stellar masses, and star-formation rates (SFRs) for the majority of the galaxies \citep{Q1-TP005}.
This catalogue was recently used to constrain the correlation between the stellar mass ($M_{\ast}$) and SFR of star-forming galaxies (known as the galaxy star-forming main sequence, or MS) out to $z\,{=}\,3$ \citep{Q1-SP031}. The correlation is related to universal processes that have been converting cold gas reservoirs into stars since at least $z\,{=}\,6$. Because the bulk of the galaxies in the Universe follow the star-forming MS, determining its evolution is crucial for understanding galaxy evolution in general. For example, it is known that the amplitude of the MS increases with redshift for galaxies of all stellar mass \citep{speagle2014,daddi2022,popesso2023}, implying that the specific SFRs (sSFRs) of all galaxies were higher in the early Universe. There is also a deviation from a linear trend at high stellar mass, meaning that there is a maximum average SFR at a given epoch, and this characteristic bending mass also increases with redshift. This has been attributed to a change in environments suppressing cold gas accretion with redshift and quenching \citep[e.g.][]{dekel2006,daddi2022}.

While optical and near-infrared (IR) light observed from the Earth is primarily sensitive to the stellar emission from galaxies at all redshifts (with the longer wavelengths being weighted to higher-redshift galaxies), far-IR light (from tens to around $1000\,\mu$m; note that this also includes wavelengths often described as submillimetre at the longer end) is sensitive to the thermal emission from warm dust grains in galaxies at all redshifts. Since these dust grains are primarily heated by hot young stars in star-forming regions, there is a tight correlation between the far-IR luminosity (often defined as the integral of the luminosity density between 8 and 1000\,$\mu$m) and the SFR \citep[e.g.][]{Kennicutt1998}, suggesting a connection to the galaxy MS. It is therefore of interest to check the consistency between the SFRs derived from optical and near-IR photometry to those derived (independently) from far-IR photometry, and to see if there is evolution in the dust properties that can provide insight into the processes changing the MS.

There have been many large extragalactic surveys carried out by far-IR and submillimetre observatories, particularly by the \Herschel Photodetector Array Camera and Spectrometer \citep[PACS;][]{poglitsch2010} at 70--160\,$\mu$m, the \Herschel Spectral and Photometric Imaging REceiver \citep[SPIRE;][]{Griffin2010} at 250, 350, and 500\,$\mu$m, and the Submillimetre Common-User Bolometer Array 2 \citep[SCUBA-2;][]{Holland2013} at 450 and 850\,$\mu$m. \Herschel effectively measures the peak of the thermal SEDs of most \Euclid galaxies, providing effective constraints on dust temperatures (which are proportional to the peak frequency), while SCUBA-2 probes the Rayleigh--Jeans tail of the thermal SED, whose slope is related to the dust emissivity index $\beta$. Moreover, \Herschel has surveyed roughly 1000\,deg$^2$ of extragalactic sky (and SCUBA-2 about 10\,deg$^2$), essentially all of which will ultimately overlap with \Euclid by the final data release.

Compared to the exquisite angular resolution of \Euclid (approximately \ang{;;0.2}), the angular resolution of \Herschel and SCUBA-2 is much coarser, ranging between about \ang{;;10} and \ang{;;30}. Far-IR maps therefore do not individually detect most of the galaxies that \Euclid sees, but rather blend these galaxies together into a coherent pattern that traces the large-scale structure. Making progress requires stacking, where the average value of the pixels at the locations of many objects is calculated, rather than trying to measure the flux densities of each object directly from the map. More precisely, this operation calculates the covariance between a catalogue and a map \citep[see e.g.][]{Marsden2009,wang2015}. The \Herschel properties of optical and near-IR catalogues have been investigated via stacking in other fields such as the UKIDSS Ultra-Deep Survey \citep[UDS;][$1\,$deg$^2$]{Viero2013}, the Cosmic Evolution Survey \citep[COSMOS;][$2\,$deg$^2$]{simpson2019,Duivenvoorden2020}, both UDS and COSMOS \citep{koprowski2024}, and the Galaxy And Mass Assembly (GAMA) fields and Stripe 82 region \citep{wang2016}. However, the fields used in these studies are either small and suffer from sample variance to some degree, or the optical catalogues do not go beyond redshift 1. With \Euclid we can dramatically expand these results to include millions of star-forming galaxies across tens of square degrees. Moreover, future \Euclid data releases will continue to overlap with existing \Herschel observations, eventually amounting to a billion galaxies over 1000\,deg$^2$.

We therefore focus on stacking the entire Q1 catalogue on \Herschel and SCUBA-2 maps, making this the largest study yet of this kind. In \cref{sec:data} we describe the \Euclid catalogues and far-IR maps used in the analysis. In \cref{sec:method} we present our stacking pipeline. In \cref{sec:results} we show our results, and in \cref{sec:discussion} we discuss our findings. The paper concludes in \cref{sec:conclusion}. Throughout this paper we assume the cosmological parameters from \citet{planck2016-l06}.

\section{Data}
\label{sec:data}

The \Euclid Q1 release is split into three fields: the EDF-F (12.1\,deg$^2$); the EDF-N (22.9\,deg$^2$); and the EDF-S (28.1\,deg$^2$). Of these three fields, EDF-F and EDF-N have overlapping coverage from SPIRE. The SPIRE field overlapping with the EDF-F is known as `CDFS-SWIRE', and the field overlapping with the EDF-N is known as the `AKARI-NEP'. Here we describe the multiwavelength data in these two fields.

\subsection{\Euclid catalogues and masks}
\label{subsec:euclid_cat}

The \Euclid merging (MER) Q1 catalogue \citep{Q1-TP004} contains the photometry of all VIS- and NISP-detected galaxies, as well as ground-based photometry in the $u$, $g$, $r$, $i$, and $z$ bands from various telescopes (see Tereno et al.\ in prep.\ for details). Here we make use of the \Euclid catalogue described in \citet{Q1-SP031}, which includes additional photometry from the Infrared Array Camera (IRAC) on board the {\it Spitzer\/} Space Telescope \citep{Fazio2004} at 3.6 and 4.5\,$\mu$m. This catalogue also includes refitting of SEDs with the additional IRAC data in order to derive photometric redshifts, stellar masses, and SFRs. The final catalogue contains 2\,884\,906 objects in the EDF-F and 6\,221\,146 objects in the EDF-N. We note that this contains somewhat fewer objects than the full Q1 catalogue, since stars and other image artefacts were removed before cross-matching to IRAC.

Since in this study we are interested in the average properties of MS galaxies, we used the same colour cuts to remove quiescent galaxies. Specifically, galaxies were removed with NUV$\,{-}\,r^+\,{>}\,3(r^+\,{-}\,J)\,{+}\,1$ and NUV$\,{-}\,r^+\,{>}\,3.1$, leaving 1\,318\,898 objects in the EDF-F and 2\,658\,118 objects in the EDF-N.

To ensure an accurate stacking analysis, we must account for the fact that some regions within the \Euclid footprint do not contain any extragalactic objects due to contamination from bright stars. This can be done by calculating a mask that is associated with the \Euclid catalogue, and propagating that mask through all the calculations.

For each of our far-IR images, we therefore calculated the number of \Euclid objects in each pixel, then lightly smoothed the map using a Gaussian kernel. We then defined the \Euclid catalogue mask to be the regions where this smoothed map has a value less than a given threshold, determined visually by ensuring that the mask agreed well with the actual galaxy distribution. Defined this way, the \Euclid mask removes far-IR pixels that have no extragalactic \Euclid objects across a sufficiently large scale.

\subsection{Far-IR imaging}
\label{subsec:spire_im}

\begin{figure*}[htbp!]
    \centering
    \includegraphics[width=\textwidth]{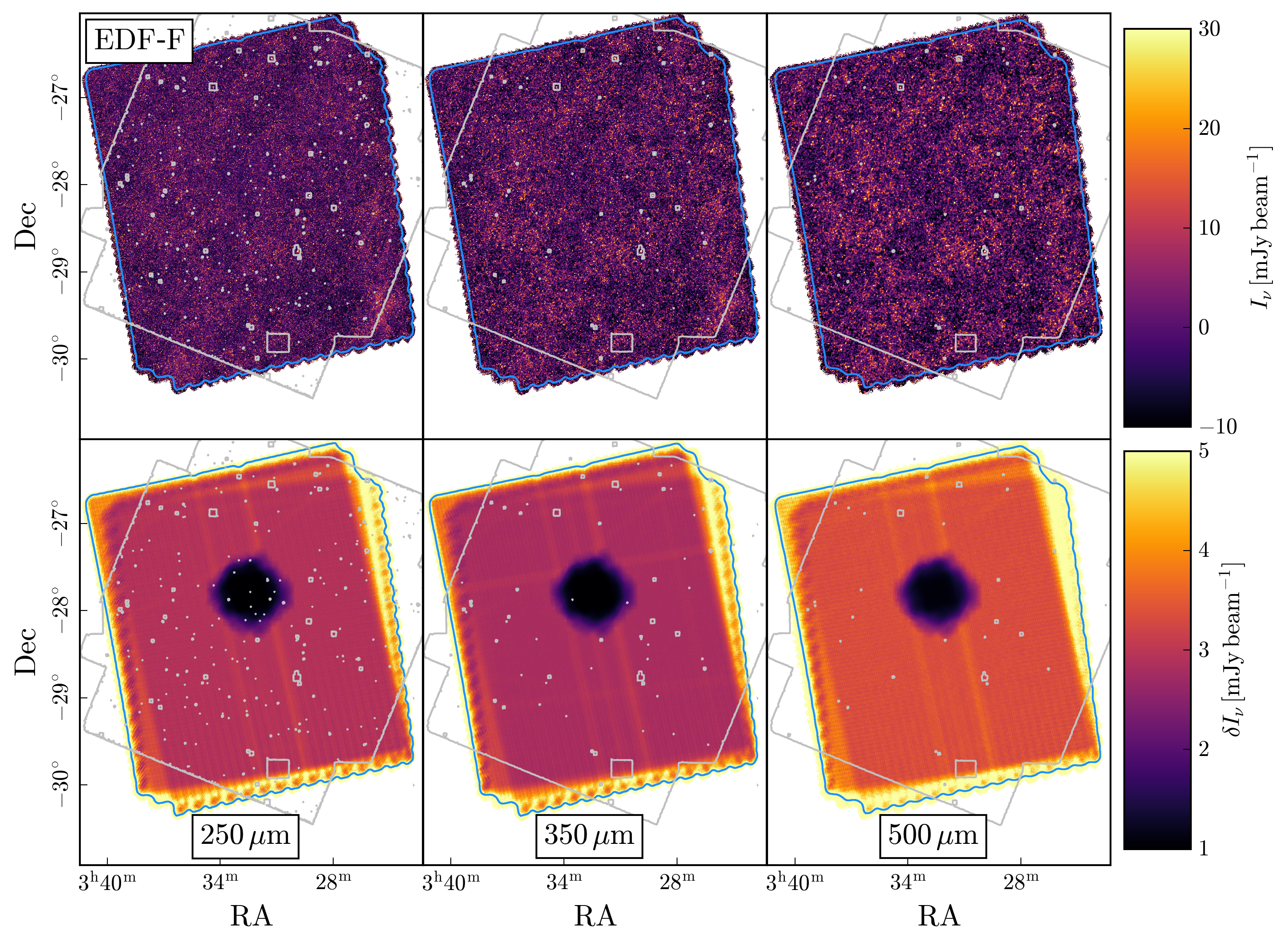}
    \caption{{\it Top:} \Herschel-SPIRE data covering the CDFS-SWIRE field (overlapping with the EDF-F) at 250, 350, and 500\,$\mu$m. The blue contour shows the mask applied to the SPIRE images to remove bad edge pixels. The grey contours show the corresponding \Euclid catalogue mask, where masked rectangles designate the locations of bright stars in the field contaminating source extraction. {\it Bottom:} Same as the top panel, but  showing the RMS of the \Herschel-SPIRE data.  Coordinates are conventional RA and Dec.} 
    \label{fig:field_summary_edff}
\end{figure*}

\begin{figure*}[htbp!]
    \centering
    \includegraphics[width=\textwidth]{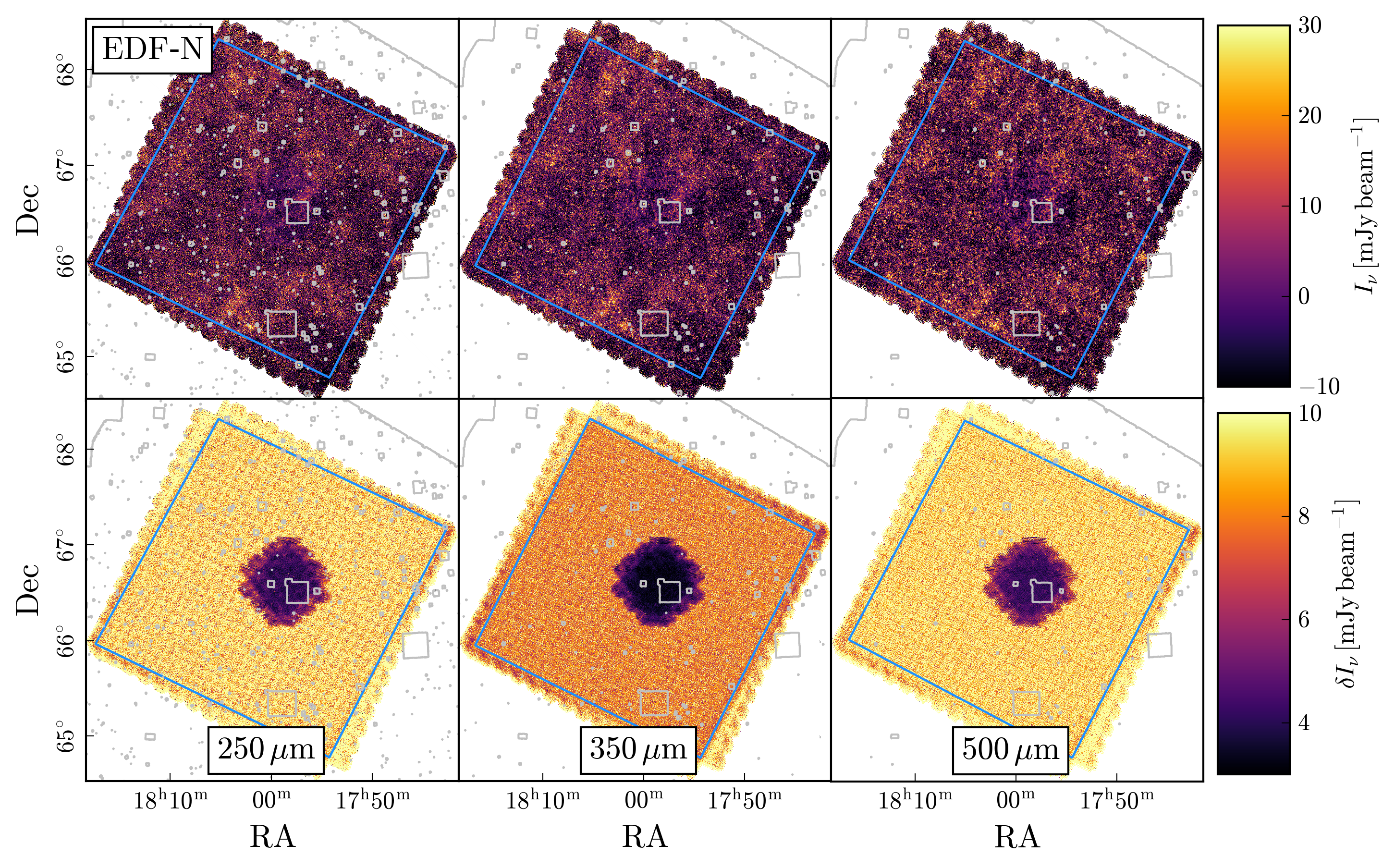}
    \caption{Same as \cref{fig:field_summary_edff}, but for the \Herschel-SPIRE AKARI-NEP field (overlapping with the EDF-N).} 
    \label{fig:field_summary_edfn}
\end{figure*}

The SPIRE maps of the EDF-F (or CDFS-SWIRE) and the EDF-N (or AKARI-NEP) were obtained from the \Herschel Extragalactic Legacy Project (HELP) archive \citep{Shirley2021}.\footnote{\url{https://hedam.lam.fr/HELP/}} The HELP data products include unfiltered maps and matched-filtered maps, along with their corresponding noise (or RMS) maps. Throughout this paper we use the unfiltered products because we do not expect these maps to have significant fluctuations caused by instrumental effects, nor significant contamination from dust in the Milky Way. The CDFS-SWIRE image covers 12.8\,deg$^2$, while the AKARI-NEP image covers 9.0\,deg$^2$ (although not all of this area overlaps with \Euclid). The raw SPIRE data, RMS maps and masks for each of the two fields are shown in Figs.~\ref{fig:field_summary_edff} and \ref{fig:field_summary_edfn}. In both maps extra observations were taken near the centres of the fields to decrease the noise and create small deep fields and wide shallow fields, and the HELP data products combined all of the observations into a single image.

In addition to SPIRE maps, the HELP archive also provides PACS maps at 100 and 160\,$\mu$m, wherever the data were taken. For the EDF-F, the full area observed by SPIRE was also observed by PACS (albeit to a much shallower depth), which we make use of here. The AKARI-NEP was also observed, covering 0.6\,deg$^2$ of the EDF-N. For these data only the output from the standard PACS processing pipeline is available; some filtering was done to the raw data in order to reduce strong $1/f$ noise and remove artefacts from bright sources; however, the maps were not matched-filtered. The PACS data, RMS maps and masks are shown in \cref{app:1}.

SCUBA-2 was used to map the AKARI-NEP (EDF-N) field as part of the SCUBA-2 Cosmology Legacy Survey \citep[S2CLS;][]{geach2017}, and this field was later expanded in the North Ecliptic Pole SCUBA-2 survey \citep[NEPSC2;][]{shim2020}\footnote{\url{https://dx.doi.org/10.5281/zenodo.3897405}}. The data products include the unfiltered maps and the matched-filtered maps, along with their corresponding RMS maps. As with the SPIRE images, we use the unfiltered maps here. The total area covered by SCUBA-2 in the AKARI-NEP is 2.9\,deg$^2$, and the entirety of this field has been observed by \Euclid. The SCUBA-2 data, RMS map and mask are shown alongside the PACS maps in \cref{app:1}.

Despite the fact that we can down-weight noisy regions using RMS maps, very noisy pixels near the edges of these maps can still cause significant issues, since the uncertainties for regions with few a `hits' can be underestimated. To mitigate this issue, we also created masks to remove problematic edge pixels. To make the mask for the SPIRE CDFS-SWIRE field we smoothed the noise maps using a Gaussian kernel with a standard deviation of 7.5 pixels, then masked regions where the smoothed noise map has a value ${>}\,7\,$mJy\,beam$^{-1}$; for reference, typical noise values are ${<}\,4\,$mJy\,beam$^{-1}$. For the AKARI-NEP field this strategy does not work because the noise map is too inhomogeneous, so instead we manually defined a rectangular masked region outside of which the noise pattern begins to deviate from the majority of the map. For the PACS maps of the CDFS-SWIRE field the noise is again quite inhomogeneous, so we also manually defined a central region with a consistent noise level. For the PACS AKARI-NEP maps we smoothed the noise maps using a Gaussian kernel with a standard deviation of 5 pixels and masked regions where the smoothed noise map has a value ${>}\,40\,$mJy\,beam$^{-1}$ (compared to the typical noise level of about $30\,$mJy\,beam$^{-1}$). For the SCUBA-2 data, we simply masked pixels with noise values ${>}\,30\,$mJy\,beam$^{-1}$ (compared to the typical noise level of about $10\,$mJy\,beam$^{-1}$). Lastly, the AKARI-NEP field contains NGC\,6543 (the `Cat's Eye Nebula'), which is particularly bright in the PACS and SCUBA-2 images. Since this is a Galactic object, we masked it before stacking our galaxy catalogue. These masks are also shown in Figs.~\ref{fig:field_summary_edff}, \ref{fig:field_summary_edfn}, \ref{fig:field_summary_edff_pacs}, and \ref{fig:field_summary_scuba2} along with the \Euclid catalogue masks.

\subsection{Overlap between \Euclid Q1 products and far-IR images}
\label{subsec:final_area}

After accounting for the masked regions, the total overlapping area between the EDF-F and the SPIRE CDFS-SWIRE data is 10.8\,deg$^2$, while for the EDF-N and the SPIRE AKARI-NEP data the total overlapping area is 6.8\,deg$^2$, for a total area of 17.6\,deg$^2$. Within the unmasked EDF-F region, the catalogue from \citet{Q1-SP031} contains about 1.5 million MS galaxies, while the EDF-N regions contains about 1.1 million galaxies (note that the exact values beyond the first digit depend somewhat on the SPIRE wavelength, since each SPIRE band has slightly different coverage). This brings the total number of MS galaxies overlapping with SPIRE to 2.6 million.  For PACS, the total area available for stacking, after accounting for the \Euclid mask, is 8.9\,deg$^2$ in the EDF-F and 0.4\,deg$^2$ in the EDF-N, with 1.3 million and 70\,000 MS galaxies, respectively. The total PACS area is thus 9.3\,deg$^2$, containing 1.4 million MS galaxies.  For SCUBA-2, the total area available for stacking (only in the EDF-N) is 2.4\,deg$^2$, and contains 240\,000 galaxies. 

\section{Stacking method}
\label{sec:method}

\subsection{Stacking algorithm}
\label{subsec:stack_algorithm}

With a few small adjustments, we employed the stacking method called {\tt SimStack} \citep[][]{Viero2013}, which stacks on multiple catalogue bins simultaneously. The main advantage of {\tt SimStack} is that it is not affected by galaxy clustering, since we do not need to assume that the galaxies being stacked are Poisson distributed. Briefly, after the input stacking catalogue was split into $N$ bins, we solved for the $N$ stacking amplitudes defined by the linear equation
\begin{equation}\label{eq:stack_model}
    y = S_1 X_1 + \dots + S_N X_N\,,
\end{equation}
where $y$ is the data map being stacked on (in this case a \textit{Herschel} or SCUBA-2 map at one wavelength), $S_b$ are the average flux densities of the galaxies in bin $b$, and $X_b$ are the beam-convolved images of the distribution of galaxies in the same bin. We note that in this analysis we subtracted the noise-weighted mean from the data map $y$ as well as the noise-weighted means from each of the beam-convolved images $X_b$. This ensures that we did not need to add an arbitrary constant to our stacking model \citep[see][]{Viero2013}. To construct the beam-convolved galaxy distribution images $X_b$ we simply looped through the positions of all the galaxies in a given bin, adding a 1 to each pixel in the model image where a galaxy in bin $b$ is located. Then, we convolved the image by the instrumental beam and subtracted the weighted mean. 

\Cref{eq:stack_model} is a linear system and can thus be solved via the weighted least-squares method (see section~3.1 of \citealt{Viero2013} for an explicit derivation). Moreover, we can use maps from multiple fields to fit for the stacked flux densities simultaneously. For $M$ total pixels across all of the data maps, we defined $\tens{X}$ as the $M\,{\times}\,N$ matrix where column $i$ contains the beam-convolved images $X_i$ and $\mathbf{y}$ is the $M\,{\times}\,1$ vector of the data (note that $X_i$ and $y$ must be flattened from a 2D image to a 1D vector in the same way, and that images from multiple fields must be added to the vectors and matrices in the same order). We also incorporated the weights (and masks) in the diagonal $M\,{\times}\,M$ matrix $\tens{W}$, where $W_{ii}\,{=}\,1/\sigma_{ii}^2$ if the pixel is not masked, and {\tt nan} otherwise. The best-fit stacking amplitudes are then just the solution to the weighted linear least-squares problem,
\begin{equation}\label{eq:lstsq_eq}
    \mathbf{\hat{S}} = (\tens{X}^{\sf{T}}\tens{W}\tens{X})^{-1}\tens{X}^{\sf{T}}\tens{W}\mathbf{y}\,.
\end{equation}

In traditional stacking, creating small equally sized cutouts around each source and averaging these cutouts together is common. Ideally, for far-IR images of unresolved galaxies, these should all resemble the instrumental point-spread function (PSF), and so these 2D stacks are excellent tools for checking for systematic errors. We can easily generalise the 2D stacking method from regular stacking to {\tt SimStack} by noting that while the central pixel in a regular 2D stack is the covariance between the map and the catalogue, the offset pixels represent the shifted cross-correlations. We can achieve the same result by calculating cross-correlations between our data $y$ and our model $S_1 X_1 + \dots + S_N X_N$. In practice, we offset the data images and corresponding weights by integer pixel values $(\Delta x, \Delta y)$ relative to the model image, then re-solved \cref{eq:lstsq_eq} for a new set of stacking flux densities $\mathbf{\hat{S}}$.

\subsection{Stacking bins}
\label{subsec:stack_bins}

To perform the stacking we defined a set of $N$ bins in stellar mass and redshift, produced from the \Euclid catalogue. \citet{Q1-SP031} focused their study of the galaxy MS on galaxies with $0.2\,{<}\, z\,{<}\,3.0$, with the understanding that most catastrophic outliers have either mistakenly very low or very high redshifts. For a similar reason they also limited their sample to galaxies with $\logten(M_{\ast}/M_{\odot})\,{\lesssim}\,11.5\,$, since the galaxies with larger stellar masses are likely to be outliers with very large redshifts. Finally, they estimated the stellar mass completeness of the sample as a function of redshift, finding ${>}\,95\%$ completeness at the lowest redshift ($z\,{=}\,0.2$) around $\logten(M_{\ast}/M_{\odot})\,{\simeq}\,8$. We therefore split the \Euclid catalogue into redshift bins between $z\,{=}\,0.2$ and 3.0, with a spacing of $\Delta z\,{=}\,0.2$, and into stellar mass bins between $\logten(M_{\ast}/M_{\odot})\,{=}\,8.3$ and 11.5, with a spacing of $\Delta \logten(M_{\ast}/M_{\odot})\,{=}\,0.4\,$.

We next added an additional layer to the stacking model that includes all remaining catalogued galaxies that do not fall into any redshift or stellar mass bin. We also included the quiescent galaxies in this additional bin. We finally added one more layer to the stacking model to account for the \Euclid mask, required for reducing biases in the stack \citep[see][for details]{Duivenvoorden2020}. For this mask layer we inverted the \Euclid mask, convolved it with the instrumental beam, and then restored the masked pixels. This layer is intended to capture far-IR surface brightness leakage from near-IR galaxies that \Euclid was unable to detect due to contamination from, for example, bright stars (although in practice this effect is small).

\subsection{Instrumental beams}
\label{subsec:stack_psf}

The final input required for stacking is a model of the \Herschel and SCUBA-2 beams. For PACS, we used the model empirical beams provided by the HELP archive \citep{Shirley2021}. These were measured by stacking WISE-selected galaxies on the PACS maps and fitting an elliptical 2D Gaussian profile to the stack. The ellipticity is due to the fact that the PACS beam is not circularly symmetric, and so for these data we used the best-fit beams for each PACS map separately.

For SPIRE, since the input maps are not filtered, we used the instrumental beams approximated as 2D Gaussians with full width half maximum (FWHM) values of \ang{;;18.15} at 250\,$\mu$m, \ang{;;25.15} at 350\,$\mu$m, and \ang{;;36.3} at 500\,$\mu$m \citep[see][]{Griffin2010}. For the SCUBA-2 image, we used the updated beam profile from \citet{mairs2021}, which is the sum of two Gaussians.  The first Gaussian (the main beam) has $\textrm{FWHM}= \ang{;;11.0}$ and a relative amplitude of 0.98, while the second Gaussian (the error beam) has $\textrm{FWHM}= \ang{;;49.1}$ and a relative amplitude of 0.02.

\subsection{Estimating uncertainties}
\label{subsec:stack_error}

We estimated the uncertainties in the best-fit stack flux densities following \citet{Viero2013} by both propagating the weight matrix and by performing bootstrap resampling to determine the overlap between neighbouring bins. The statistical covariance matrix from solving the weighted linear least-squares system is analytic and can be calculated as 
\begin{equation}\label{eq:lstsq_error}
    \tens{\Sigma_{\hat{S}}} = (\tens{X}^{\sf{T}}\tens{W}\tens{X})^{-1}\,,
\end{equation}
and so the statistical uncertainty in $\hat{S}_i$ is just $\sqrt{\Sigma_{\hat{S},ii}}$. We added this in quadrature with the uncertainty from bootstrap resampling (which dominates the error budget); for this contribution, we generated 100 random catalogues by drawing stellar masses and redshifts for each galaxy from their probability distributions. In principle the stellar masses and redshifts are correlated with each other and would require a full 2D posterior distribution for each galaxy; however, we simplified the procedure by assuming that the distributions are Gaussian with a standard deviation equal to half the 68\% confidence interval. We then re-binned each random catalogue, re-solved \cref{eq:lstsq_eq}, and calculated the standard deviation of the 100 random estimates of $\mathbf{\hat{S}}$.

\section{Results}
\label{sec:results}

\subsection{Average far-IR flux densities of main-sequence \Euclid galaxies}
\label{subsec:results_flux_core}

We ran our stacking algorithm (essentially {\tt SimStack} generalised to provide 2D cross-correlations) on the \Herschel and SCUBA-2 maps described in \cref{sec:data}. We set the 2D stacking cutout size for each of the SPIRE wavelengths to be 200$^{\prime\prime}$ (meaning that we solved Eq.~\ref{eq:lstsq_eq} for offsets in a 200$^{\prime\prime}\,{\times}\,200^{\prime\prime}$ grid), while for SCUBA-2 we set the cutout size to be 70$^{\prime\prime}$, and for PACS we set the cutout size to be 40$^{\prime\prime}$. The full 2D cross-correlations are shown in \cref{app:2}. We find significant detections in most bins above $\logten(M_{\ast}/M_{\odot})\,{=}\,10\,$, and the detections are generally consistent with the instrumental PSFs. 

More precisely, for perfect point sources the 2D profile produced by our algorithm is the cross-correlation of the beam with itself; for a Gaussian, this increases the FWHM by a factor of $\sqrt{2}$. We tested this by computing the averaged 1D radial profiles of our 2D cross-correlations and comparing these to the expected PSF profiles. We find good agreement between the stacked signals and the PSFs for most bins with $\logten(M_{\ast}/M_{\odot})\,{>}\,9.9$. Below this stellar mass we find that the 2D profiles can be more extended than the beam, with the largest effect seen at 500\,$\mu$m where the PSF is largest. This could be caused by a combination of the beam size and incompleteness in the \Euclid catalogue. For example, dust galaxies bright at far-IR wavelengths are likely to be undetected by \Euclid, and these galaxies could have different clustering properties. However, as detailed in the following sections, we do not expect this extended emission to affect our results and so we do not attempt to correct for it. 

As discussed previously, the central pixel values shown in the 2D stacks are the best-fit mean flux densities of the galaxies in the given bin (i.e., the covariances between the catalogue and the maps). For the high S/N detections, we checked that the central pixel in our 2D stacks was consistently the brightest pixel, meaning that there were no astrometric offsets between the \Euclid catalogue and our stacking images. We provide these central values and the corresponding uncertainties in \cref{app:2}.

\subsection{Average SEDs of main-sequence \Euclid galaxies}
\label{subsec:results_sed_core}

Our best-fit stacked flux densities contain information about the average far-IR SEDs of the \Euclid-selected galaxies in each stellar mass and redshift bin. We modelled the lower frequencies of these SEDs using a standard modified blackbody for $\nu(1+z)\,{<}\,\nu_{\alpha}$ (explained below):
\begin{equation}\label{eq:mbb}
    S_{\nu}\,[\mathrm{mJy}] = A \left[ \frac{\nu(1+z)}{\nu_0} \right]^{\beta} (1+z)\,B_{\nu(1+z)}(T_{\rm d})\, \left[\frac{10^{29}\,\mathrm{mJy}}{\mathrm{W\,m^{-2}\,Hz^{-1}}}\right]\,,
\end{equation}
where $B_{\nu}(T_{\rm d})$ is the blackbody function
\begin{equation}\label{eq:bb}
    B_{\nu}(T_{\rm d}) = \frac{2 h \nu^3}{c^2} \frac{1}{{\rm e}^{h \nu / kT_{\rm d}}-1}\,.
\end{equation}
Since we dealt with galaxies at $z\,{<}\,3$, we ignored corrections related to the cosmic microwave background \citep[CMB; see][]{dacunha2013}. In these equations $\nu$ is the observed frequency, $z$ is the redshift, $h$ is the Planck constant, $k$ is the Boltzmann constant, $c$ is the speed of light, $T_{\rm d}$ is the dust temperature, and the factor of $10^{29}$ converts the observed flux density units from SI to mJy. The quantity $\nu_0$ is not a free parameter of the model but sets the dust mass normalisation (as explained in \cref{subsec:results_params}), and we used $\nu_0\,{=}\,$353\,GHz. We note that this simple parametrisation assumes the dust is optically thin, so we did not need to include the characteristic frequency where the optical depth is unity \citep[e.g.][]{draine2006,drew2022}. This left the amplitude $A$, dust temperature $T$, and dust emissivity index $\beta$ as the three free parameters.

At rest-frame mid-IR frequencies $\nu(1+z)\,{>}\,\nu_{\alpha}$, i.e. around a few terahertz, the Wien side of the thermal SED falls off less steeply than an exponential and is typically modelled as a power law with a slope of $-\alpha$ \citep[e.g.][]{blain2003,roseboom2013,casey2014,reuter2020}. Since our PACS data cover this region of the SED, we needed to include this phenomenological feature. To do so, we modelled the SED for $\nu(1+z)>{\nu_{\alpha}}$ as 
\begin{equation}\label{eq:midir_powerlaw}
    S_{\nu}\,[\mathrm{mJy}] = A_{\alpha}\left[ \frac{\nu(1+z)}{\nu_{\alpha}} \right]^{-\alpha}\!\!(1+z)\, \left[\frac{10^{29}\,\mathrm{mJy}}{\mathrm{W\,m^{-2}\,Hz^{-1}}}\right]\,.
\end{equation}
Here, $A_{\alpha}$ and $\nu_{\alpha}$ are determined by matching the amplitude and slope of \cref{eq:mbb,eq:midir_powerlaw}. This gives $A_{\alpha}\,{=}\,A(\nu_{\alpha}/\nu_0)^{\beta}B_{\nu_{\alpha}}(T)$, and $\nu_{\alpha}$ solves the equation $3\,{+}\,\beta\,{+}\,\alpha\,{=}x{\rm e}^x/({\rm e}^x-1)$ where $x\,{=}\,h\nu_{\alpha}/kT$. 

\begin{figure*}[htbp!]
    \centering
    \includegraphics[width=\textwidth]{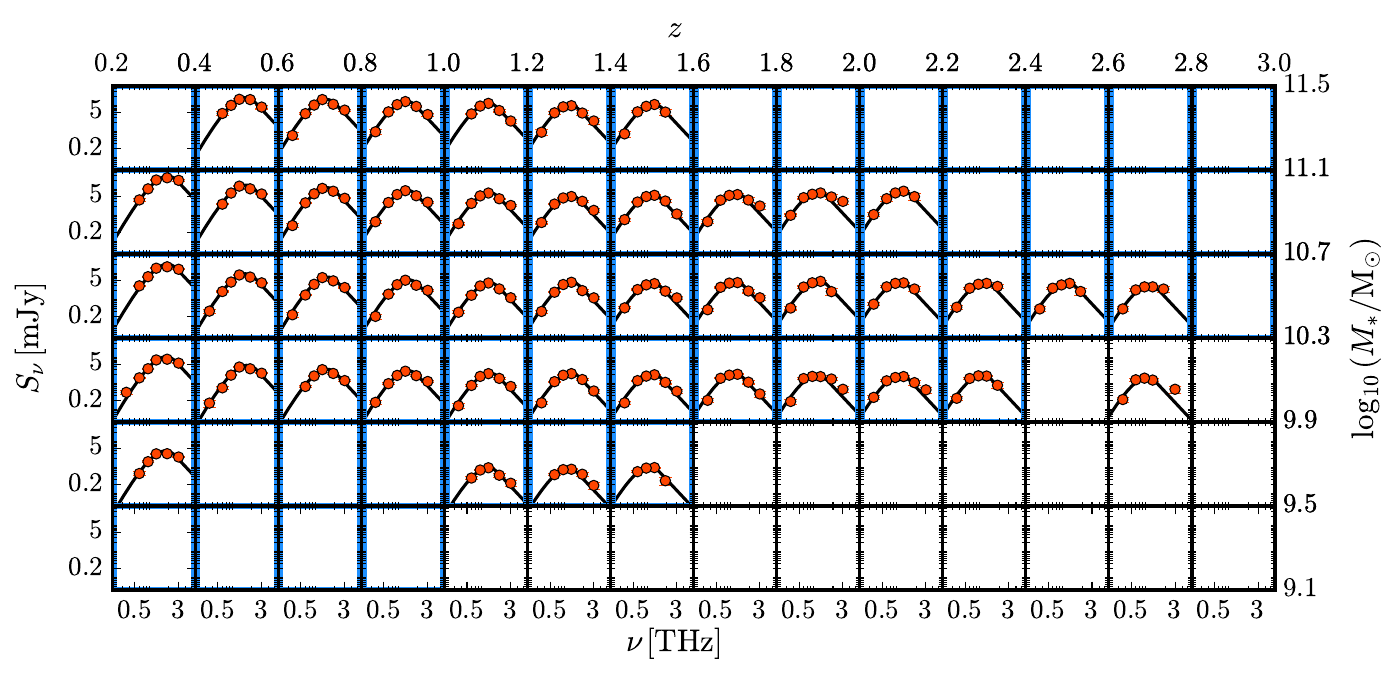}
    \caption{Modified blackbody SEDs (with $\beta=1.96$ and $\alpha=2.3$) fitted to the stacked \Herschel and SCUBA-2 flux densities. The redshift and stellar mass of each bin are indicated by the top and right axis labels, respectively. The best-fit parameters are given in Table~\ref{table:bestfit_sed}. The SEDs were fitted to bins where all three SPIRE flux densities are detected with S/N$\,{>}\,3$ and at least one PACS flux density is detected with S/N$\,{>}\,3$. The panels are otherwise blank. Bins that are ${>}\,95\%$ complete in stellar mass are highlighted in blue.} 
    \label{fig:stack_sed_fits}
\end{figure*}

Since we did not have enough photometry information to constrain $\beta$ on the Rayleigh--Jeans sides or $\alpha$ on the Wien sides of the SEDs, we fixed these values to $\beta\,{=}\,1.96$ and $\alpha\,{=}\,2.3$ according to the mean values found for local far-IR-selected galaxies \citep{drew2022}. Fixing $\beta$ and $\alpha$ to these values, we performed our SED fits on all redshift and stellar mass bins where all three SPIRE flux densities were detected in the stack with S/N$\,{>}\,3$, and where there is also a PACS detection at 100 or 160\,$\mu$m with S/N$\,{>}\,3$. The reason for these constraints is simply that we needed the observed photometry to bracket the peak of the SED to properly constrain the dust temperature (which is directly proportional to the peak frequency). We included a 4\% absolute calibration uncertainty and a 1.5\% relative calibration uncertainty in the SPIRE bands, a 5\% absolute calibration uncertainty in the PACS bands, and a 15\% absolute calibration uncertainty in the SCUBA-2 band. The resulting best-fit SEDs are shown in \cref{fig:stack_sed_fits}, and the best-fit parameters are given in \cref{app:3}. In this figure we do not show the lowest stellar mass bins as we do not have enough far-IR photometry to fit SEDs. For two of the highest redshift and stellar mass bins, $z\,{=}\,2.7$, $\logten(M_{\ast}/M_{\odot})\,{=}\,10.9$ and 11.3, the fits provide unphysically high dust temperatures (${>}\,60\,$K), potentially due to contamination in the \Euclid catalogue. We discard these two bins for the remainder of this work. As shown, it is mainly the $\logten(M_{\ast}/M_{\odot})\,{>}\,9.9$ bins where we have enough far-IR photometry to fit SEDs, and these bins show good agreement between the 2D cross-correlation profiles and the instrumental beams.

\subsection{Derived parameters for main-sequence \Euclid galaxies}
\label{subsec:results_params}

From the best-fit parameters to \cref{eq:mbb} we can derive certain average physical properties for the galaxies in the \Euclid catalogue. First, the best-fit temperatures $T_{\rm d}$ are already the rest-frame dust temperatures because we included the redshifts in the fits. Next, the best-fit amplitudes are directly proportional to the dust mass, $M_{\rm d}$, and can be calculated using \citep[e.g.][]{reuter2020,Eales2024,jolly2025}
\begin{equation}\label{eq:dust_mass}
    M_{\rm d} = \frac{D_{\rm L}^2(z)A}{\kappa_0}\,,
\end{equation}
where $\kappa_0$ is the calibration factor that scales the specific luminosity at the rest-frame frequency of $\nu_0$ (the same reference frequency in our SED fit, see Eq.~\ref{eq:mbb}) to a dust mass and $D_{\rm L}$ is the luminosity distance. Here, we used the factor $\kappa_0\,{=}\,0.077\,$m$^2$\,kg$^{-1}$, calibrated to the frequency $\nu_0\,{=}\,353\,$GHz \citep{dunne2000,dacunha2008,dunne2011}, and we included an uncertainty of $\pm0.02\,$m$^2$\,kg$^{-1}$ \citep[from][]{james2002}. This approach uses the best-fit model to estimate the rest-frame specific luminosity at the frequency $\nu_0$, as opposed to using a single measured flux density. There are many other values of $\kappa_0$ used throughout the literature, typically resulting in dust mass differences of a factor of a few; however, picking a different $\kappa_0$ only changes the absolute value of the dust mass, rather than trends in stellar mass or redshift (although $\kappa_0$ could in principle vary with redshift). On the other hand, the dust emissivity index $\beta$ can have an effect on the best-fit SEDs, and therefore any stellar mass and redshift trends. While we do not expect $\beta$ to vary by much compared to the fiducial value of about 2, this could be investigated in the future if more far-IR wavelengths could be added.

Lastly, we can estimate far-IR SFRs using the linear relation
\begin{equation}\label{eq:sfr}
    \mathrm{SFR}\,[\mathrm{M_{\odot}\,yr^{-1}}] = 1.49 \times 10^{-10} L_{\rm FIR}\,[\mathrm{L_{\odot}}]
\end{equation}
(assuming a Kroupa initial mass function as performed with the \Euclid catalogue; see \citealt{Q1-TP005}), where
\begin{equation}\label{eq:lfir}
    L_{\rm FIR} = 4 \pi D_{\rm L}^2(z) \left[ \int_{\nu_1}^{\nu_{\alpha}}\!\!A \left( \frac{\nu}{\nu_0} \right)^{\beta}\!B_{\nu}(T_{\rm d})\,{\rm d}\nu+\!\int_{\nu_{\alpha}}^{\nu_2}\!\!A_{\alpha} \left( \frac{\nu}{\nu_{\alpha}} \right)^{-\alpha}\!{\rm d}\nu \right]
\end{equation}
is the area under the far-IR SED between $\nu_1\,{=}\,c/1000\,\mu\mathrm{m}$ and $\nu_2\,{=}\,c/8\,\mu\mathrm{m}$ in the rest frame. Again, we used $\beta\,{=}\,1.96$ and $\alpha\,{=}\,2.3$. We note that this derived physical parameter SFR is not independent of $T_{\rm d}$ and $M_{\rm d}$, but instead these three quantities are correlated.

In \cref{app:3} we show the resulting physical parameters $T_{\rm d}$, $M_{\rm d}$, and SFR in each of the stellar mass and redshift bins where we have sufficient stacked photometry to derive these physical parameters, and we also provide the derived physical quantities. We find a clear trend of increasing dust temperature with redshift, and an increase in the dust mass from redshifts to around $z\,{\simeq}\,2$. For the SFRs, we see an increase towards high-$z$, with a dependence on the stellar mass (as expected from the galaxy MS). These trends are discussed further in \cref{sec:discussion}.

\subsection{The mean brightness of the CIB from stacking}
\label{subsec:results_cib}

We now turn to estimating the fraction of the cosmic infrared background (CIB) resolved by \Euclid-selected galaxies in Q1. The average brightness of the extragalactic sky at far-IR wavelengths was measured by the Cosmic Background Explorer (COBE) satellite, which carried two relevant instruments: the Far Infrared Absolute Spectrophotometer \citep[FIRAS;][]{mather1993}; and the Diffuse Infrared Background Experiment (DIRBE; see \citealt{boggess1992}). Analysis of these data provide the best available estimates of the absolute value of the surface brightness of the sky at far-IR wavelengths. \Herschel and SCUBA-2, on the other hand, have no sensitivity to the monopole on the sky, but instead measure differences between sources and an unknown background, amounting to fluctuations caused by individual galaxies. The fraction of the CIB can be estimated by taking the sum of detected sources and dividing by the COBE-estimated background.

To do this, for each bin in the stack we multiplied the mean flux density by the total number of SPIRE galaxies in the bin, then summed the contribution from all of the bins (including surface brightness leakage from the mask and the galaxies that do not fall in any redshift or stellar mass bin or are classified as quiescent). Finally, we divided this total flux density by the area of the SPIRE map after applying the masks (these are the areas given in \cref{subsec:final_area}). The results are shown in \cref{table:resolved_cib}. We note that ${>}\,70\%$ of our measured CIB surface brightness from star-forming galaxies comes from $\logten(M_{\ast}/M_{\odot})\,{>}\,9.9$ bins which show good agreement between the 2D cross-correlation signals and the instrumental beams.

\begin{table*}[htbp!]
\centering
\caption{Contribution to the CIB from stacking the \Euclid catalogue.}
\label{table:resolved_cib}
\begingroup
\setlength{\tabcolsep}{5pt}
\begin{tabular}{ccccc@{\hskip 30pt}cccc}
\hline\hline
\noalign{\vskip 2pt}
Band & {\pz}Main gals. & Mask & Remaining gals. & {\bf Total} & Fixsen & {\pz}Odegard & {\pz}Watts & Casandjian\\
$[\mu$m] & \pz[Jy\,deg$^{-2}$] & [Jy\,deg$^{-2}$] & [Jy\,deg$^{-2}$] & $\mathbf{[\mathrm{Jy\,deg^{-2}}]}$ & [Jy\,deg$^{-2}$] & \pz[Jy\,deg$^{-2}$] & \pz[Jy\,deg$^{-2}$] & [Jy\,deg$^{-2}$]\\
\hline
\noalign{\vskip 2pt}
100 & \pz34.3$\pm$1.1 & 0.088$\pm$0.035 & 32.8$\pm$0.6 & \pz$\mathbf{67.2\pm1.3}$ & \dots & \pz\dots & \pz80$\pm$17 & 249$\pm$40\\
160 & \pz62.5$\pm$1.6 & 0.131$\pm$0.038 & 50.8$\pm$0.9 & $\mathbf{113.4\pm1.8}$ & \dots & \pz\dots & \pz158$\pm$106 & 271$\pm$86\\
250 & 101.0$\pm$0.7 & 0.621$\pm$0.004 & 78.2$\pm$0.1 & $\mathbf{179.7\pm0.7}$ & 260$\pm$80 & \pz\dots & 144$\pm$72 & 164$\pm$48\\
350 & \pz76.5$\pm$0.7 & 0.500$\pm$0.003 & 58.5$\pm$0.2 & $\mathbf{135.5\pm0.7}$ & 200$\pm$61 & \pz175$\pm$10 & \pz\dots & 114$\pm$11\\
500 & \pz42.8$\pm$0.7 & 0.340$\pm$0.003 & 35.2$\pm$0.2 & \pz$\mathbf{78.4\pm0.7}$ & 102$\pm$31 & 133$\pm$6 & \pz\dots & 99$\pm$7\\
850 & \pz\pz8.8$\pm$0.4 & 0.075$\pm$0.013 & \pz6.2$\pm$0.2 & \pz$\mathbf{15.0\pm0.4}$ & \pz43$\pm$13 & \pz45$\pm$5 & \pz\dots & 35$\pm$4\\
\hline
\end{tabular}
\endgroup
\footnotesize
 \tablefoot{The `Main gals.' column is the value of the CIB from star-forming galaxies with reliable optical/near-IR SED fits, $8.3\,{<}\,\logten(M_{\ast}/M_{\odot})\,{<}\,11.5$, $0.2\,{<}\,z\,{<}\,3.0$, the `Mask' column is the contribution to the CIB from the flux leakage through the \Euclid mask, the 'Remaining gals.' column is the value of the CIB from galaxies outside this stellar mass and redshift range or classified as quiescent, and the `Total' column (in bold) is the sum of these three contributions. The estimated absolute values of the CIB are in the right four columns, with references being \citet{fixsen1998}, \citet{odegard2019}, \citet{watts2024}, and \citet{Casanjian2024}.}
\end{table*}

To determine the fraction of the CIB resolved by \Euclid, we estimated the absolute value of the CIB measured by FIRAS and DIRBE. FIRAS measurements were recalculated using an improved Galactic emission removal procedure at \textit{Planck} wavelengths by \citet{odegard2019}, which should be more accurate than FIRAS alone at 350, 500, and 850\,$\mu$m. From DIRBE, we used the results from a re-analysis of the data from the Cosmoglobe project \citep{watts2024}, which provides measurements of the CIB intensity at 100, 140, and 240\,$\mu$m. Additionally, \citet{Casanjian2024} used a combination of DIRBE and {\it Planck} data to estimate the CIB at 100, 140, 240, 350, 500, and 850\,$\mu$m. Lastly, we used the result from FIRAS alone at 250, 350, 500, and 850\,$\mu$m by taking the best-fit spectral shape from \citep{fixsen1998}. Here, the fit was performed using a modified blackbody function, with $\beta\,{=}\,0.64$ and $T\,{=}\,18.5\,$K. For the uncertainty we used the uncertainty in the best-fit amplitude. To correct the above 140\,$\mu$m intensities to 160\,$\mu$m, the 240\,$\mu$m intensities to 250\,$\mu$m, and the 550\,$\mu$m intensities to 500\,$\mu$m, we used this same best-fit spectral shape -- these corrections range from 1 to 24\%. All the CIB estimates are given in \cref{table:resolved_cib}.

\section{Discussion}
\label{sec:discussion}

\subsection{Redshift trends of mean physical properties}
\label{subsec:z_trends}

In our stacking analysis, we found significant redshift trends for the dust temperatures, dust masses, and SFRs of average \Euclid-selected galaxies. To interpret these results, we used the 95\% stellar mass completeness limits from \citet{Q1-SP031} to select stacking bins that are ${>}\,95\%$ complete. These bins are highlighted in blue in \cref{fig:stack_sed_fits}.

\begin{figure}[htbp!]
    \centering
    \includegraphics[width=\columnwidth]{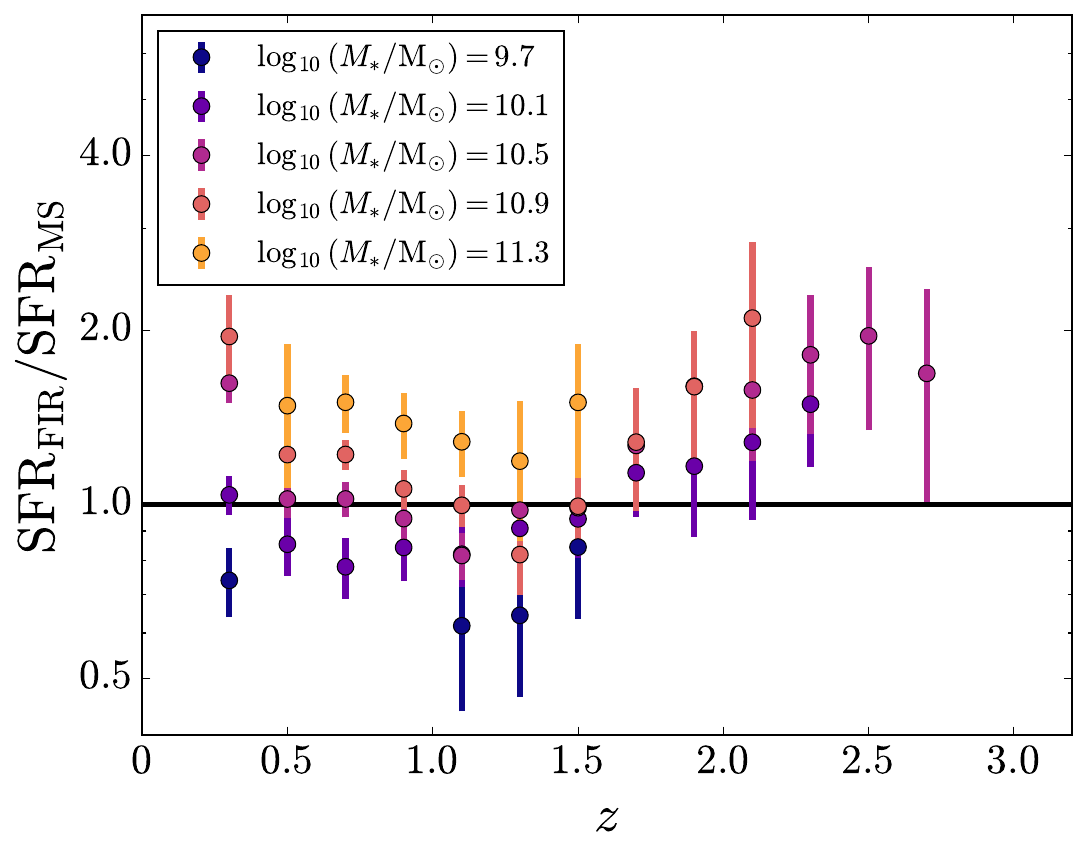}
    \caption{Ratio of our SFRs measured from far-IR photometry to a parametrisation  of the star-forming MS, shown as a function of redshift and split into different stellar mass stacking bins. Only redshift and stellar mass bins with ${>}\,95\%$ completeness are shown. We use the MS parametrisation from \citet{popesso2023}, which is a continuous function of $z$/$t$ and is in good agreement with \citet{Q1-SP031}.} 
    \label{fig:SFR_difference}
\end{figure}

Focusing only on the stellar mass and redshift bins where the stellar masses are ${>}\,95\%$ complete, we first compared our far-IR-derived SFRs to that predicted from the star-forming MS. The \Euclid SFRs were estimated by modelling solely optical and near-IR photometry from \Euclid and its supporting observations, and the star-forming galaxies were split into four redshift bins between $z\,{=}\,0.2$ and $z\,{=}\,3.0$. In each bin the MS was parametrised as
\begin{equation}\label{eq:galaxy_ms}
    \mathrm{SFR} = \frac{\mathrm{SFR}_{\mathrm{max}}}{1+(M_0/M_{\ast})^{\gamma}}\,,
\end{equation}
with $\gamma$ fixed to 1 and SFR$_{\mathrm{max}}$ and $M_0$ left as free parameters. Since our redshift bins are much smaller than those used by \citet{Q1-SP031}, we instead used the fit to \cref{eq:galaxy_ms} from \citet{popesso2023}, which was found to be in good agreement with the \Euclid results. \citet{popesso2023} also fixed $\gamma\,{=}\,1$, letting $\logten(\mathrm{SFR}_{\mathrm{max}})(t)\,{=}\,a_0\,{+}\,a_1 t$ and $\logten(M_0)(t)\,{=}\,a_2\,{+}\,a_3 t$. The resulting ratio of far-IR SFRs to the prediction from the MS are shown in \cref{fig:SFR_difference}. We note that below $\logten(M_{\ast}/M_{\odot})\,{=}\,9.7$ we do not have any far-IR detections in the stellar mass-complete bins, so we only plot the results for the five most massive bins. We find good agreement between the two estimates, with a weighted mean ratio of 1.0 and a weighted standard deviation of 0.2. 

However, we do see a systematic trend where more massive galaxies have higher SFRs than expected from the MS (and vice versa for less massive galaxies), as well as a trend where we estimate overall higher SFRs than the MS for $z\,{\gtrsim}\,1.5$. It is worth noting that the SFRs used in \citet{popesso2023} are a compilation from the literature and include estimates using ultraviolet luminosities, H$\alpha$, and far-IR luminosities, and the authors do not note any systematic differences between the estimators. We also compared our SFRs to the best-fit MS from \citet{Q1-SP031} after interpolating between the redshift bins, and directly to the SFRs from the \Euclid catalogue after averaging over the same redshift and stellar mass bins, finding similar results. It is thus more likely that the systematic differences we observe here are due to differences in the assumed far-IR SED shapes (for example the dust emissivity index $\beta$, the slope $\alpha$, whether an evolving dust temperature with redshift was assumed, or if the dust temperature depended on the stellar mass), or biases in the stellar masses and photometric redshifts in the \Euclid catalogue. However, a more complete understanding of these issues is beyond the scope of this paper.

\begin{figure}[htbp!]
    \centering
    \includegraphics[width=\columnwidth]{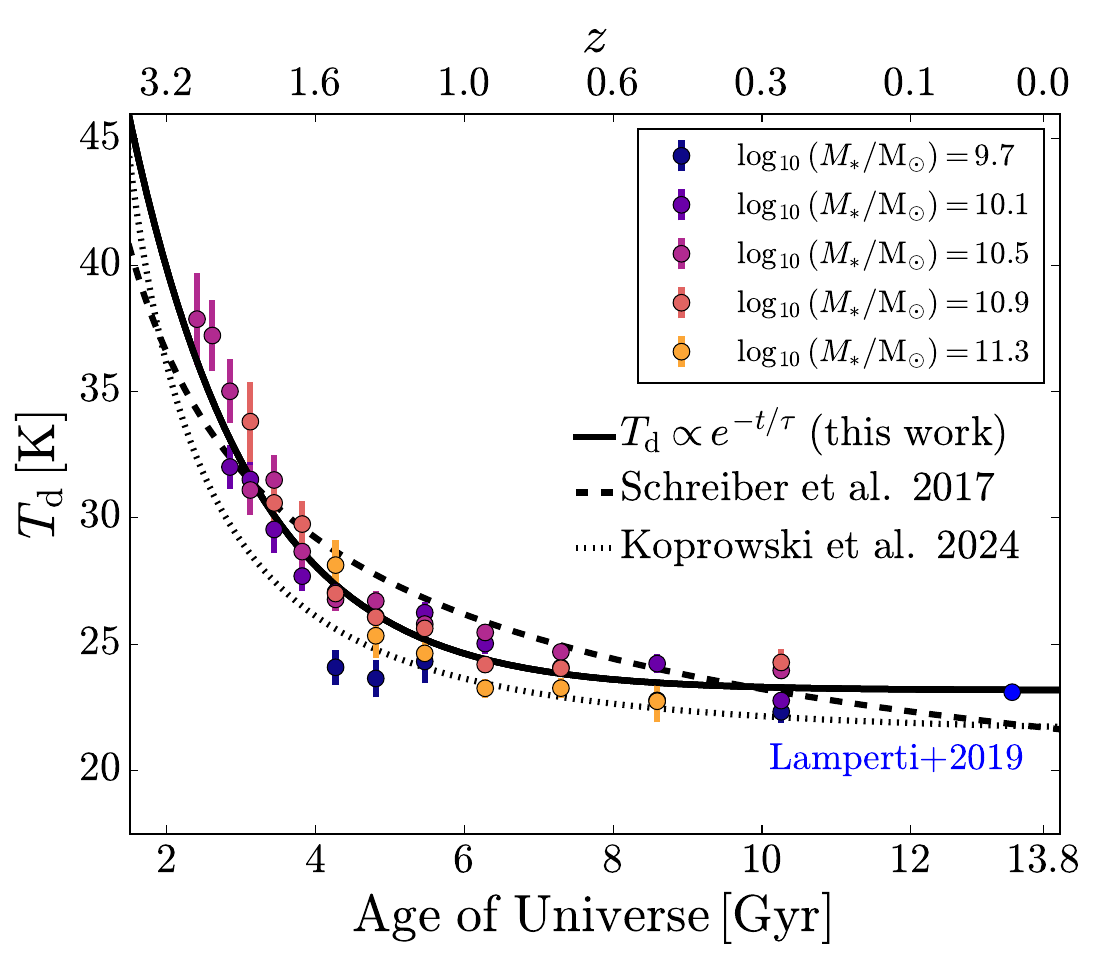}
    \caption{Best-fit dust temperatures from our SED fitting, $T_{\rm d}$, as a function of time (bottom axis) and redshift (top axis), considering only the redshift and stellar mass bins that are ${>}\,95\%$ complete. We show the dust temperature evolution for five different stellar mass bins, with the stellar mass values of the centres of the bins given in the legend. The solid curve is a fit to the simple form $T_2\,{+}\,(T_1\,{-}\,T_2)\,e^{-t/\tau}$; the dotted line is the quadratic-in-redshift fit from \citet{koprowski2024} and the dashed line is the linear-in-redshift fit from \citet{schreiber2017}. We also show published mean temperature estimates for star-forming galaxies at low redshifts \citep[$0.01\,{<}\,z\,{<}\,0.05$;][]{lamperti2019} in blue.} 
    \label{fig:Td_z}
\end{figure}

Next, in \cref{fig:Td_z} we show $T_{\rm d}$ versus time (bottom axis) and $z$ (top axis), split by the different stellar mass bins used in our stacking analysis. We find that average dust temperatures increase with redshift from about 20 to 35\,K, without any obvious trend in stellar mass. Looking at the dust temperature trend plotted linearly as a function of time, we see that the data decrease steeply between 2 and 6\,Gyr ($z\,{=}\,3$ and 1), then plateau to a constant dust temperature until the present day. To capture this behaviour, we fitted a phenomenological function of the form
\begin{equation}\label{eq:T_decay}
    T_{\rm d}(t) = T_2+(T_1 - T_2)\,{\rm e}^{-t/\tau}\,,
\end{equation}
where $T_1$ is the mean dust temperature of all star-forming galaxies at $t\,{=}\,0\,$Gyr, $T_2$ is the dust temperature all star-forming galaxies approach as $t$ becomes large, and $\tau$ is the characteristic timescale. We find best-fit values of $T_1\,{=}\,(79.7\pm7.4)\,$K, $T_2\,{=}\,(23.2\pm0.1)\,$K, and $\tau\,{=}\,(1.6\pm0.1)\,$Gyr. 

For comparison, in \cref{fig:Td_z} we also show a fit to the average dust temperature of star-forming galaxies as a function of redshift from \citet{schreiber2017}, which was later used to generate a simulated \Euclid catalogue with far-IR photometry (see \cref{subsec:simulation}). Here, a similar stacking analysis was used to estimate far-IR photometry in bins of redshift and stellar mass. However, as opposed to fitting a modified blackbody to the photometry (as done here), an empirical template was used to fit the data. They found that the dust temperatures increased linearly as a function of redshift, but as seen in \cref{fig:Td_z} this approach appears to overestimate the dust temperatures relative to our analysis around $z\,{\approx}\,1.5$. We also include the results from \citet{koprowski2024}, who stacked optically selected catalogues on far-IR images and fitted a modified blackbody to the photometry (with $\beta$ and $\alpha$ fixed to the same values as used here). They found that the dust temperatures follow a quadratic polynomial in redshift, which agrees well with our exponential decay function. We note that neither of these studies found any trend in stellar mass. In \cref{fig:Td_z} we also show the average dust temperature derived from a sample of local ($0.01\,{<}\,z\,{<}\,0.05$) star-forming galaxies from \citet{lamperti2019}. Since this sample includes bright galaxies, no stacking was required to estimate far-IR flux densities. The SEDs were fitted to the same modified blackbody function used here, although wavelengths ${>}\,100\,\mu$m were not used in the fit. Therefore, the transition to a power-law was not needed, and $\beta$ was kept as a free parameter with best-fit values ranging from 1--2. After averaging over all of the dust temperatures in the sample, we find a mean value of $23.1\pm0.1\,$K, consistent with our fit.

\begin{figure*}[htbp!]
    \centering
    \includegraphics[width=0.45\textwidth]{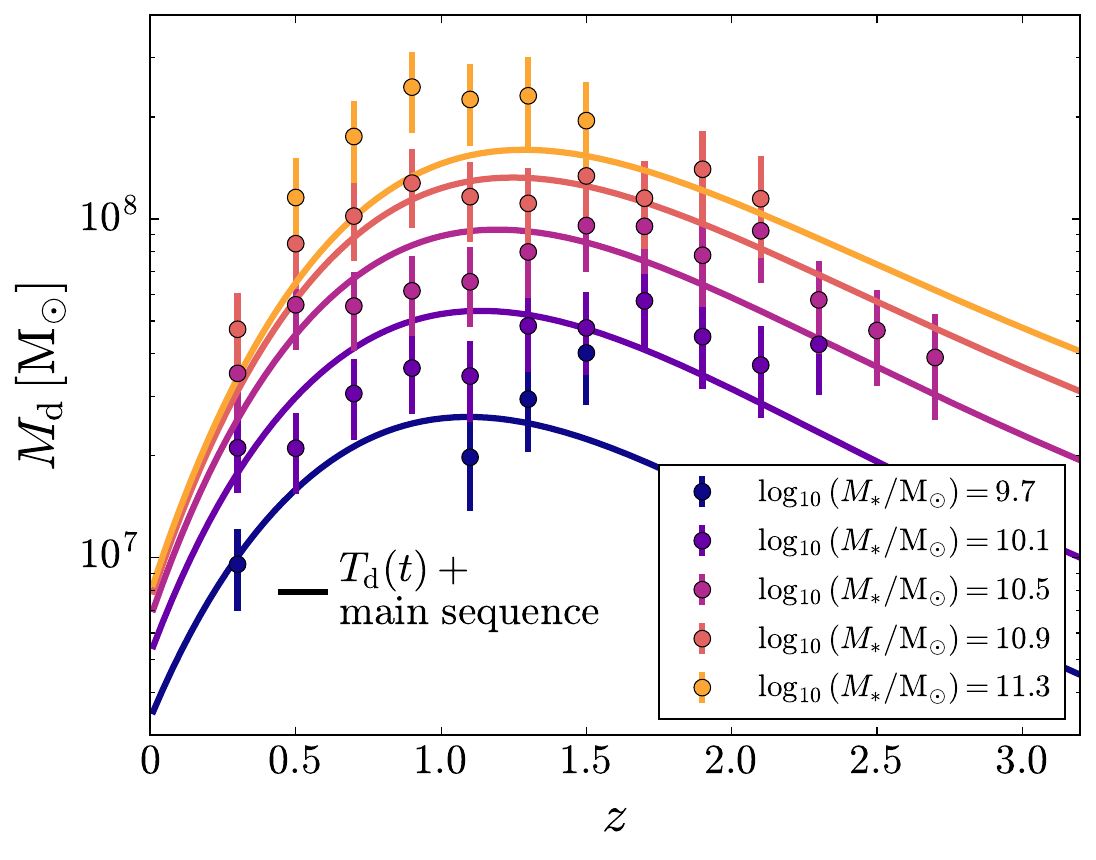}
    \includegraphics[width=0.46\textwidth]{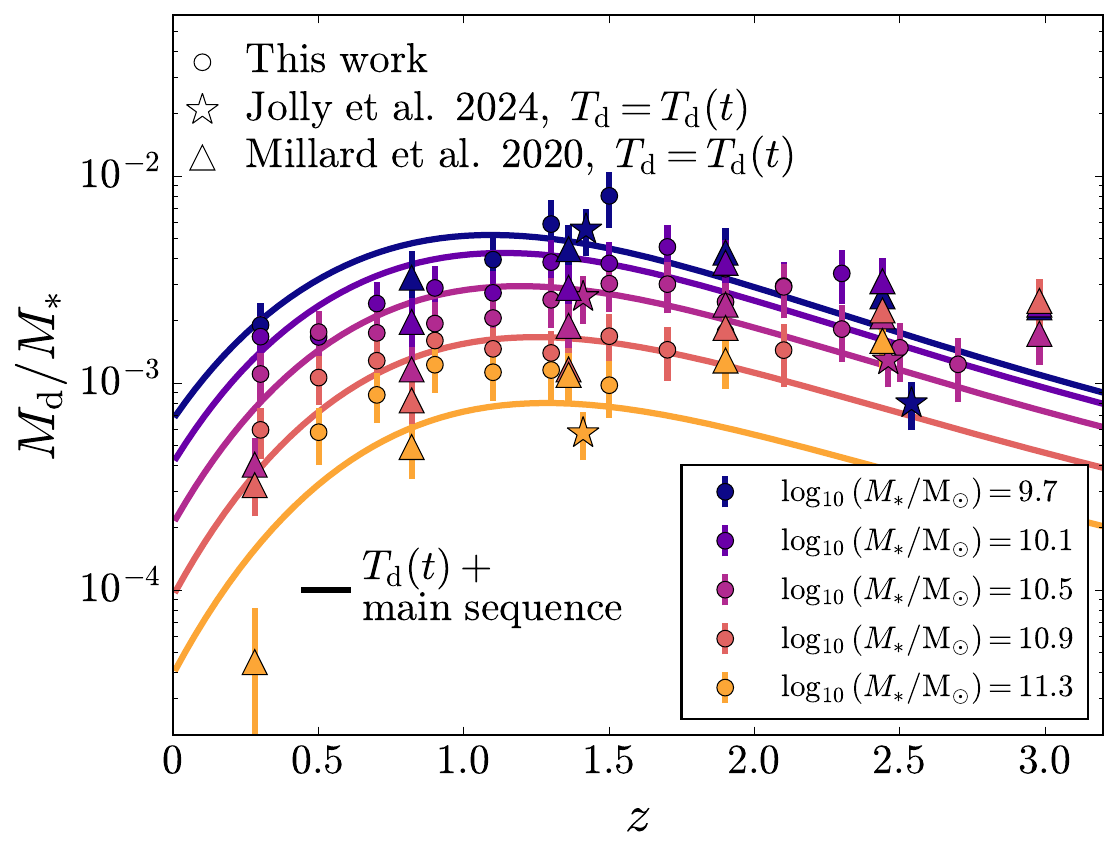}
    \caption{{\it Left:} Best-fit dust masses from our SED fits, $M_{\rm d}$, as a function of redshift, considering only the redshift and stellar mass bins that are ${>}\,95\%$ complete. {\it Right:} Same as the left panel but showing the best-fit dust mass-to-stellar mass ratio, where the stellar masses are from the \Euclid catalogue and are used to define the stacking bins. We also show SCUBA-2 stacking results from \citet{millard2020}, scaled to dust mass assuming the same modified blackbody SEDs used here and our best-fit dust temperature as a function of time, and ALMA 1.2\,mm stacking results from \citet{jolly2025} scaled in the same way. In both panels, the solid-coloured curves show the predicted trends for the corresponding stellar masses by combining the star-forming MS from \citet{popesso2023} with our fit to the dust temperature as a function of time.} 
    \label{fig:Md_z}
\end{figure*}

We next plotted the dust mass, $M_{\rm d}$, and dust-to-stellar mass ratio, $M_{\rm d}/M_{\ast}$, in the same way for the same stellar-mass-complete bins. The results are shown in \cref{fig:Md_z}. Here, we find an increase in dust mass (and the dust mass ratio) from $z\,{=}\,0.2$ to around $z\,{=}\,1$, followed by a plateau, for all stellar mass bins. We also find that the dust-to-stellar mass fraction decreases with increasing stellar mass at all redshifts.

These trends can be explained using our fit to \cref{eq:T_decay} and the fit to \cref{eq:galaxy_ms} from \citet{popesso2023}, or indeed any fit to the MS that agrees with our far-IR-measured SFRs. By combining these two equations we can solve for the dust mass as a function of $t$ (and therefore $z$); see Appendix \ref{app:4} for details. The resulting curves for the dust mass and dust-to-stellar mass ratio as a function of redshift are shown in \cref{fig:Md_z} for each stellar mass bin. It is important to emphasise that these are not fits to the data, but predictions based on fits to the dust temperatures, SFRs, and stellar masses. We find an increase in dust mass and dust-to-stellar mass ratio up to $z\,{=}\,1$. Beyond this redshift, the curves predict a decrease in these quantities as opposed to a plateau. However, we note that the actual functional forms are very sensitive to the best-fit parameters to the dust temperature and SFR as a function of $t$, and we do not have enough stacked photometry beyond $z\,{=}\,1.5$ to properly constrain the model. We also see the same trend of decreasing dust-to-stellar mass fraction with increasing stellar mass. The decrease in the dust-to-stellar mass ratio at redshifts ${<}\,1$ has been observed and discussed by \citet{bethermin2015} as a consequence of decreasing gas mass (although increasing metallicity works against this). At high redshift galaxies have high gas-to-stellar mass ratios as material is accreted from the large-scale environment, while at low redshift this gas has been consumed by star formation or ejected and the accretion rate is much lower. Since the dust mass is roughly proportional to the gas mass, we expect the dust mass to be lower in low-redshift galaxies as well. While metallicity competes against this trend (and it is important to keep in mind that high-redshift galaxies at a given stellar mass are not the same objects as low-redshift galaxies with the same stellar mass), the lower gas mass dominates the overall evolution and we observe decreasing dust-to-stellar mass ratios for $z\,{<}\,1$. Similarly, more massive galaxies have lower sSFRs (SFR$/M_{\ast}$), so they produce less dust per unit stellar mass, explaining the decrease in the dust-to-stellar mass fraction with stellar mass.

For comparison, we show the results from \citet{millard2020} and \citet{jolly2025}. Both studies stacked on all optically selected galaxies (i.e. star-forming and quiescent) rather than just star-forming galaxies as done here, and both only stacked one far-IR band (SCUBA-2 850\,$\mu$m for \citealt{millard2020}, and ALMA 1.2\,mm for \citealt{jolly2025}). To convert the single far-IR photometry points to dust masses, we used the same modified blackbody function fit to our data and our dust temperature model as a function of $t$ to scale the flux densities. The results are shown in the right panel of \cref{fig:Md_z}, where there is general agreement with our findings, although the dust mass-to-stellar mass ratios are lower at $z\,{<}\,1.5$, likely due to the presence of quiescent galaxies in the comparison stacking catalogues. Interestingly, at higher redshifts ($z\,{>}\,2$) where there is less contamination from quiescent galaxies, the stacked flux densities from \citet{millard2020} suggest a plateauing mass ratio as opposed to a drop-off. This can be further investigated with future \Euclid data releases once the deep fields get deeper, allowing us to probe higher redshifts with better statistical power.

\begin{figure}[htbp!]
    \centering
    \includegraphics[width=\columnwidth]{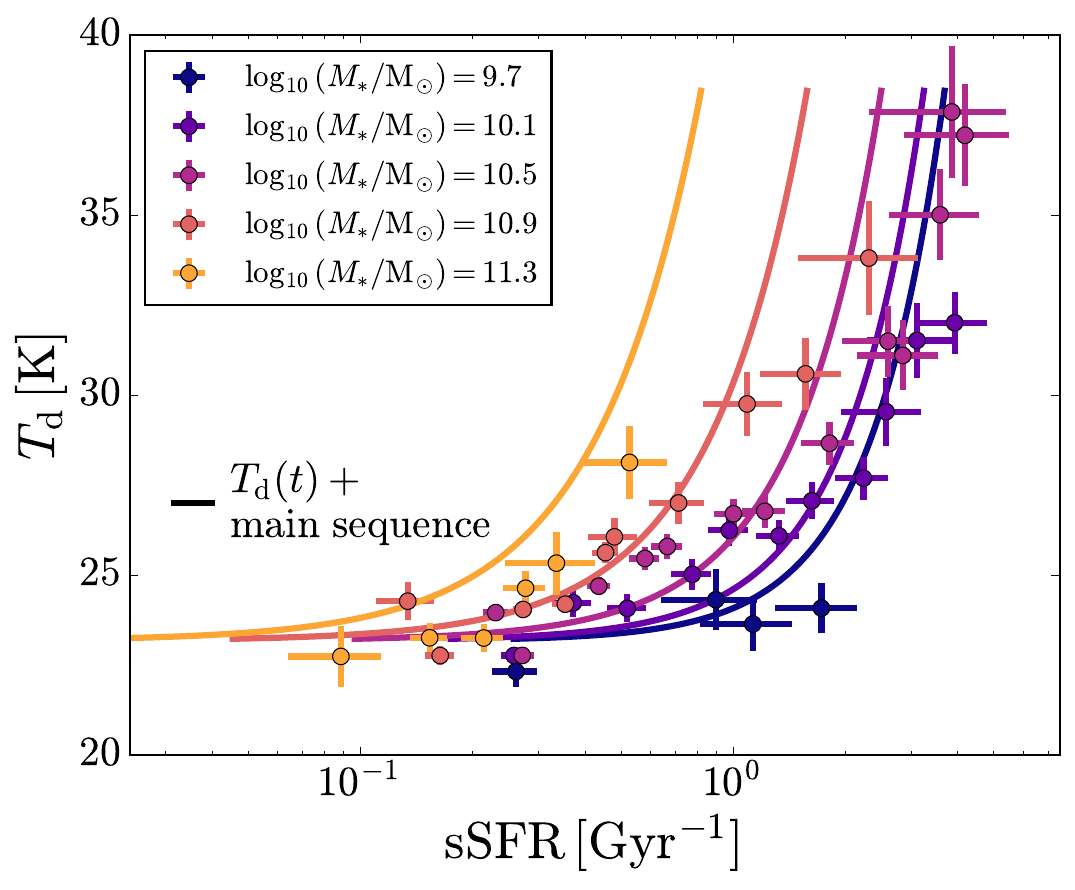}
    \caption{Dust temperature as a function of sSFR from our best-fit far-IR SEDs, colour-coded to show the measured stellar mass dependence. Here, we only show physical properties for bins that are ${>}\,95\%$ complete. The coloured curves show the predicted trends for the corresponding stellar masses by combining the star-forming MS from \citet{popesso2023} with our fit to the dust temperature as a function of time, ranging from $z\,{=}\,0.2$ (starting at the bottom-left of each curve) to $z\,{=}\,3.0$ (ending at the top-right of each curve).} 
    \label{fig:sSFR-T_relation}
\end{figure}

The increasing dust temperatures with redshift have been attributed to the fact that MS galaxies have higher sSFRs at higher redshifts \citep[e.g.][]{liang2019,koprowski2024} and therefore contain more massive, young hot stars. This can be seen in our stacking results in \cref{fig:sSFR-T_relation}, where we plot the dust temperature as a function of sSFR, separated by stellar mass in the same way as the previous plots. Galaxies with higher sSFRs clearly tend to have higher dust temperatures, although the trend depends on stellar mass and redshift. To understand why, we note that we can predict the sSFR--$T_{\rm d}$ relation for each stellar mass and redshift bin by combining the star-forming MS (Eq.~\ref{eq:galaxy_ms}) and our fit to the dust temperature-$t$ relation (Eq.~\ref{eq:T_decay}); see Appendix \ref{app:4} for details. In \cref{fig:sSFR-T_relation} we show these predictions, where for each stellar mass bin we calculated the relation between $z\,{=}\,0.2$ (bottom-left of each curve) and $z\,{=}\,3.0$ (top-right of each curve). Again, we emphasise that these are not fits to the data shown in the plot, but predictions based on other fits. At a given redshift, the dust temperatures of all star-forming galaxies are independent of stellar mass. However, higher stellar mass galaxies have lower sSFRs (due to the bending of the MS), which leads to the mass dependence seen in \cref{fig:sSFR-T_relation}. We note that the apparent convergence towards a constant sSFR at high redshift is simply a selection effect, since we measured the dust temperatures of lower mass galaxies out to higher redshifts than higher mass galaxies (see Fig.~\ref{fig:stack_sed_fits}).

One feature worth noting is that according to the star-forming MS, the SFRs (and sSFRs) of galaxies continuously decrease to zero as a function of time, yet the dust temperatures do not. Instead, the average dust temperatures appear to be converging to a constant value of about 23\,K, independent of stellar mass. Indeed, it seems reasonable to expect that the dust temperatures of MS galaxies are not strictly proportional to the SFR or sSFR, since this would imply that galaxies with no star formation would have zero dust temperature, which is impossible. Instead, \cref{eq:T_decay} implies that as the SFRs of galaxies fall below a certain threshold, the dust is no longer heated by hot young stars but by the existing cooler and older stellar population, which changes on timescales much longer than the current age of the Universe. Indeed, \citet{chapman2003} showed that IRAS-selected galaxies are bi-modal in luminosity-versus-far-IR colour (a proxy for temperature) with a break around 10\,$M_{\odot}\,$yr$^{-1}$, and discussed a possible transition from cirrus-dominated SEDs to SFR-dominated SEDs. Similarly, \citet{groves2012} found that the mean dust temperature within M31 increases from about 17\,K in the disc to 35\,K in the bulge, despite the lack of any strong star-forming regions throughout the galaxy, suggesting that the heating is driven primarily by the density of the old stellar population.

\subsection{\Euclid's contribution to the CIB}
\label{subsec:discussion_cib}

Looking at the CIB estimates in \cref{table:resolved_cib}, the latest estimates cannot be said to be in close agreement; hence these CIB determinations are still dominated by systematic effects. This highlights the difficulty in subtracting zodiacal light and emission from the Milky Way, i.e. assessing the level of appropriate zero points when trying to determine the level of the extragalactic background. We can therefore only roughly estimate the fraction of the CIB resolved by the current Q1 \Euclid catalogue: about 30--80\% at 100\,$\mu$m; 40--70\% at 160\,$\mu$m; 70--120\% at 250\,$\mu$m; 70--120\% at 350\,$\mu$m; 60--80\% at 500\,$\mu$m; and 30--40\% at 850\,$\mu$m. 

A similar calculation was previously carried out using catalogues from the COSMOS field stacked on SPIRE images using the same stacking algorithm used here \citep{Duivenvoorden2020}. The authors found that $r$-band catalogues down to about magnitude 26 or $K$-band catalogues down to about magnitude 24 can recover essentially all of the CIB measured by FIRAS at all three SPIRE wavelengths. This provides a good benchmark for \Euclid\ -- the depth of the VIS images is about 24.7 in VIS and 23.2 in NISP, which approaches the depth of the COSMOS catalogues, and with a substantially larger numbers of catalogued objects.

In particular, at the SPIRE wavelengths we recovered ${>}\,60\%$ of the CIB. These results are in line with \citet{Duivenvoorden2020}, considering the $r$ and $K$ bands they used to create near-IR catalogues in their stacking study do not match perfectly with \Euclid's VIS and NISP instruments. It is reasonable to conclude that while we have not yet resolved the entire CIB with \Euclid in Q1, ${>}\,60\%$ is consistent. As the Euclid Deep Fields get deeper, we can therefore expect to approach a more complete recovery of the CIB through \Euclid-selected galaxies.

\subsection{Comparison to the MAMBO simulation}
\label{subsec:simulation}

A number of simulations able to reproduce \Euclid observations have been investigated, including the MAMBO (Mocks with Abundance Matching in BOlogna; \citealt{girelli2020}) mock catalogue. The MAMBO catalogue is based on the Millennium Simulation \citep{springel2005} with physical properties prescribed using Empirical Galaxy Generator (\texttt{EGG}) code \citep{schreiber2017} across 3.14\,deg$^2$. In particular, magnitudes in the \Euclid bands, as well as far-IR flux densities at the \Herschel and SCUBA-2 bands used here, were calculated from the simulation by \citet{parmar2026} and used to construct a mock \Euclid catalogue at the depth of the Q1 catalogue (equal to the depth of the Euclid Wide Survey). Simulated stacked far-IR flux densities of star-forming MS galaxies were then calculated within the same redshift and stellar mass bins used here, from which we fitted the same modified blackbody SEDs.

\begin{figure}[htbp!]
    \centering
    \includegraphics[width=\columnwidth]{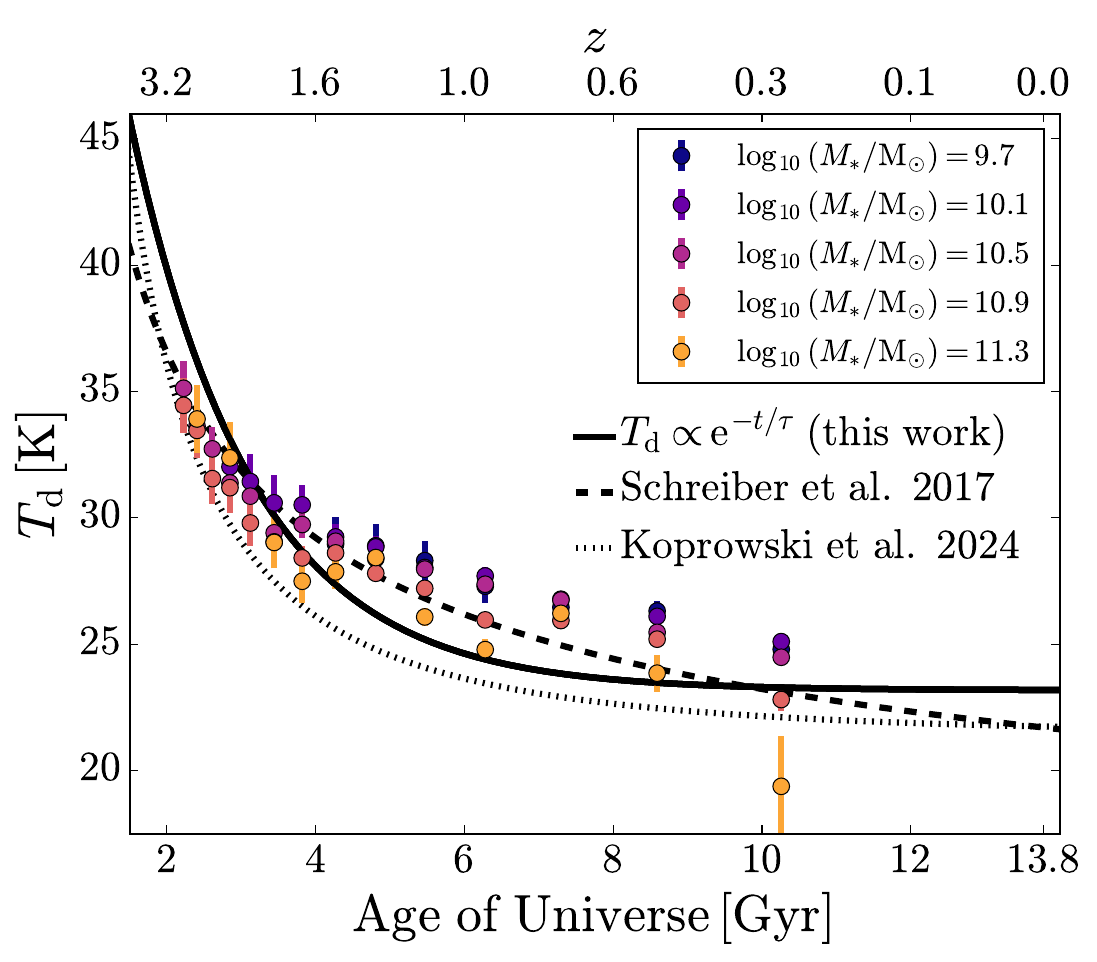}
    \caption{Same as \cref{fig:Td_z}, but with the data points derived from fits to simulated stacked far-IR photometry in the MAMBO simulation \citep{parmar2026}.} 
    \label{fig:Td_z_sim}
\end{figure}

From this simulation we found slightly higher far-IR SFRs (although by a factor ${<}\,2$) and dust masses (also larger by a factor ${<}\,2$), with no significantly different redshift trends compared to our measurements. Interestingly, the simulated dust temperatures as a function of time are well-recovered by our pipeline, as show in \cref{fig:Td_z_sim}. The simulation does not show the same plateauing behaviour as with our measurements, but instead decrease monotonically with time across the entire redshift range investigated here. This difference can be attributed to details of \texttt{EGG}, which assigns dust temperatures using an equation linear in redshift (Eq. 14 in \citealt{schreiber2017}) as opposed to quadratic in redshift, which better matches our observations.

\subsection{Star-formation rate density}

\begin{figure}[htbp!]
    \centering
    \includegraphics[width=\columnwidth]{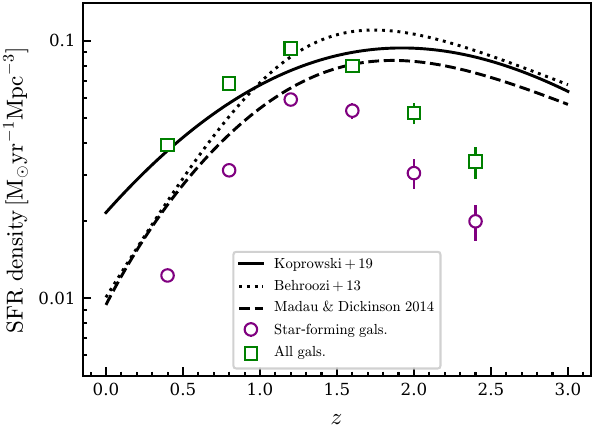}
    \caption{SFRD as a function of redshift solely for star-forming galaxies (purple) and the full \Euclid catalogue (green). The black curves show published SFRD fits from  \cite{koprowski2017}, \cite{Behroozi2013}, and \cite{Madau2014}, with the latter scaled by $0.63$ to convert from Salpeter to Chabrier initial mass functions. Beyond redshift 1.5 our estimates of the SFRD are incomplete because we are not able to recover enough far-IR photometry to fit SEDs.} 
    \label{fig:sfr_density}
\end{figure}

Lastly, we calculated the far-IR-derived SFR density (SFRD) as a function of redshift coming from the \Euclid catalogue. We multiplied the number of SPIRE stacking galaxies by the mean SFR, then summed the contributions from each stellar mass bin at a given redshift, and divided them by the volume of the redshift slice in the maps. Lastly, we averaged over every second redshift bin. In \cref{fig:sfr_density} we show the results for star-forming galaxies in purple, compared to several published estimates \citep{Behroozi2013,Madau2014,koprowski2017}. Since these literature curves include all galaxies (not just star-forming ones), we re-ran our stacking pipeline on the full \Euclid catalogue using the same stellar mass and redshift bins, then re-fitted the same SEDs to derive the mean SFRs of all galaxies in each bin. We show the resulting SFRD for all \Euclid galaxies as the green points.

We find that at $z<1.5$ our total SFRD agrees well with the literature, but past this redshift our results are incomplete. This makes sense looking at \cref{fig:stack_sed_fits}, which shows that we are not able to recover enough far-IR photometry for high-stellar-mass galaxies at high redshifts to fit SEDs and derive SFRs. Future \Euclid data releases will include more of these galaxies overlapping with more SPIRE data, allowing us to measure the complete SFRD past $z\,{=}\,1.5$. These SFRD values should be regarded as lower bounds, since stacking only recovers the mean flux of cataloged galaxies and can miss contributions from galaxies that were not detected or heavily obscured galaxies. 

\section{Conclusions}
\label{sec:conclusion}

We stacked over 2 million star-forming galaxies from the \Euclid Q1 catalogue across 17.6\,deg$^2$ of far-IR imaging, providing robust statistics for their mean far-IR flux densities. We performed our stacking on \Herschel-PACS 100- and 160-$\mu$m maps, \Herschel-SPIRE 250-, 350-, and 500-$\mu$m maps, and SCUBA-2 850-$\mu$m maps. To avoid biases related to clustering, we used the {\tt SimStack} algorithm, which simultaneously fits flux densities to far-IR beam-convolved model images of galaxy distributions in different bins.

Given the large number of galaxies available for stacking, we split the \Euclid star-forming catalogue into eight stellar mass bins from $\logten(M_{\ast}/M_{\odot})\,{=}\,8.3\,$ to $\logten(M_{\ast}/M_{\odot})\,{=}\,11.5\,$, and 14 redshift bins from $z\,{=}\,0.2$ to $z\,{=}\,3.0$. Beyond these ranges, the \Euclid stellar masses and redshifts are no longer reliable. We find significant stacked detections in most bins at all wavelengths.

Using these average flux densities, we modelled the average far-IR SEDs of \Euclid galaxies using a modified blackbody function transitioning to a power law at high frequencies, finding good fits where we are able to measure the stacked flux densities. From our fits we derived average dust temperatures, dust masses, and far-IR-derived SFRs, and we find significant redshift evolution in all of these parameters. In particular, we investigate the difference between our far-IR-derived SFRs and the SFRs predicted from the star-forming MS. We find consistent values between the two estimates, with no significant trend in redshift. The average dust temperature decreases as a function of time following a functional form $T_2\,{+}\,(T_1\,{-}\,T_2)\,{\rm e}^{-t/\tau}$, with no stellar mass dependence. We argue that the dust temperatures of MS galaxies below $z\,{=}\,1$ have converged to a constant value ($T\,{\simeq}\,23\,$K) because the dust is now primarily heated by existing cooler and older stellar populations as opposed to hot young stars in star-forming regions. We also find that the average dust-to-stellar mass ratio increases for galaxies of all stellar mass up to $z\,{\simeq}\,1$, and decreases with increasing stellar mass. This is a consequence of higher gas mass-to-stellar mass ratios at higher redshifts, although decreasing metallicity competes against this trend. Similarly, more massive galaxies have lower dust-to-stellar mass ratios due to their lower sSFRs compared to low mass galaxies. Lastly, we showed that the correlation between dust temperature and SFR (and therefore sSFR) is stellar mass-dependent since dust temperatures are stellar mass-independent. 

We compared our results to a recent mock \Euclid catalogue with derived far-IR photometry, finding good agreement for the SFRs and dust masses. However, we showed that the simulated catalogue predicts consistently decreasing dust temperatures below $z\,{=}\,1$, in disagreement with our observation. We attribute this discrepancy to the model used to produce the dust temperatures, which assigns dust temperatures using a monotonically decreasing function of redshift as opposed to a functional form which converges to a constant value at low $z$.

In the future, \Euclid will observe more area overlapping with far-IR surveys and will obtain deeper VIS and NISP imaging in the Euclid Deep Fields, where some of the best existing far-IR and submillimetre imaging lies. These advances will provide even better statistics than are currently available, allowing our stacking analysis to extend to higher redshifts and lower stellar masses. In addition, upcoming far-IR facilities such as the Cerro Chajnantor Atacama Telescope \citep{CCAT2023} will play an important role in better constraining the Rayleigh--Jeans tail of the average far-IR SEDs, improving the SED constraints and derived physical parameters.

\begin{acknowledgements}
\AckQone
\AckEC
%Canada
This work was supported by the Natural Sciences and Engineering Research Council of Canada and the Canadian Space Agency. 
This research used the Canadian Advanced Network For Astronomy Research (CANFAR) operated in partnership by the Canadian Astronomy Data Centre and The Digital Research Alliance of Canada with support from the National Research Council of Canada, the Canadian Space Agency, CANARIE, and the Canadian Foundation for Innovation.
%\herschel\
The \textit{Herschel} spacecraft was designed, built, tested, and launched under a contract to ESA managed by the \textit{Herschel}/\textit{Planck} Project team by an industrial consortium under the overall responsibility of the prime contractor Thales Alenia Space (Cannes), and including Astrium (Friedrichshafen), Thales Alenia Space (Turin), and Astrium (Toulouse), with in excess of a hundred subcontractors.
PACS was developed by a consortium of institutes led by MPE (Germany) and including: UVIE (Austria); KU Leuven, CSL, IMEC (Belgium); CEA, LAM (France); MPIA (Germany); INAF-IFSI/OAA/OAP/OAT, LENS, SISSA (Italy); and IAC (Spain). This development has been supported by the funding agencies BMVIT (Austria), ESA-PRODEX (Belgium), CEA/CNES (France), DLR (Germany), ASI/INAF (Italy), and CICYT/MCYT (Spain).
SPIRE was developed by a consortium of institutes led by Cardiff University (UK) and including: Univ.\ Lethbridge (Canada); NAOC (China); CEA, LAM (France); IFSI, Univ.\ Padua (Italy); IAC (Spain); Stockholm Observatory (Sweden); Imperial College London, RAL, UCL-MSSL, UKATC, Univ. Sussex (UK); and Caltech, JPL, NHSC, Univ.\ Colorado (USA). This development has been supported by national funding agencies: CSA (Canada); NAOC (China); CEA, CNES, CNRS (France); ASI (Italy); MCINN (Spain); SNSB (Sweden); STFC, UKSA (UK); and NASA (USA).
%JCMT
We acknowledge use of data obtained with the JCMT, which is operated by the EAO on behalf of NAOJ, ASIAA, KASI, and CAMS as well as the National Key R\&D Program of China (No.~2017YFA0402700). Additional funding support is provided by the STFC and participating universities in the UK and Canada.  Additional funds for the construction of SCUBA-2 were provided by the Canada Foundation for Innovation.

\end{acknowledgements}

\bibliographystyle{aa}
\bibliography{Euclid.bib,spire_stacking.bib}

\appendix

\section{PACS and SCUBA-2 field overviews}
\label{app:1}

\begin{figure}[htbp!]
    \centering
    \includegraphics[width=\columnwidth]{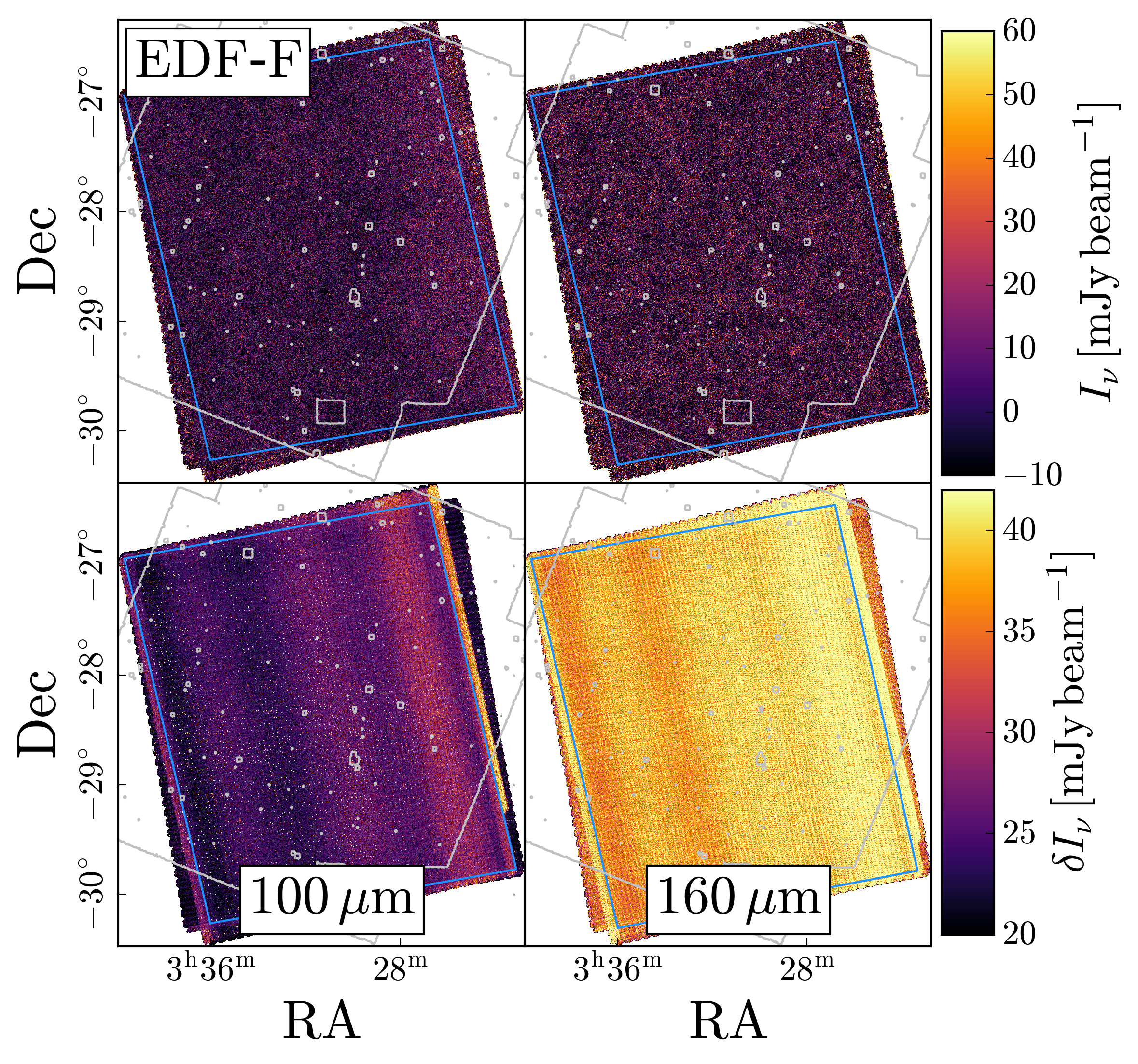}
    \includegraphics[width=\columnwidth]{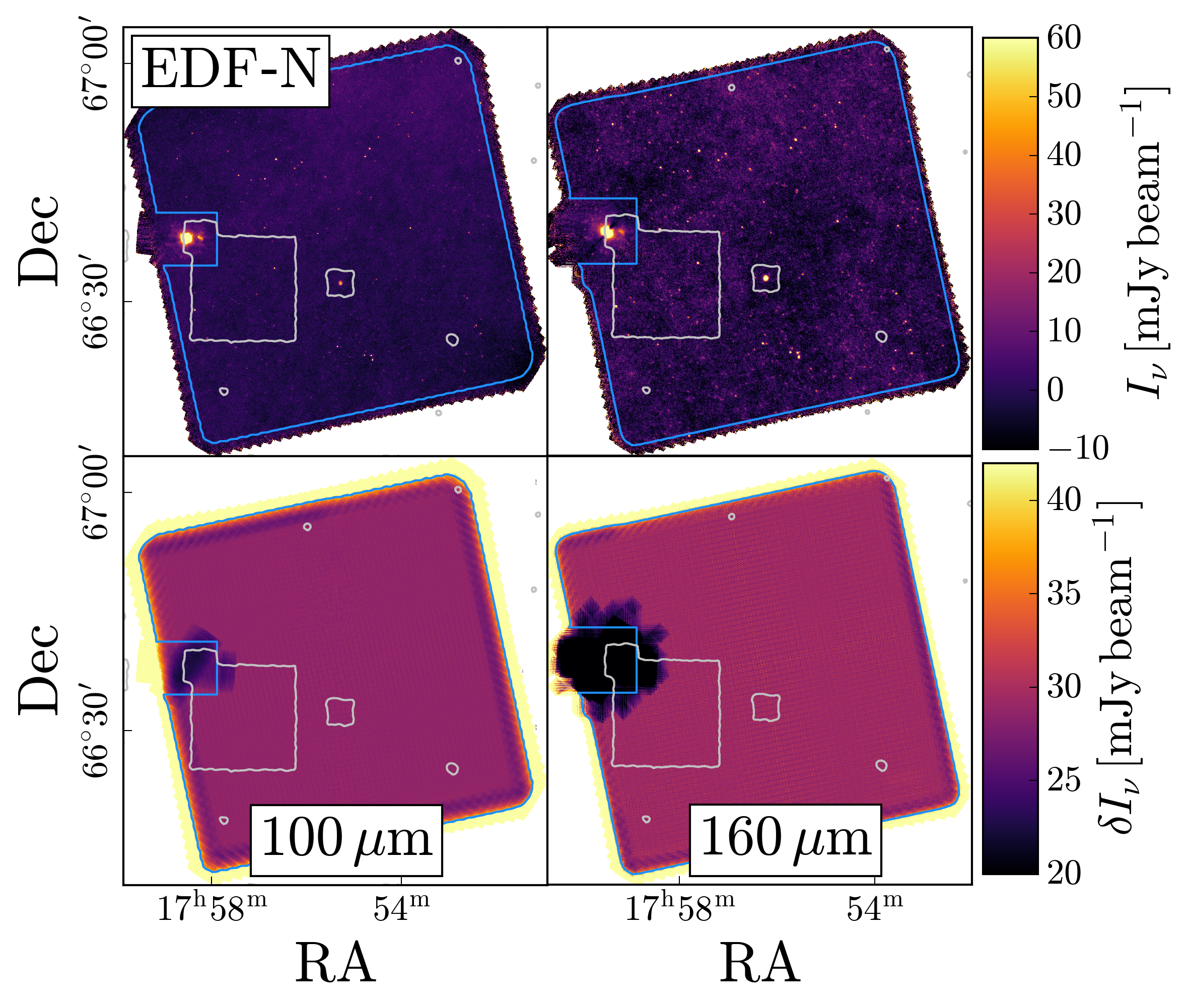}
    \caption{{\it Top:} Same as \cref{fig:field_summary_edff}, but for the PACS observations of the CDFS-SWIRE (EDF-F) field. {\it Bottom:} Same as the above panel, but for the PACS observations of the AKARI-NEP (EDF-N) field.} 
    \label{fig:field_summary_edff_pacs}
\end{figure}

\begin{figure}[htbp!]
    \centering
    \includegraphics[width=\columnwidth]{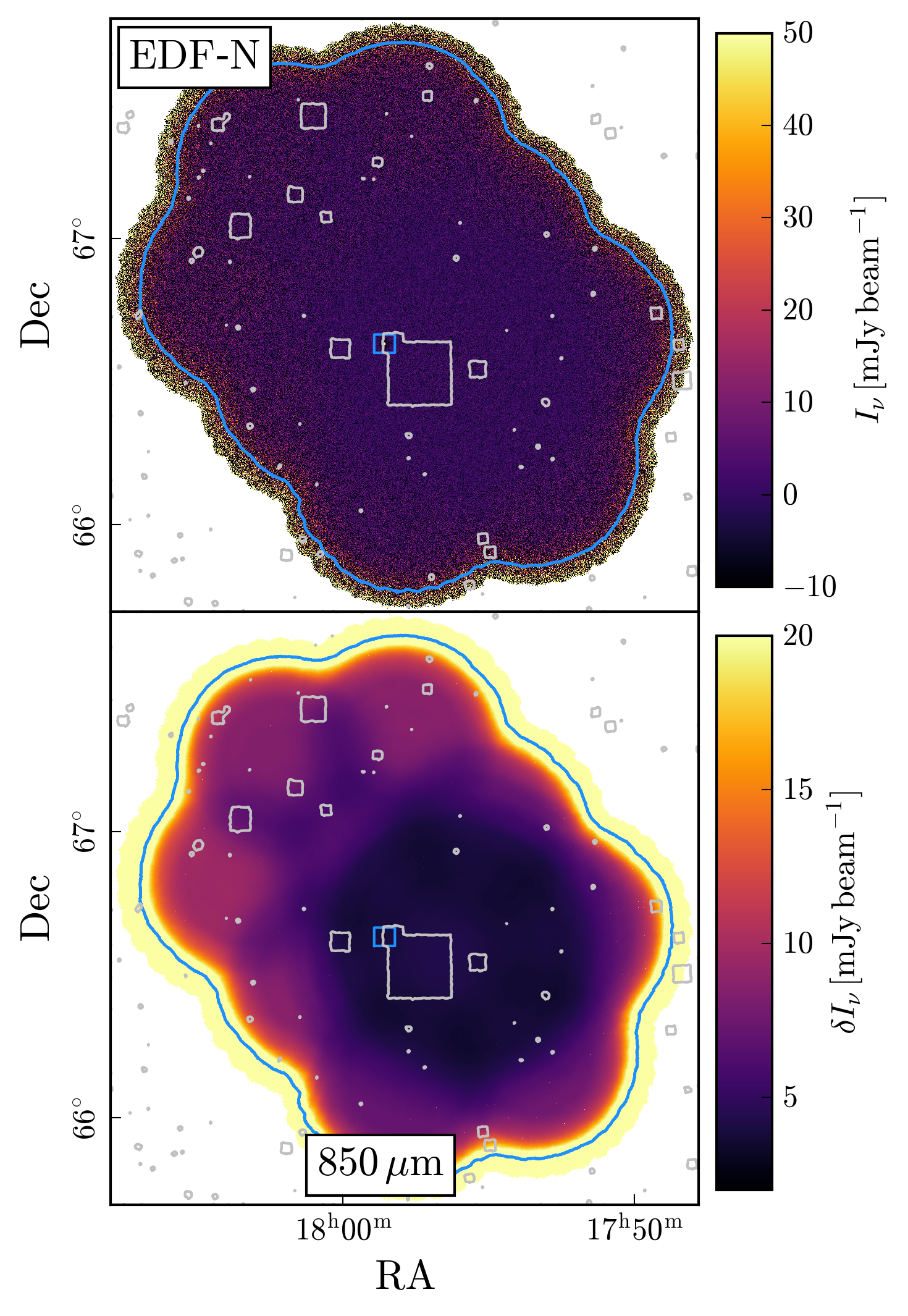}
    \caption{Same as \cref{fig:field_summary_edff} but for the SCUBA-2 observations of the AKARI-NEP (EDF-N) field.} 
    \label{fig:field_summary_scuba2}
\end{figure}

In \cref{fig:field_summary_edff_pacs} we show the \Herschel-PACS 100- and 160-$\mu$m data and RMS maps overlapping with the EDF-F and EDF-N. We also show the SCUBA-2 data and RMS map overlapping with the EDF-N in \cref{fig:field_summary_scuba2}. In both figures the \Euclid mask is shown as the grey contours. The bright resolved source in the EDF-N is NGC\,6543 (the `Cat's Eye Nebula'), which we mask prior to the stacking.

\section{Stacking cutouts and flux densities}
\label{app:2}

\begin{figure*}[htbp!]
    \centering
    \includegraphics[width=0.48\textwidth]{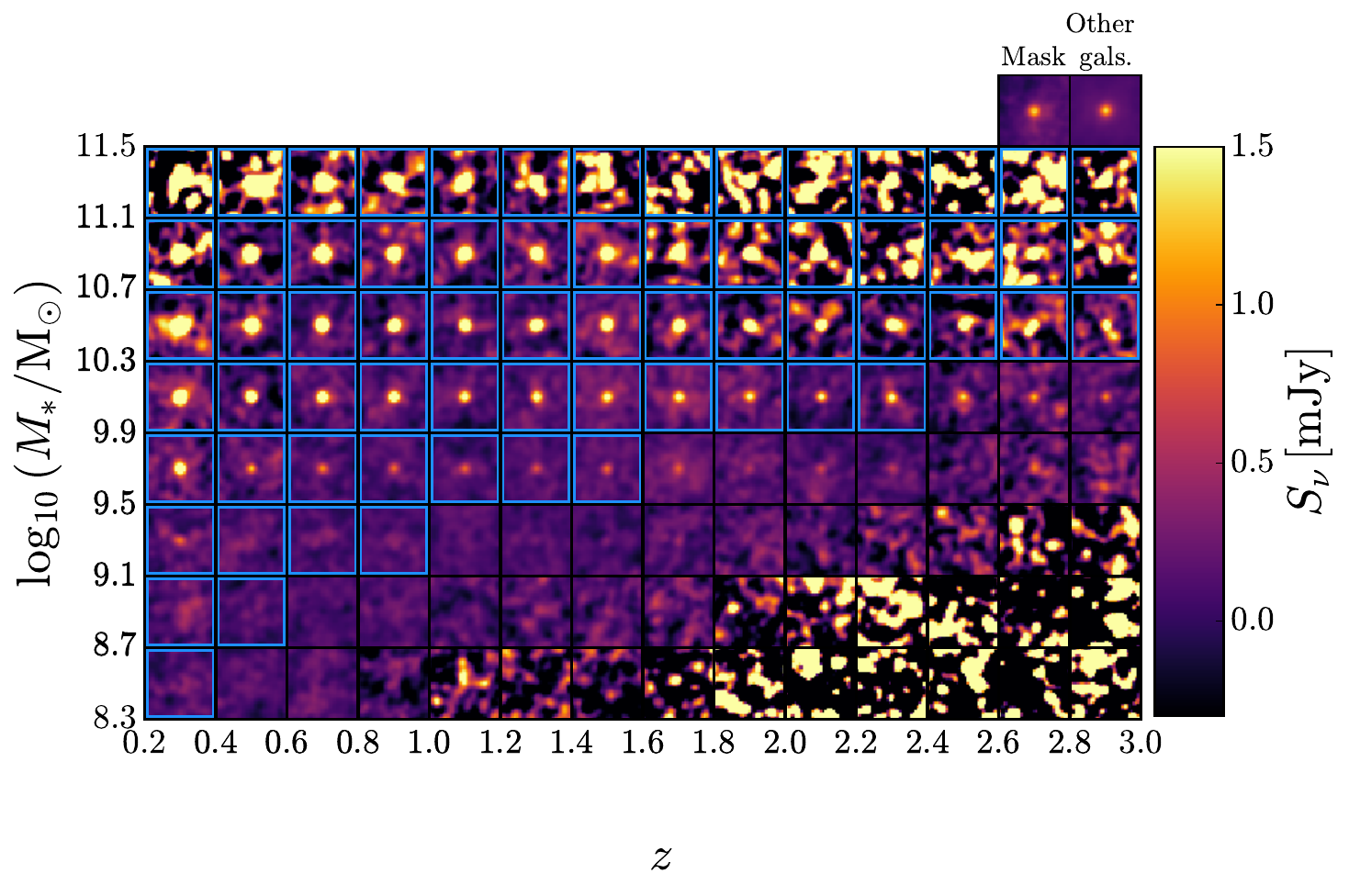}
    \includegraphics[width=0.48\textwidth]{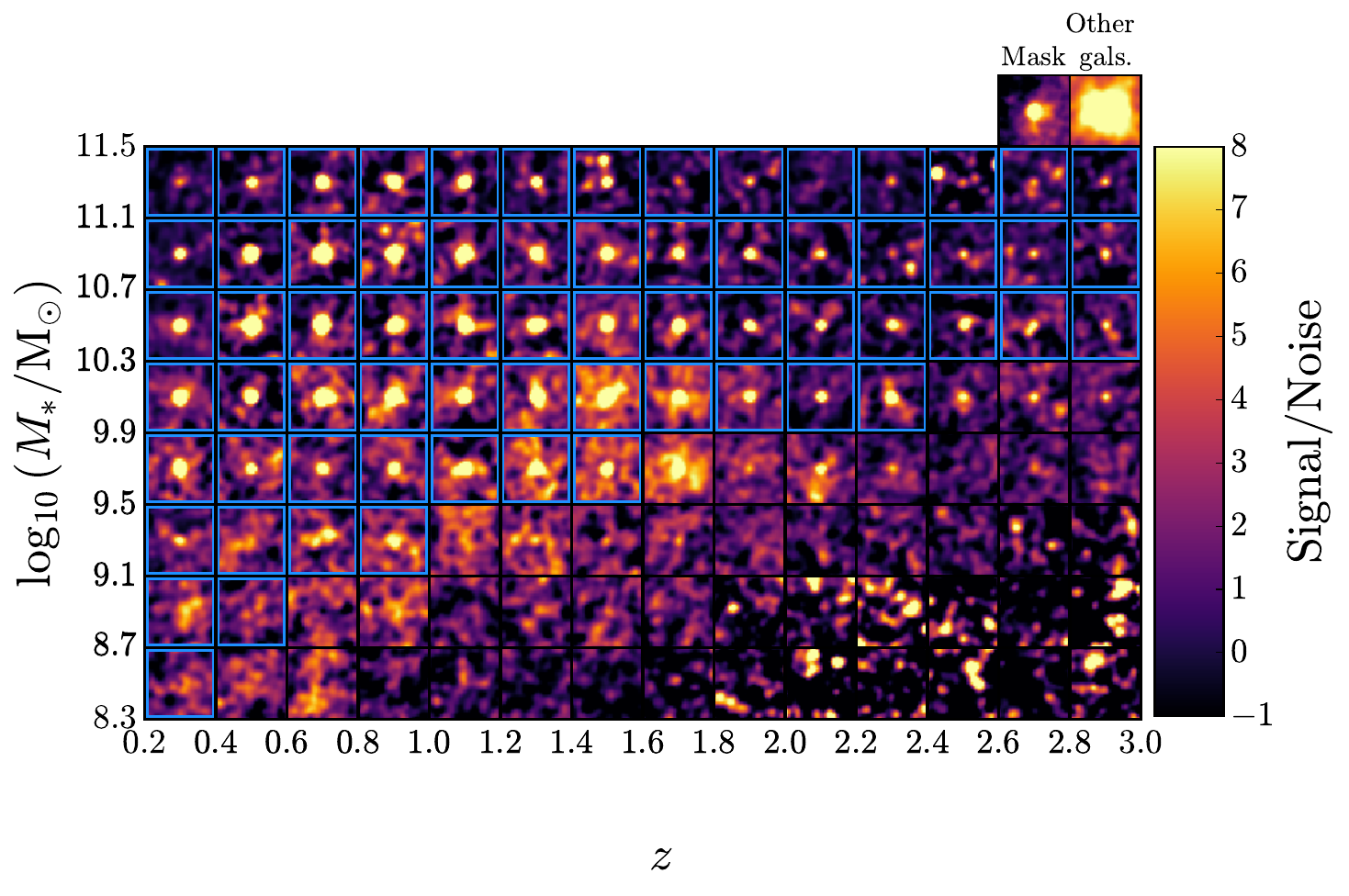}
    \includegraphics[width=0.48\textwidth]{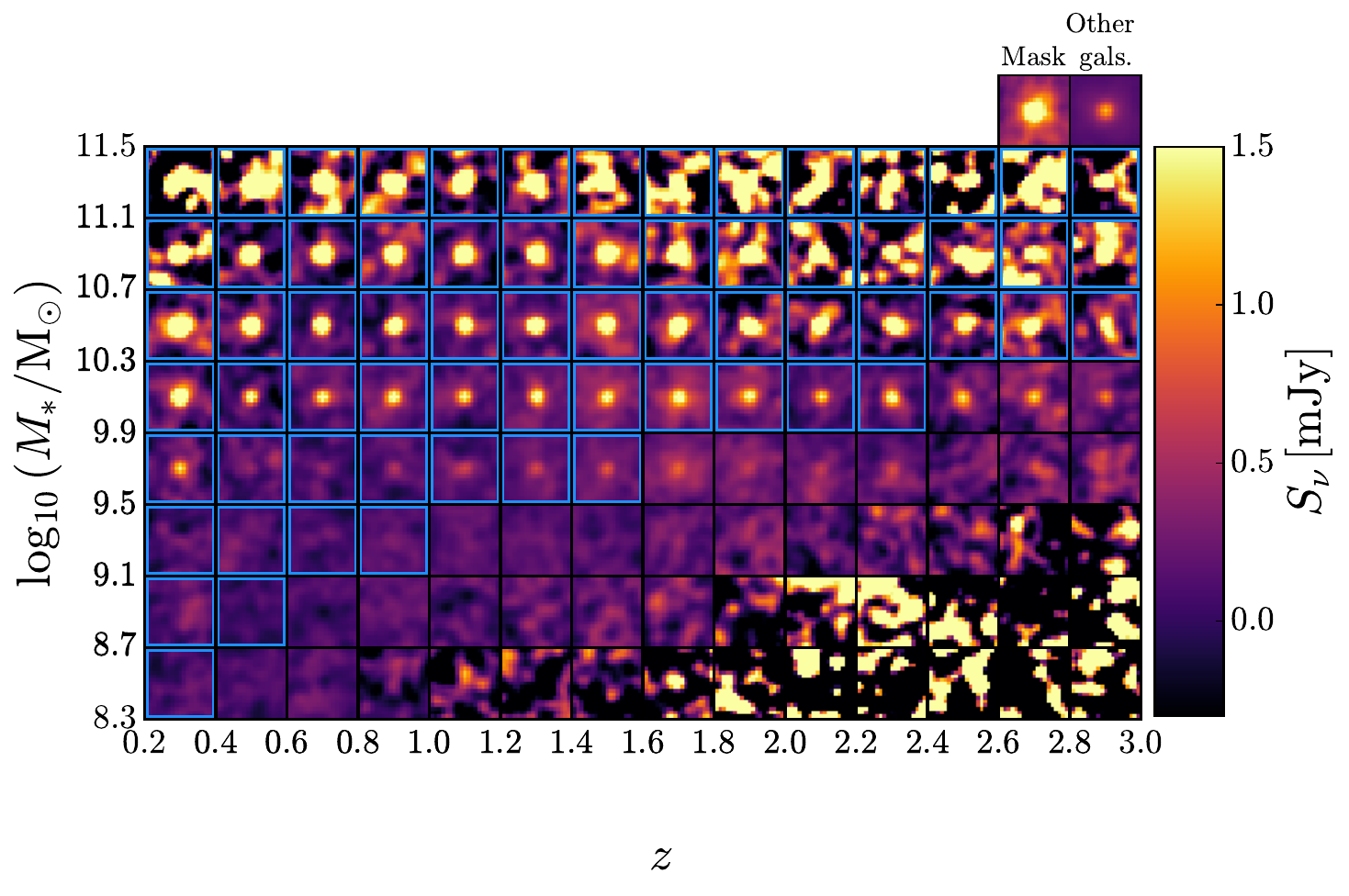}
    \includegraphics[width=0.48\textwidth]{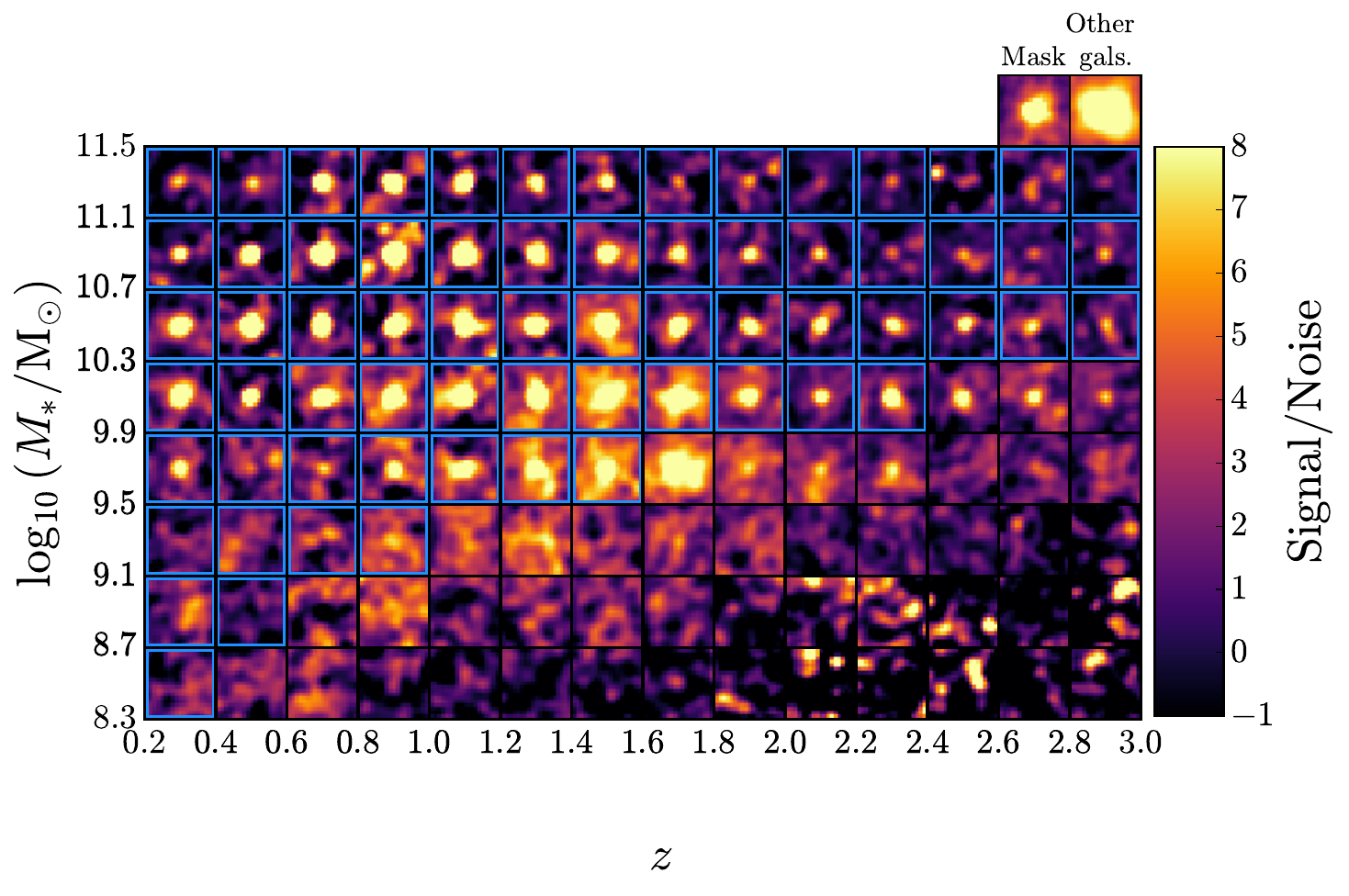}
    \includegraphics[width=0.48\textwidth]{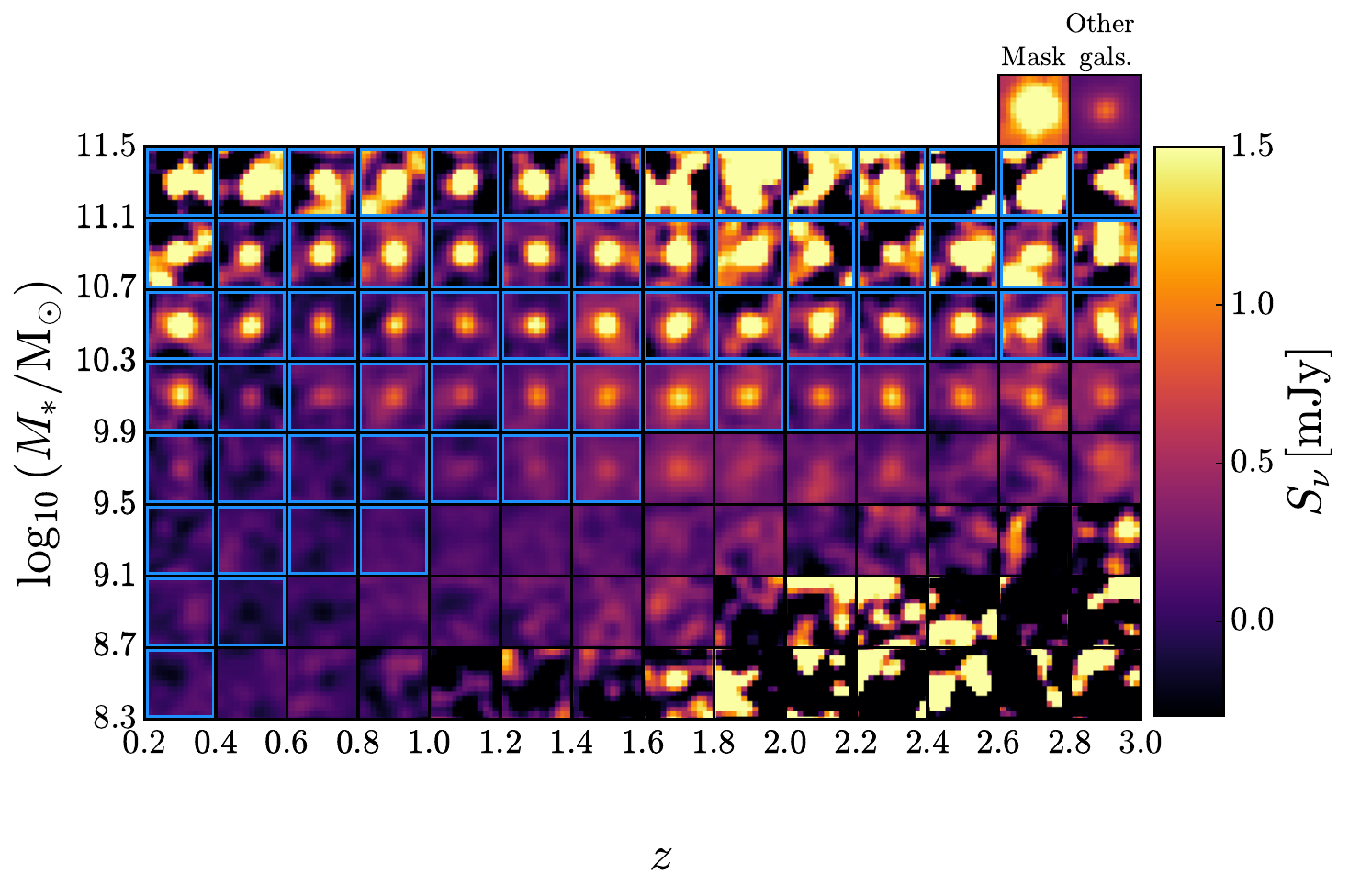}
    \includegraphics[width=0.48\textwidth]{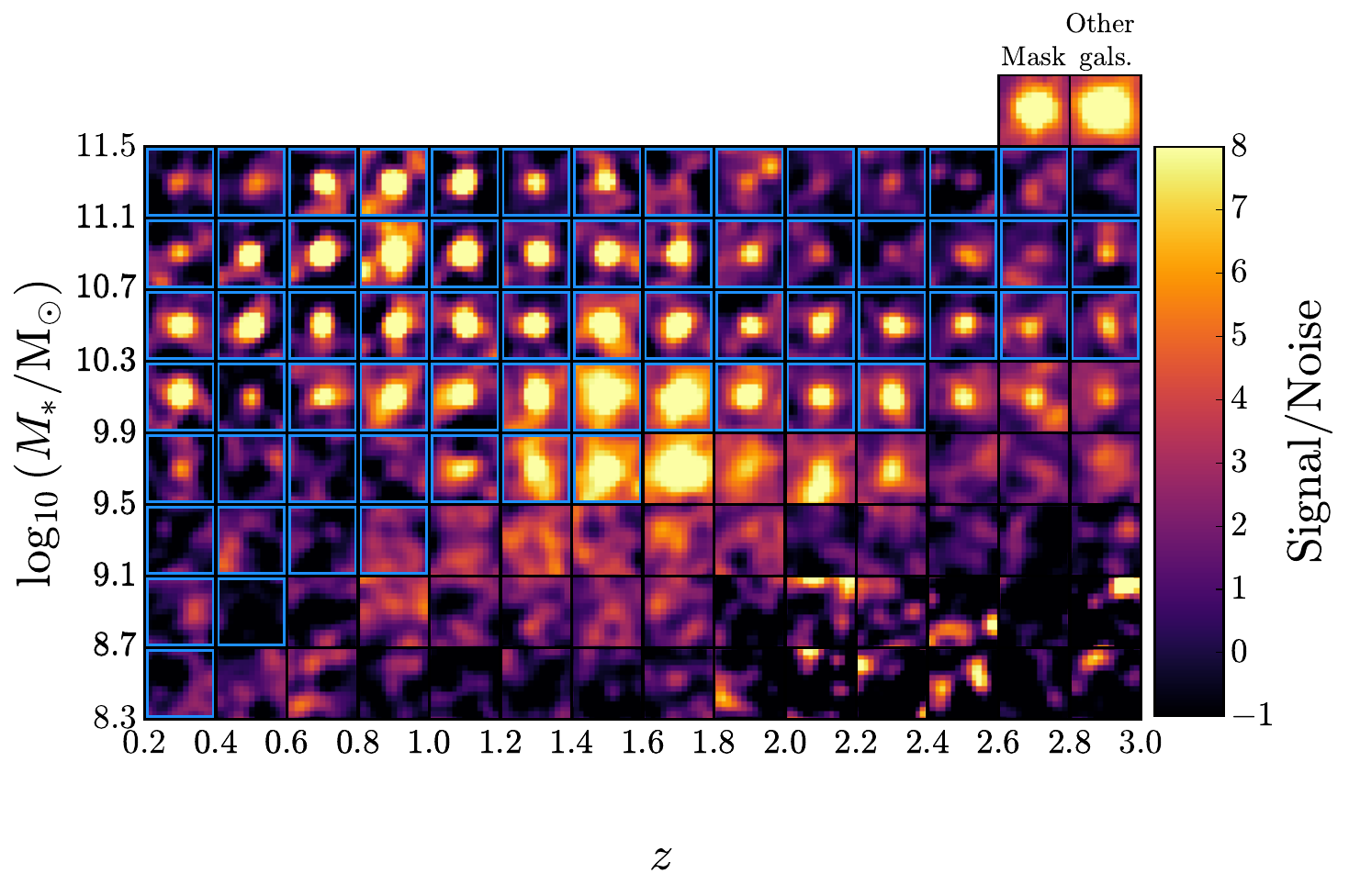}
    \caption{\Herschel-SPIRE results from stacking star-forming MS galaxies from the \Euclid catalogue \citep{Q1-SP031} with redshifts between 0.2 and 3.0, and stellar masses between $\logten(M_{\ast}/M_\odot)\,{=}\,8.3$ and 11.5. Each 2D cutout is $200^{\prime\prime}\,{\times}\,200^{\prime\prime}$. Bins that are ${>}\,95\%$ complete in stellar mass \citep{Q1-SP031} are highlighted in blue. {\it Left column:} Stacking signal in units of mJy. {\it Right column:} The S/N of the stacked flux densities. The `Mask' and `Other gals.' panels differ from the colour bar and range from S/N$\,{=}\,-1$ to S/N$\,{=}\,60$. {\it Top row:} SPIRE 250-$\mu$m. {\it Middle row:} SPIRE 350-$\mu$m. {\it Bottom row:} SPIRE 500-$\mu$m.} 
    \label{fig:stack_results_main}
\end{figure*}

\begin{figure*}[htbp!]
    \centering
    \includegraphics[width=0.48\textwidth]{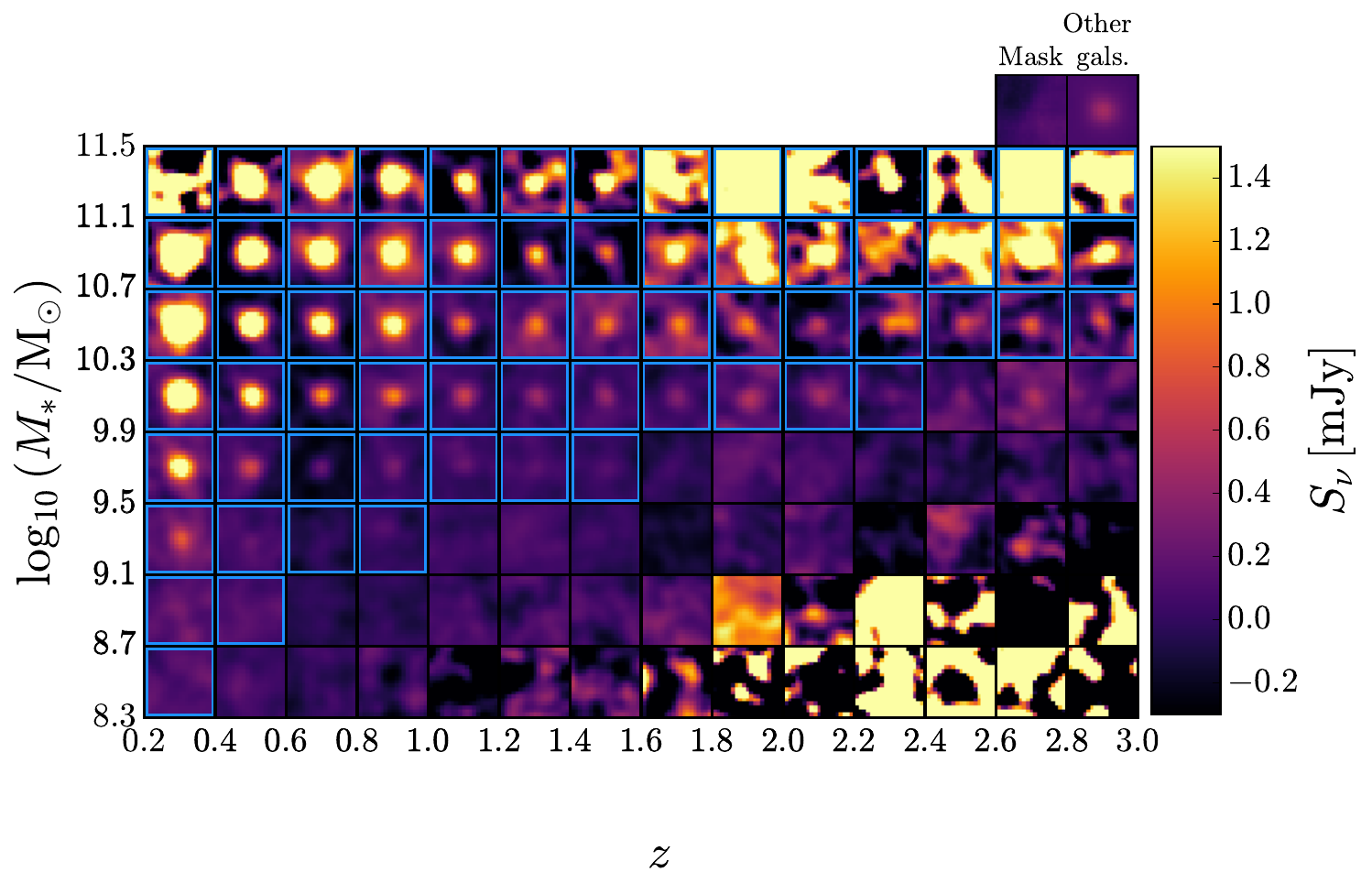}
    \includegraphics[width=0.48\textwidth]{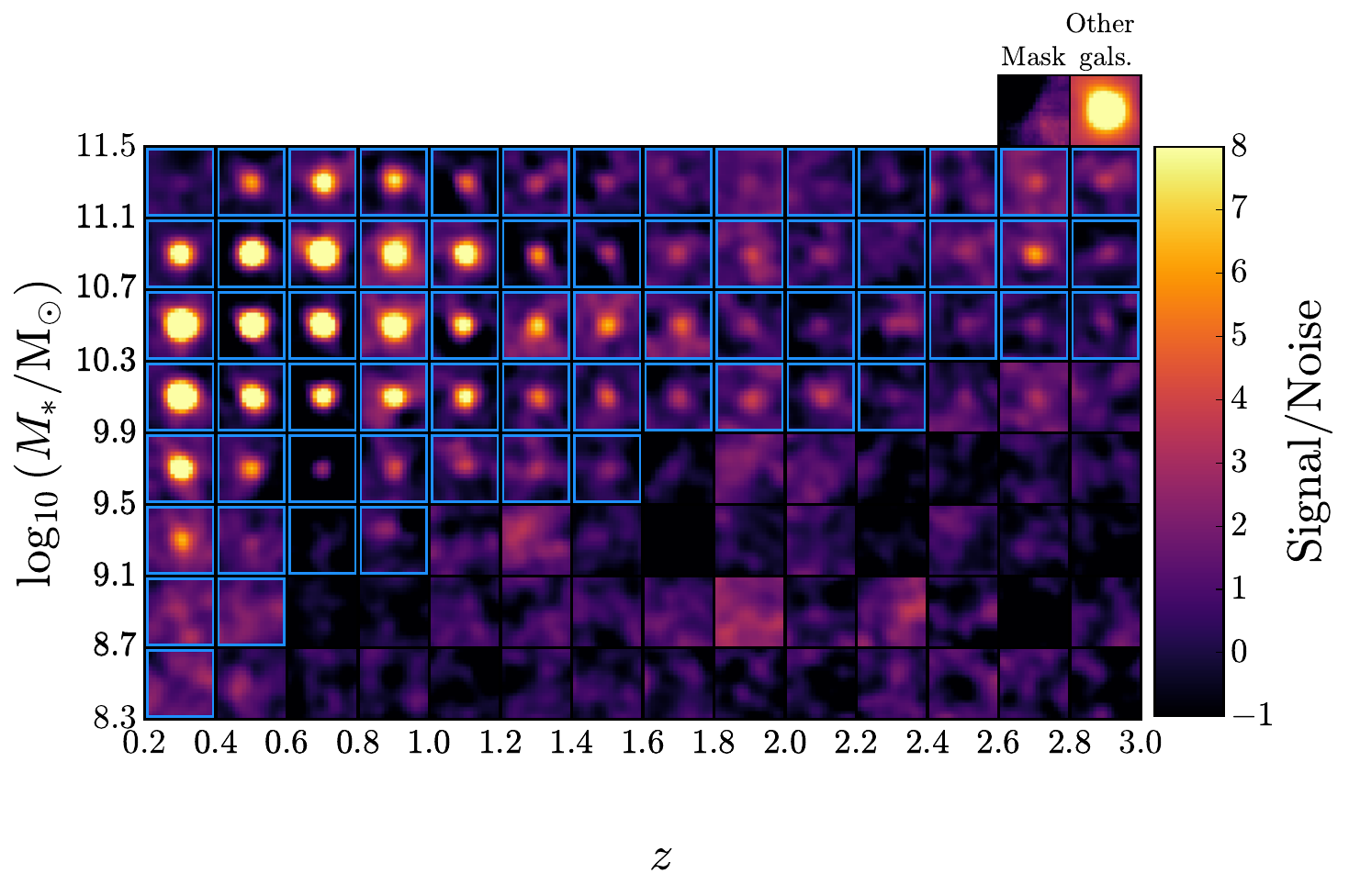}
    \includegraphics[width=0.48\textwidth]{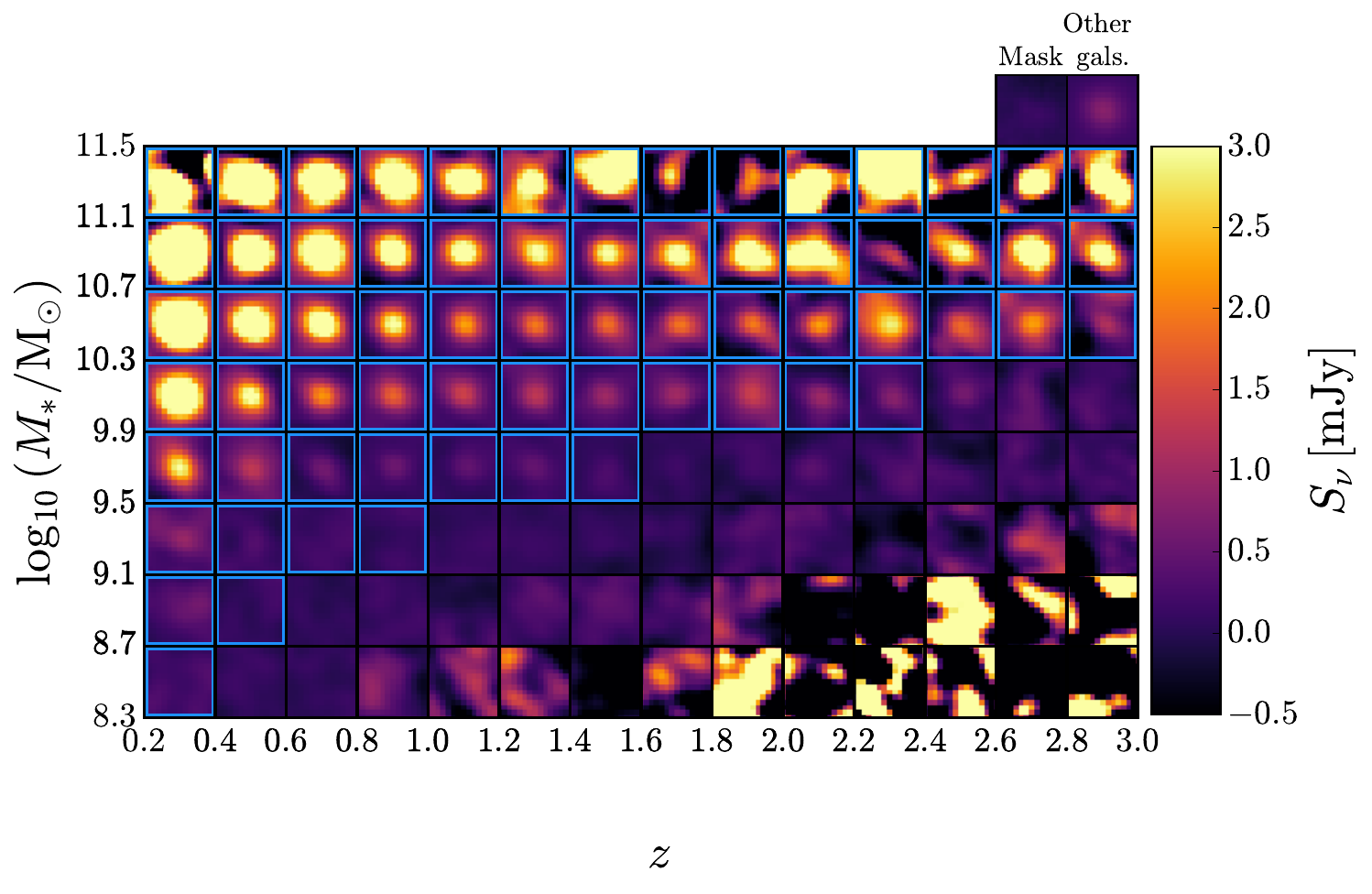}
    \includegraphics[width=0.48\textwidth]{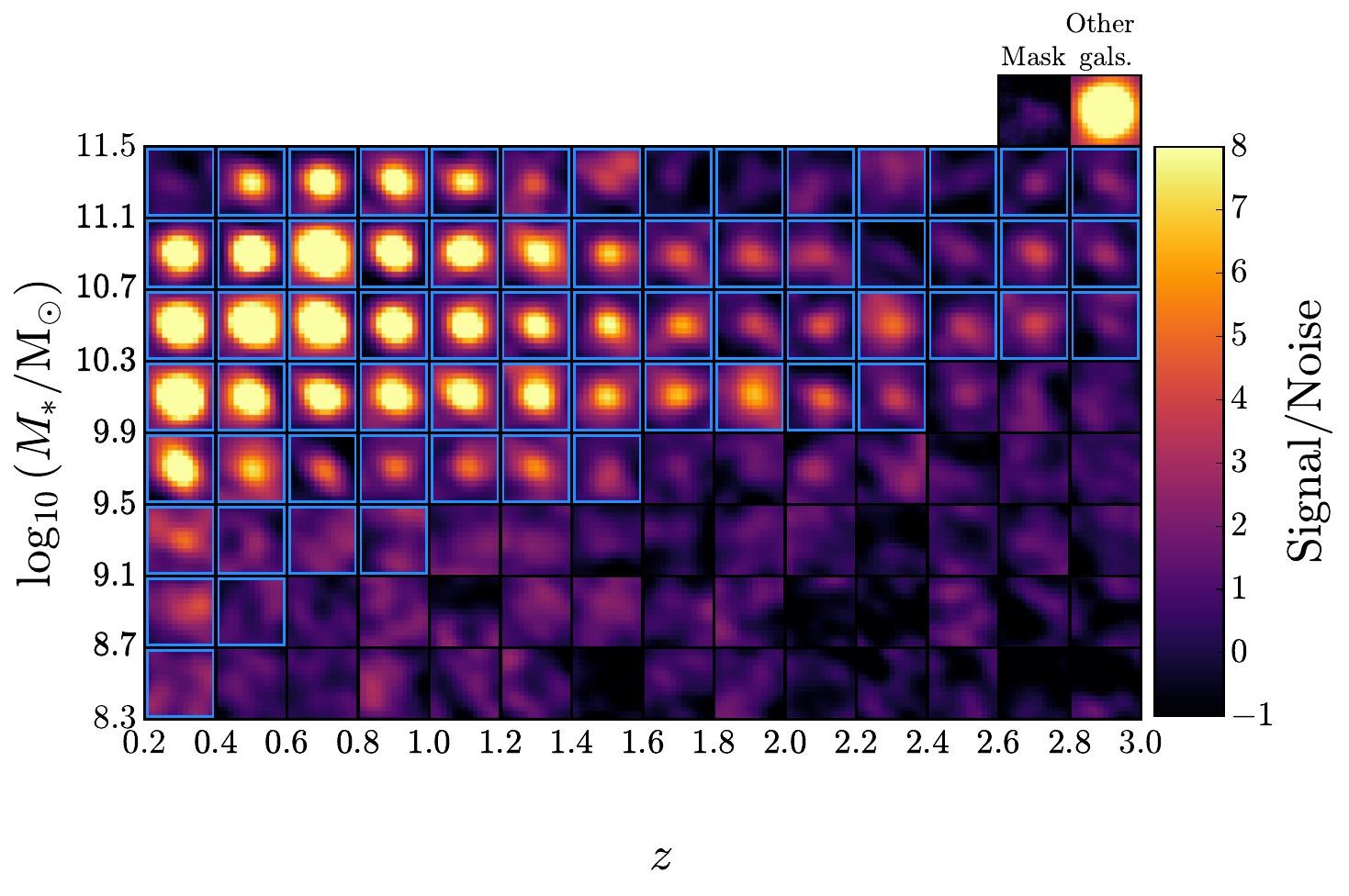}
    \caption{Same as \cref{fig:stack_results_main} but for PACS 100 (top row) and 160\,$\mu$m (bottom row). Here, each 2D cutout measures $40^{\prime\prime}\,{\times}\,40^{\prime\prime}$. The `Mask' and `Other gals.' panels differ from the colour bar and range from S/N$\,{=}\,-1$ to S/N$\,{=}\,20$.} 
    \label{fig:stack_results_main_pacs}
\end{figure*}

\begin{figure*}[htbp!]
    \centering
    \includegraphics[width=0.48\textwidth]{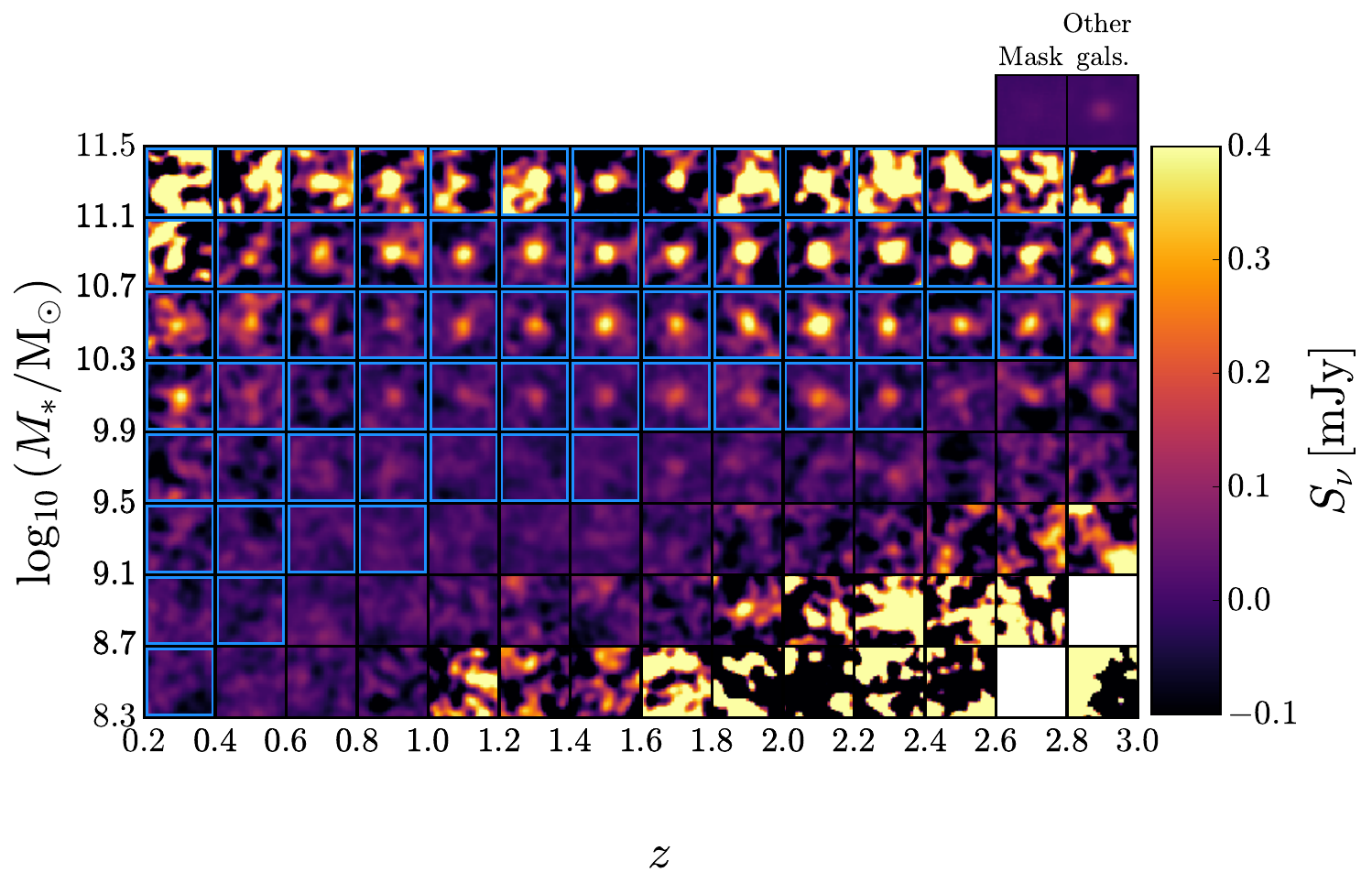}
    \includegraphics[width=0.48\textwidth]{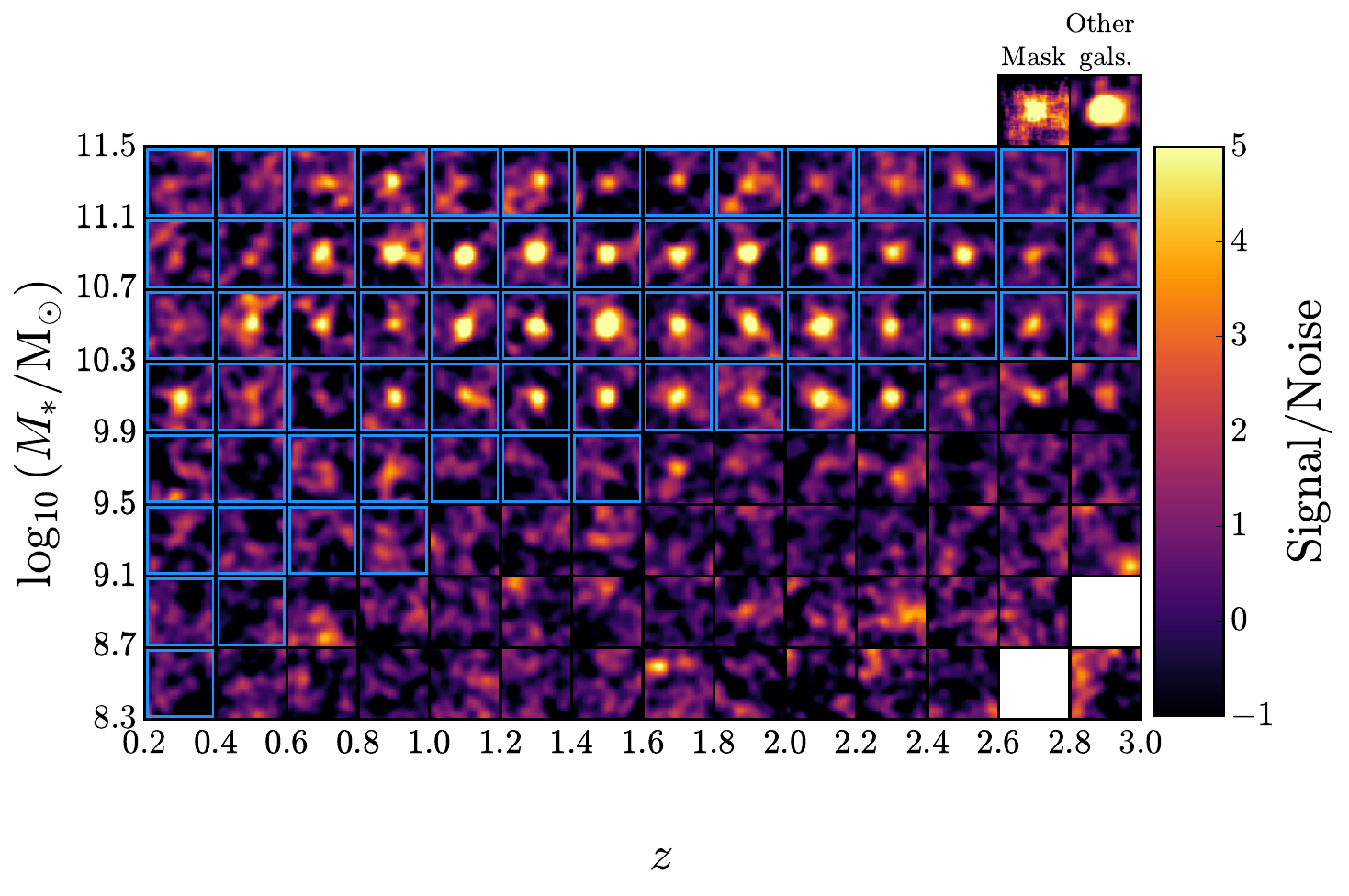}
    \caption{Same as \cref{fig:stack_results_main}, but for SCUBA-2 850\,$\mu$m. Here, each 2D cutout measures $70^{\prime\prime}\,{\times}\,70^{\prime\prime}$.} 
    \label{fig:stack_results_main_s2}
\end{figure*}

Here we show the 2D cross-correlations from our stacking algorithm and provide a table of stacked flux densities. The 2D stacking results for the star-forming galaxies in the \Euclid catalogue are shown in \cref{fig:stack_results_main} (for SPIRE), \cref{fig:stack_results_main_pacs} (for PACS), and \cref{fig:stack_results_main_s2} (for SCUBA-2). We show both the signal (in units of mJy) in the left column and the S/N in the right column, and the results for the mask and the remaining galaxies in the \Euclid catalogue (i.e., non-star-forming and outside the main redshift and stellar ranges we considered here) are shown in the top row. The corresponding stacked flux densities (the value of the central pixel in the 2D cross-correlations) are provided in \cref{table:stack_flux}.  Bins that are ${>}\,95\%$ complete in stellar mass \citep{Q1-SP031} are highlighted in blue. The 2D cross-correlation profiles are expected to follow the autocorrelations of the instrumental beams. We tested this by computing the averaged 1D radial profiles and compared these to the expected PSF profiles. We found good agreement between the stacked signals and the PSFs for most bins with $\logten(M_{\ast}/M_{\odot})\,{>}\,9.9$. Below this stellar mass we found that the 2D profiles can be more extended than the beam, with the largest effect seen at 500\,$\mu$m where the PSF is largest. However, this does not affect the calculations and results in this work so we do not attempt to correct for it. 

\clearpage
\onecolumn
\begin{center}
\setlength{\tabcolsep}{4pt}
\begin{longtable}{c c r r r r r r r r r}
\captionsetup{width=\textwidth}
\caption{Results from stacking the \Euclid catalogue galaxies with reliable redshifts and stellar masses on the \Herschel and SCUBA-2 images at 100, 160, 250, 350, 500, and 850\,$\mu$m.} \label{table:stack_flux} \\

\hline\noalign{\vskip 2pt}\hline\noalign{\vskip 2pt}
\multicolumn{1}{c}{$\logten$} & \multicolumn{1}{c}{$z$} & \multicolumn{1}{c}{\pz$S_{100}$} & \multicolumn{1}{c}{\pz$S_{160}$} & \multicolumn{1}{c}{\pz $S_{250}$} & \multicolumn{1}{c}{\pz$S_{350}$} & \multicolumn{1}{c}{\pz$S_{500}$} & \multicolumn{1}{c}{\pz$S_{850}$} & \multicolumn{1}{c}{$N_{\mathrm{PACS}}$} & \multicolumn{1}{c}{$N_{\mathrm{SPIRE}}$} & \multicolumn{1}{c}{\pz$N_{\mathrm{S2}}$} \\ \multicolumn{1}{c}{($M_{\ast}/M_\odot)$} & \multicolumn{1}{c}{} & \multicolumn{1}{c}{$\pz[\mathrm{mJy}]$} & \multicolumn{1}{c}{\pz$[\mathrm{mJy}]$} & \multicolumn{1}{c}{\pz$[\mathrm{mJy}]$} & \multicolumn{1}{c}{\pz$[\mathrm{mJy}]$} & \multicolumn{1}{c}{\pz$[\mathrm{mJy}]$} & \multicolumn{1}{c}{\pz$[\mathrm{mJy}]$} & \multicolumn{1}{c}{} & \multicolumn{1}{c}{} & \multicolumn{1}{c}{} \\ \noalign{\vskip 1pt}\hline\noalign{\vskip 2pt}
\endfirsthead

\multicolumn{11}{c}
{{\tablename\ \thetable{} -- continued from previous page.}} \\
\hline\noalign{\vskip 2pt}\hline\noalign{\vskip 2pt}
\multicolumn{1}{c}{$\logten$} & \multicolumn{1}{c}{$z$} & \multicolumn{1}{c}{\pz$S_{100}$} & \multicolumn{1}{c}{\pz$S_{160}$} & \multicolumn{1}{c}{\pz$S_{250}$} & \multicolumn{1}{c}{\pz$S_{350}$} & \multicolumn{1}{c}{\pz$S_{500}$} & \multicolumn{1}{c}{\pz$S_{850}$} & \multicolumn{1}{c}{\pz$N_{\mathrm{PACS}}$} & \multicolumn{1}{c}{$N_{\mathrm{SPIRE}}$} & \multicolumn{1}{c}{$N_{\mathrm{S2}}$} \\ \multicolumn{1}{c}{$(M_{\ast}/M_\odot)$} & \multicolumn{1}{c}{} & \multicolumn{1}{c}{\pz$[\mathrm{mJy}]$} & \multicolumn{1}{c}{\pz$[\mathrm{mJy}]$} & \multicolumn{1}{c}{\pz$[\mathrm{mJy}]$} & \multicolumn{1}{c}{\pz$[\mathrm{mJy}]$} & \multicolumn{1}{c}{\pz$[\mathrm{mJy}]$} & \multicolumn{1}{c}{\pz$[\mathrm{mJy}]$} & \multicolumn{1}{c}{} & \multicolumn{1}{c}{} & \multicolumn{1}{c}{} \\ \noalign{\vskip 1pt}\hline\hline\noalign{\vskip 2pt}
\endhead

\noalign{\vskip 2pt}
\multicolumn{11}{r}{{Continued on next page.}} \\
\endfoot

\hline
\endlastfoot

8.5 & 0.3 & 0.06$\pm$0.09 & 0.16$\pm$0.13 & 0.16$\pm$0.07 & 0.02$\pm$0.06 &  $-$0.11$\pm$0.06 &  $-$0.01$\pm$0.03 & 17711 & 26737 & 3098\\
8.5 & 0.5 & 0.11$\pm$0.07 & 0.10$\pm$0.11 & 0.20$\pm$0.05 & 0.08$\pm$0.05 & 0.06$\pm$0.05 & 0.03$\pm$0.03 & 22283 & 33265 & 3622\\
8.5 & 0.7 &  $-$0.04$\pm$0.08 & 0.03$\pm$0.13 & 0.03$\pm$0.05 &  $-$0.04$\pm$0.05 &  $-$0.11$\pm$0.06 &  $-$0.02$\pm$0.03 & 16637 & 26712 & 3537\\
8.5 & 0.9 &  $-$0.23$\pm$0.19 & 0.52$\pm$0.29 & 0.42$\pm$0.13 & 0.46$\pm$0.13 & 0.28$\pm$0.13 &  $-$0.04$\pm$0.07 & 3585 & 6146 & 893\\
8.5 & 1.1 &  $-$0.32$\pm$0.46 & 0.89$\pm$0.71 & 1.54$\pm$0.34 & 0.93$\pm$0.30 & 0.20$\pm$0.26 & 0.06$\pm$0.24 & 542 & 726 & 58\\
8.5 & 1.3 &  $-$0.02$\pm$0.53 & 0.04$\pm$0.81 & 0.08$\pm$0.38 &  $-$0.38$\pm$0.41 &  $-$0.49$\pm$0.39 &  $-$0.02$\pm$0.21 & 452 & 656 & 78\\
8.5 & 1.5 & 0.16$\pm$0.44 &  $-$0.98$\pm$0.65 & 0.03$\pm$0.33 & 0.14$\pm$0.29 & 0.04$\pm$0.33 & 0.01$\pm$0.18 & 729 & 1038 & 120\\
8.5 & 1.7 &  $-$0.32$\pm$1.01 & 1.04$\pm$1.55 & 1.34$\pm$0.53 & 2.80$\pm$0.66 & 1.97$\pm$0.65 & 0.06$\pm$0.46 & 97 & 176 & 19\\
8.5 & 1.9 &  $-$1.63$\pm$3.31 & 4.56$\pm$5.24 & 1.40$\pm$0.94 & 1.06$\pm$1.04 & 0.75$\pm$1.02 & 0.03$\pm$1.33 & 7 & 16 & 1\\
8.5 & 2.1 &  $-$6.57$\pm$4.18 &  $-$1.87$\pm$5.87 &  $-$4.81$\pm$1.26 &  $-$6.63$\pm$1.20 &  $-$4.88$\pm$1.22 &  $-$1.44$\pm$0.96 & 4 & 7 & 2\\
8.5 & 2.3 & 8.31$\pm$7.30 & 2.23$\pm$12.65 &  $-$0.80$\pm$1.41 &  $-$3.88$\pm$1.49 &  $-$2.64$\pm$1.47 &  $-$0.55$\pm$1.37 & 2 & 9 & 2\\
8.5 & 2.5 &  $-$6.91$\pm$4.08 &  $-$1.53$\pm$5.96 &  $-$2.75$\pm$1.33 &  $-$3.79$\pm$1.45 &  $-$1.52$\pm$1.55 &  $-$1.37$\pm$1.18 & 4 & 14 & 2\\
8.5 & 2.7 & 3.57$\pm$6.04 &  $-$7.76$\pm$9.61 &  $-$4.93$\pm$1.97 &  $-$3.70$\pm$2.04 &  $-$3.32$\pm$2.20 & $\dots$ & 2 & 4 & 0\\
8.5 & 2.9 &  $-$2.48$\pm$4.06 &  $-$11.68$\pm$6.31 &  $-$2.74$\pm$1.16 &  $-$2.84$\pm$1.43 &  $-$4.05$\pm$1.20 &  $-$5.97$\pm$5.65 & 6 & 6 & 1\\
8.9 & 0.3 & 0.30$\pm$0.10 & 0.50$\pm$0.15 & 0.43$\pm$0.08 & 0.22$\pm$0.07 & 0.07$\pm$0.07 & 0.02$\pm$0.04 & 13464 & 20832 & 2485\\
8.9 & 0.5 & 0.16$\pm$0.08 & 0.02$\pm$0.11 & 0.17$\pm$0.05 &  $-$0.06$\pm$0.06 &  $-$0.21$\pm$0.06 &  $-$0.03$\pm$0.03 & 23380 & 34784 & 3730\\
8.9 & 0.7 &  $-$0.03$\pm$0.06 &  $-$0.02$\pm$0.09 &  $-$0.03$\pm$0.04 &  $-$0.18$\pm$0.03 &  $-$0.23$\pm$0.04 & 0.04$\pm$0.02 & 32056 & 54852 & 7937\\
8.9 & 0.9 &  $-$0.07$\pm$0.07 & 0.12$\pm$0.10 & 0.25$\pm$0.04 & 0.25$\pm$0.04 & 0.18$\pm$0.05 &  $-$0.01$\pm$0.02 & 26080 & 42735 & 5554\\
8.9 & 1.1 & 0.10$\pm$0.11 &  $-$0.04$\pm$0.16 & 0.15$\pm$0.08 & 0.02$\pm$0.08 & 0.00$\pm$0.08 &  $-$0.04$\pm$0.05 & 11300 & 15511 & 1507\\
8.9 & 1.3 & 0.04$\pm$0.13 & 0.33$\pm$0.18 & 0.44$\pm$0.09 & 0.34$\pm$0.09 & 0.25$\pm$0.09 &  $-$0.02$\pm$0.06 & 8134 & 11553 & 1147\\
8.9 & 1.5 & 0.08$\pm$0.14 & 0.42$\pm$0.20 & 0.46$\pm$0.09 & 0.49$\pm$0.10 & 0.46$\pm$0.12 &  $-$0.02$\pm$0.05 & 8304 & 11868 & 1319\\
8.9 & 1.7 & 0.19$\pm$0.20 & 0.23$\pm$0.29 & 0.14$\pm$0.13 & 0.15$\pm$0.14 & 0.19$\pm$0.15 &  $-$0.03$\pm$0.07 & 3493 & 6282 & 959\\
8.9 & 1.9 & 1.17$\pm$0.42 & 0.46$\pm$0.61 &  $-$0.13$\pm$0.24 &  $-$0.16$\pm$0.27 &  $-$0.58$\pm$0.29 & 0.44$\pm$0.16 & 730 & 1164 & 156\\
8.9 & 2.1 & 0.91$\pm$0.86 &  $-$0.84$\pm$1.32 & 0.17$\pm$0.43 & 0.41$\pm$0.47 &  $-$0.16$\pm$0.47 & 0.00$\pm$0.34 & 138 & 203 & 21\\
8.9 & 2.3 & 4.47$\pm$2.60 &  $-$3.06$\pm$3.72 &  $-$1.32$\pm$0.65 &  $-$0.16$\pm$0.69 &  $-$0.39$\pm$0.76 & 0.66$\pm$0.46 & 12 & 98 & 19\\
8.9 & 2.5 &  $-$0.49$\pm$2.81 & 3.00$\pm$4.27 &  $-$2.20$\pm$0.83 &  $-$2.10$\pm$0.90 &  $-$0.90$\pm$0.87 & 0.66$\pm$0.65 & 9 & 43 & 10\\
8.9 & 2.7 &  $-$2.96$\pm$2.00 &  $-$1.46$\pm$3.13 &  $-$0.20$\pm$0.91 &  $-$1.34$\pm$1.00 &  $-$1.95$\pm$0.83 &  $-$0.46$\pm$0.66 & 22 & 56 & 12\\
8.9 & 2.9 &  $-$1.17$\pm$3.44 &  $-$9.10$\pm$5.32 &  $-$8.45$\pm$1.13 &  $-$5.46$\pm$1.28 &  $-$1.70$\pm$1.07 & $\dots$ & 7 & 10 & 0\\
9.3 & 0.3 & 0.83$\pm$0.13 & 0.91$\pm$0.18 & 0.93$\pm$0.10 & 0.33$\pm$0.09 &  $-$0.12$\pm$0.09 & 0.07$\pm$0.04 & 8614 & 13669 & 1740\\
9.3 & 0.5 & 0.21$\pm$0.09 & 0.33$\pm$0.13 & 0.41$\pm$0.06 & 0.13$\pm$0.07 &  $-$0.06$\pm$0.06 & 0.00$\pm$0.03 & 16626 & 24633 & 2664\\
9.3 & 0.7 & 0.02$\pm$0.07 & 0.18$\pm$0.11 & 0.27$\pm$0.04 & 0.12$\pm$0.04 &  $-$0.03$\pm$0.05 & 0.00$\pm$0.02 & 25346 & 43919 & 6410\\
9.3 & 0.9 & 0.03$\pm$0.06 & 0.08$\pm$0.08 & 0.35$\pm$0.04 & 0.32$\pm$0.04 & 0.10$\pm$0.04 & 0.02$\pm$0.02 & 34386 & 59915 & 8713\\
9.3 & 1.1 & 0.05$\pm$0.05 & 0.13$\pm$0.08 & 0.26$\pm$0.04 & 0.26$\pm$0.05 & 0.10$\pm$0.04 & 0.04$\pm$0.02 & 39345 & 58955 & 6907\\
9.3 & 1.3 & 0.10$\pm$0.05 & 0.15$\pm$0.08 & 0.27$\pm$0.03 & 0.26$\pm$0.04 & 0.14$\pm$0.04 & 0.03$\pm$0.02 & 44484 & 66271 & 7823\\
9.3 & 1.5 & 0.04$\pm$0.07 & 0.16$\pm$0.11 & 0.31$\pm$0.05 & 0.26$\pm$0.05 & 0.15$\pm$0.05 & 0.06$\pm$0.02 & 27452 & 50678 & 8247\\
9.3 & 1.7 &  $-$0.20$\pm$0.09 & 0.09$\pm$0.14 & 0.47$\pm$0.07 & 0.47$\pm$0.07 & 0.39$\pm$0.07 & 0.02$\pm$0.04 & 16232 & 25556 & 3196\\
9.3 & 1.9 &  $-$0.08$\pm$0.17 & 0.31$\pm$0.25 & 0.34$\pm$0.12 & 0.47$\pm$0.11 & 0.40$\pm$0.12 &  $-$0.06$\pm$0.06 & 5500 & 8718 & 1069\\
9.3 & 2.1 & 0.07$\pm$0.19 & 0.13$\pm$0.27 & 0.22$\pm$0.12 & 0.22$\pm$0.15 & 0.10$\pm$0.15 &  $-$0.02$\pm$0.08 & 4698 & 6887 & 732\\
9.3 & 2.3 &  $-$0.06$\pm$0.25 &  $-$0.12$\pm$0.39 & 0.41$\pm$0.17 & 0.28$\pm$0.21 & 0.11$\pm$0.20 &  $-$0.03$\pm$0.10 & 2001 & 3385 & 424\\
9.3 & 2.5 & 0.20$\pm$0.36 &  $-$0.19$\pm$0.54 & 0.52$\pm$0.26 & 0.19$\pm$0.32 & 0.20$\pm$0.27 &  $-$0.04$\pm$0.13 & 1060 & 1760 & 231\\
9.3 & 2.7 & 0.38$\pm$0.61 & 1.36$\pm$0.93 & 0.77$\pm$0.44 & 0.20$\pm$0.45 &  $-$0.22$\pm$0.48 & 0.06$\pm$0.17 & 330 & 651 & 121\\
9.3 & 2.9 &  $-$0.88$\pm$0.62 & 0.22$\pm$0.99 & 0.41$\pm$0.45 & 0.14$\pm$0.50 & 0.13$\pm$0.52 & 0.23$\pm$0.21 & 306 & 559 & 104\\
9.7 & 0.3 & 2.26$\pm$0.16 & 3.04$\pm$0.22 & 3.00$\pm$0.10 & 1.50$\pm$0.11 & 0.53$\pm$0.09 & 0.11$\pm$0.05 & 4893 & 8120 & 1113\\
9.7 & 0.5 & 0.75$\pm$0.11 & 1.28$\pm$0.17 & 1.28$\pm$0.08 & 0.44$\pm$0.08 &  $-$0.09$\pm$0.08 &  $-$0.05$\pm$0.04 & 9545 & 14595 & 1648\\
9.7 & 0.7 & 0.24$\pm$0.08 & 0.60$\pm$0.12 & 1.03$\pm$0.07 & 0.46$\pm$0.06 & 0.04$\pm$0.06 & 0.07$\pm$0.03 & 16956 & 27321 & 3673\\
9.7 & 0.9 & 0.30$\pm$0.07 & 0.56$\pm$0.11 & 0.90$\pm$0.05 & 0.59$\pm$0.04 & 0.14$\pm$0.06 & 0.08$\pm$0.02 & 22923 & 40925 & 6018\\
9.7 & 1.1 & 0.22$\pm$0.06 & 0.45$\pm$0.09 & 0.88$\pm$0.04 & 0.72$\pm$0.05 & 0.35$\pm$0.04 & 0.04$\pm$0.02 & 32854 & 51898 & 6737\\
9.7 & 1.3 & 0.18$\pm$0.06 & 0.49$\pm$0.09 & 0.77$\pm$0.03 & 0.74$\pm$0.04 & 0.48$\pm$0.04 &  $-$0.00$\pm$0.02 & 36846 & 59350 & 8295\\
9.7 & 1.5 & 0.15$\pm$0.06 & 0.28$\pm$0.09 & 0.89$\pm$0.04 & 0.85$\pm$0.04 & 0.62$\pm$0.04 & 0.03$\pm$0.02 & 33404 & 64649 & 11275\\
9.7 & 1.7 & 0.05$\pm$0.07 & 0.19$\pm$0.10 & 0.84$\pm$0.05 & 0.89$\pm$0.06 & 0.79$\pm$0.05 & 0.11$\pm$0.03 & 27613 & 43866 & 5555\\
9.7 & 1.9 & 0.20$\pm$0.10 & 0.22$\pm$0.15 & 0.55$\pm$0.08 & 0.65$\pm$0.08 & 0.58$\pm$0.08 & 0.02$\pm$0.04 & 14264 & 24471 & 3590\\
9.7 & 2.1 & 0.13$\pm$0.09 & 0.37$\pm$0.14 & 0.69$\pm$0.06 & 0.65$\pm$0.08 & 0.59$\pm$0.07 & 0.07$\pm$0.04 & 17092 & 27020 & 3492\\
9.7 & 2.3 &  $-$0.09$\pm$0.10 & 0.28$\pm$0.15 & 0.60$\pm$0.08 & 0.76$\pm$0.09 & 0.70$\pm$0.09 & 0.04$\pm$0.04 & 12021 & 18856 & 2400\\
9.7 & 2.5 &  $-$0.09$\pm$0.14 & 0.10$\pm$0.22 & 0.26$\pm$0.13 & 0.35$\pm$0.12 & 0.31$\pm$0.13 &  $-$0.02$\pm$0.06 & 7219 & 11203 & 1453\\
9.7 & 2.7 &  $-$0.04$\pm$0.19 & 0.11$\pm$0.28 & 0.96$\pm$0.17 & 0.80$\pm$0.16 & 0.54$\pm$0.15 & 0.04$\pm$0.06 & 4156 & 7323 & 1141\\
9.7 & 2.9 & 0.04$\pm$0.19 & 0.12$\pm$0.28 & 0.50$\pm$0.17 & 0.61$\pm$0.16 & 0.67$\pm$0.16 & 0.00$\pm$0.08 & 4087 & 6550 & 922\\
10.1 & 0.3 & 5.43$\pm$0.21 & 7.67$\pm$0.32 & 7.18$\pm$0.16 & 3.31$\pm$0.12 & 1.48$\pm$0.11 & 0.41$\pm$0.07 & 2877 & 4714 & 593\\
10.1 & 0.5 & 2.23$\pm$0.15 & 3.29$\pm$0.22 & 3.84$\pm$0.09 & 1.88$\pm$0.09 & 0.61$\pm$0.08 & 0.16$\pm$0.05 & 5557 & 8823 & 1097\\
10.1 & 0.7 & 1.16$\pm$0.10 & 2.18$\pm$0.15 & 3.05$\pm$0.06 & 1.75$\pm$0.07 & 0.69$\pm$0.07 & 0.09$\pm$0.04 & 11363 & 17957 & 2305\\
10.1 & 0.9 & 1.07$\pm$0.09 & 1.79$\pm$0.14 & 2.65$\pm$0.06 & 1.83$\pm$0.07 & 0.88$\pm$0.07 & 0.17$\pm$0.03 & 13094 & 23321 & 3540\\
10.1 & 1.1 & 0.69$\pm$0.08 & 1.41$\pm$0.12 & 2.14$\pm$0.06 & 1.68$\pm$0.05 & 0.76$\pm$0.06 & 0.13$\pm$0.03 & 18424 & 30049 & 4223\\
10.1 & 1.3 & 0.46$\pm$0.08 & 1.27$\pm$0.12 & 2.14$\pm$0.06 & 1.87$\pm$0.06 & 1.05$\pm$0.06 & 0.16$\pm$0.03 & 16999 & 29549 & 4702\\
10.1 & 1.5 & 0.45$\pm$0.09 & 0.99$\pm$0.13 & 1.94$\pm$0.06 & 1.73$\pm$0.07 & 1.16$\pm$0.06 & 0.16$\pm$0.03 & 16666 & 33086 & 5827\\
10.1 & 1.7 & 0.36$\pm$0.11 & 1.01$\pm$0.15 & 2.03$\pm$0.08 & 1.87$\pm$0.08 & 1.42$\pm$0.08 & 0.20$\pm$0.04 & 12959 & 22961 & 3490\\
10.1 & 1.9 & 0.54$\pm$0.14 & 1.33$\pm$0.21 & 1.65$\pm$0.11 & 1.68$\pm$0.11 & 1.39$\pm$0.11 & 0.18$\pm$0.04 & 6590 & 13943 & 2519\\
10.1 & 2.1 & 0.51$\pm$0.13 & 0.97$\pm$0.19 & 1.66$\pm$0.11 & 1.53$\pm$0.12 & 1.14$\pm$0.10 & 0.26$\pm$0.04 & 8978 & 16621 & 2796\\
10.1 & 2.3 & 0.22$\pm$0.14 & 0.77$\pm$0.21 & 1.73$\pm$0.09 & 1.76$\pm$0.11 & 1.40$\pm$0.11 & 0.24$\pm$0.04 & 7396 & 14279 & 2502\\
10.1 & 2.5 & 0.34$\pm$0.16 & 0.61$\pm$0.24 & 1.38$\pm$0.13 & 1.45$\pm$0.12 & 1.18$\pm$0.13 & 0.15$\pm$0.05 & 6408 & 12037 & 2043\\
10.1 & 2.7 & 0.54$\pm$0.17 & 0.50$\pm$0.25 & 1.21$\pm$0.13 & 1.44$\pm$0.13 & 1.24$\pm$0.15 & 0.22$\pm$0.05 & 4815 & 10184 & 1896\\
10.1 & 2.9 & 0.08$\pm$0.16 & 0.23$\pm$0.24 & 1.01$\pm$0.12 & 1.08$\pm$0.14 & 0.97$\pm$0.15 & 0.17$\pm$0.06 & 6051 & 10639 & 1787\\
10.5 & 0.3 & 12.95$\pm$0.45 & 16.11$\pm$0.66 & 14.24$\pm$0.44 & 6.73$\pm$0.25 & 3.00$\pm$0.22 & 0.35$\pm$0.14 & 1126 & 1873 & 194\\
10.5 & 0.5 & 3.90$\pm$0.18 & 6.50$\pm$0.26 & 7.84$\pm$0.11 & 4.18$\pm$0.10 & 1.81$\pm$0.09 & 0.32$\pm$0.06 & 3398 & 5476 & 725\\
10.5 & 0.7 & 2.58$\pm$0.13 & 4.65$\pm$0.18 & 6.20$\pm$0.09 & 3.43$\pm$0.09 & 1.33$\pm$0.08 & 0.23$\pm$0.04 & 6785 & 10793 & 1358\\
10.5 & 0.9 & 2.01$\pm$0.13 & 3.16$\pm$0.20 & 4.92$\pm$0.09 & 3.32$\pm$0.09 & 1.44$\pm$0.08 & 0.20$\pm$0.04 & 6464 & 11564 & 1728\\
10.5 & 1.1 & 1.04$\pm$0.11 & 2.28$\pm$0.16 & 3.85$\pm$0.08 & 2.97$\pm$0.07 & 1.31$\pm$0.07 & 0.29$\pm$0.04 & 8955 & 15278 & 2356\\
10.5 & 1.3 & 1.04$\pm$0.14 & 1.96$\pm$0.20 & 4.10$\pm$0.09 & 3.37$\pm$0.10 & 1.70$\pm$0.10 & 0.31$\pm$0.04 & 6475 & 12375 & 2230\\
10.5 & 1.5 & 1.03$\pm$0.15 & 1.88$\pm$0.23 & 3.64$\pm$0.12 & 3.27$\pm$0.10 & 2.16$\pm$0.11 & 0.43$\pm$0.04 & 5409 & 12493 & 2515\\
10.5 & 1.7 & 1.04$\pm$0.21 & 1.90$\pm$0.30 & 3.99$\pm$0.14 & 3.81$\pm$0.17 & 2.64$\pm$0.16 & 0.36$\pm$0.05 & 2905 & 7341 & 1540\\
10.5 & 1.9 & 1.07$\pm$0.37 & 1.80$\pm$0.49 & 4.46$\pm$0.26 & 3.92$\pm$0.23 & 2.69$\pm$0.26 & 0.42$\pm$0.07 & 1030 & 3848 & 940\\
10.5 & 2.1 & 0.64$\pm$0.32 & 2.27$\pm$0.49 & 3.88$\pm$0.27 & 3.73$\pm$0.25 & 2.74$\pm$0.24 & 0.59$\pm$0.08 & 1276 & 3520 & 827\\
10.5 & 2.3 & 1.06$\pm$0.38 & 2.85$\pm$0.56 & 3.72$\pm$0.28 & 3.45$\pm$0.29 & 2.43$\pm$0.23 & 0.45$\pm$0.08 & 892 & 3264 & 822\\
10.5 & 2.5 & 0.72$\pm$0.37 & 1.83$\pm$0.51 & 3.74$\pm$0.26 & 3.25$\pm$0.28 & 2.43$\pm$0.26 & 0.39$\pm$0.08 & 1025 & 2968 & 717\\
10.5 & 2.7 & 0.90$\pm$0.39 & 2.26$\pm$0.56 & 2.74$\pm$0.28 & 2.73$\pm$0.31 & 2.09$\pm$0.31 & 0.39$\pm$0.08 & 834 & 2857 & 715\\
10.5 & 2.9 & 0.78$\pm$0.38 & 1.11$\pm$0.56 & 2.41$\pm$0.27 & 2.35$\pm$0.31 & 2.27$\pm$0.30 & 0.40$\pm$0.10 & 918 & 2498 & 558\\
10.9 & 0.3 & 20.03$\pm$1.63 & 25.09$\pm$1.58 & 20.58$\pm$1.28 & 9.42$\pm$0.70 & 3.55$\pm$0.49 & 0.52$\pm$0.31 & 171 & 289 & 31\\
10.9 & 0.5 & 5.99$\pm$0.29 & 9.46$\pm$0.44 & 12.01$\pm$0.22 & 6.51$\pm$0.21 & 2.45$\pm$0.17 & 0.22$\pm$0.10 & 1215 & 2106 & 270\\
10.9 & 0.7 & 4.10$\pm$0.19 & 7.65$\pm$0.28 & 9.82$\pm$0.12 & 6.09$\pm$0.12 & 2.74$\pm$0.11 & 0.37$\pm$0.06 & 2703 & 4678 & 666\\
10.9 & 0.9 & 2.90$\pm$0.21 & 5.14$\pm$0.30 & 8.02$\pm$0.13 & 5.72$\pm$0.10 & 2.83$\pm$0.10 & 0.52$\pm$0.06 & 2578 & 4614 & 693\\
10.9 & 1.1 & 2.18$\pm$0.17 & 3.93$\pm$0.26 & 6.52$\pm$0.13 & 5.08$\pm$0.10 & 2.61$\pm$0.10 & 0.44$\pm$0.05 & 3453 & 6049 & 982\\
10.9 & 1.3 & 1.41$\pm$0.24 & 3.14$\pm$0.34 & 4.81$\pm$0.17 & 4.23$\pm$0.16 & 2.46$\pm$0.16 & 0.48$\pm$0.06 & 2029 & 4407 & 872\\
10.9 & 1.5 & 1.03$\pm$0.30 & 3.21$\pm$0.45 & 5.34$\pm$0.21 & 4.78$\pm$0.21 & 2.94$\pm$0.19 & 0.62$\pm$0.07 & 1353 & 3595 & 766\\
10.9 & 1.7 & 2.07$\pm$0.59 & 3.52$\pm$0.74 & 5.69$\pm$0.36 & 5.31$\pm$0.40 & 3.48$\pm$0.25 & 0.52$\pm$0.08 & 443 & 2027 & 588\\
10.9 & 1.9 & 3.12$\pm$0.81 & 4.62$\pm$1.06 & 6.57$\pm$0.44 & 5.87$\pm$0.50 & 4.33$\pm$0.48 & 0.92$\pm$0.11 & 191 & 1144 & 369\\
10.9 & 2.1 & 2.83$\pm$1.06 & 4.76$\pm$1.37 & 7.68$\pm$0.72 & 6.57$\pm$0.73 & 3.97$\pm$0.79 & 0.98$\pm$0.15 & 118 & 709 & 223\\
10.9 & 2.3 & 0.88$\pm$1.26 & 1.42$\pm$1.57 & 3.81$\pm$0.73 & 4.09$\pm$0.72 & 2.31$\pm$0.65 & 0.94$\pm$0.18 & 83 & 647 & 189\\
10.9 & 2.5 & 3.26$\pm$1.19 & 3.16$\pm$1.54 & 6.64$\pm$0.79 & 5.82$\pm$0.82 & 4.68$\pm$0.78 & 0.93$\pm$0.16 & 122 & 638 & 188\\
10.9 & 2.7 & 6.06$\pm$0.94 & 5.56$\pm$1.34 & 6.26$\pm$0.77 & 4.15$\pm$0.74 & 2.77$\pm$0.67 & 0.71$\pm$0.18 & 153 & 668 & 171\\
10.9 & 2.9 & 2.52$\pm$0.99 & 4.05$\pm$1.37 & 6.77$\pm$0.68 & 6.47$\pm$0.84 & 4.66$\pm$0.66 & 0.58$\pm$0.20 & 149 & 510 & 124\\
11.3 & 0.3 & 6.57$\pm$5.51 & 6.64$\pm$6.06 & 18.83$\pm$2.87 & 12.51$\pm$1.62 & 6.45$\pm$1.17 & 1.00$\pm$0.64 & 14 & 33 & 7\\
11.3 & 0.5 & 7.76$\pm$1.21 & 15.02$\pm$1.80 & 15.42$\pm$1.04 & 9.06$\pm$1.08 & 4.42$\pm$0.81 & 0.64$\pm$0.41 & 125 & 217 & 27\\
11.3 & 0.7 & 5.92$\pm$0.54 & 9.96$\pm$0.80 & 15.07$\pm$0.48 & 9.26$\pm$0.38 & 4.34$\pm$0.31 & 0.63$\pm$0.17 & 399 & 704 & 113\\
11.3 & 0.9 & 4.01$\pm$0.52 & 8.21$\pm$0.72 & 12.59$\pm$0.33 & 9.63$\pm$0.28 & 5.07$\pm$0.27 & 0.87$\pm$0.16 & 413 & 776 & 121\\
11.3 & 1.1 & 2.26$\pm$0.44 & 5.70$\pm$0.67 & 10.75$\pm$0.35 & 8.48$\pm$0.31 & 4.23$\pm$0.21 & 0.40$\pm$0.16 & 523 & 938 & 119\\
11.3 & 1.3 & 2.27$\pm$0.65 & 4.58$\pm$1.01 & 8.77$\pm$0.53 & 7.91$\pm$0.51 & 4.67$\pm$0.51 & 0.84$\pm$0.21 & 248 & 542 & 102\\
11.3 & 1.5 & 2.72$\pm$0.94 & 4.98$\pm$1.32 & 9.73$\pm$0.61 & 8.29$\pm$0.59 & 5.09$\pm$0.46 & 0.73$\pm$0.18 & 141 & 435 & 112\\
11.3 & 1.7 & 3.17$\pm$1.76 & 1.81$\pm$2.17 & 6.57$\pm$1.10 & 5.94$\pm$0.85 & 3.43$\pm$0.80 & 0.78$\pm$0.19 & 46 & 294 & 99\\
11.3 & 1.9 & 5.47$\pm$2.39 & 2.08$\pm$2.82 & 9.92$\pm$1.25 & 8.13$\pm$1.12 & 6.25$\pm$1.10 & 1.08$\pm$0.26 & 26 & 176 & 52\\
11.3 & 2.1 & 3.26$\pm$3.34 & 5.98$\pm$4.06 & 5.98$\pm$2.14 & 6.81$\pm$1.62 & 5.67$\pm$1.74 & 1.17$\pm$0.46 & 12 & 106 & 24\\
11.3 & 2.3 & 4.16$\pm$2.90 & 6.33$\pm$3.28 & 12.13$\pm$1.79 & 9.21$\pm$1.52 & 6.34$\pm$1.35 & 1.34$\pm$0.42 & 21 & 129 & 36\\
11.3 & 2.5 & 2.10$\pm$2.33 & 2.67$\pm$3.13 & 6.84$\pm$1.23 & 5.78$\pm$1.33 & 3.54$\pm$1.29 & 1.72$\pm$0.44 & 26 & 121 & 29\\
11.3 & 2.7 & 9.51$\pm$2.36 & 6.82$\pm$2.85 & 11.01$\pm$1.64 & 7.45$\pm$1.52 & 5.06$\pm$1.49 & 0.80$\pm$0.44 & 34 & 138 & 30\\
11.3 & 2.9 & 7.20$\pm$2.01 & 7.09$\pm$2.72 & 9.80$\pm$1.24 & 5.35$\pm$1.64 & 3.44$\pm$1.41 & 0.32$\pm$0.48 & 34 & 104 & 20\\

\end{longtable}

\footnotesize
 \tablefoot{The central stellar mass and redshift of each bin are given in the first two columns. The last three columns give the total number of \Euclid galaxies in the stacking bin for the PACS, SPIRE, and SCUBA-2 images, respectively.}

\end{center}
\clearpage

\section{Best-fit far-IR SED parameters and derived quantities}
\label{app:3}

\begin{figure}[htbp!]
    \onecolumn
    \centering
    \includegraphics[width=0.65\textwidth]{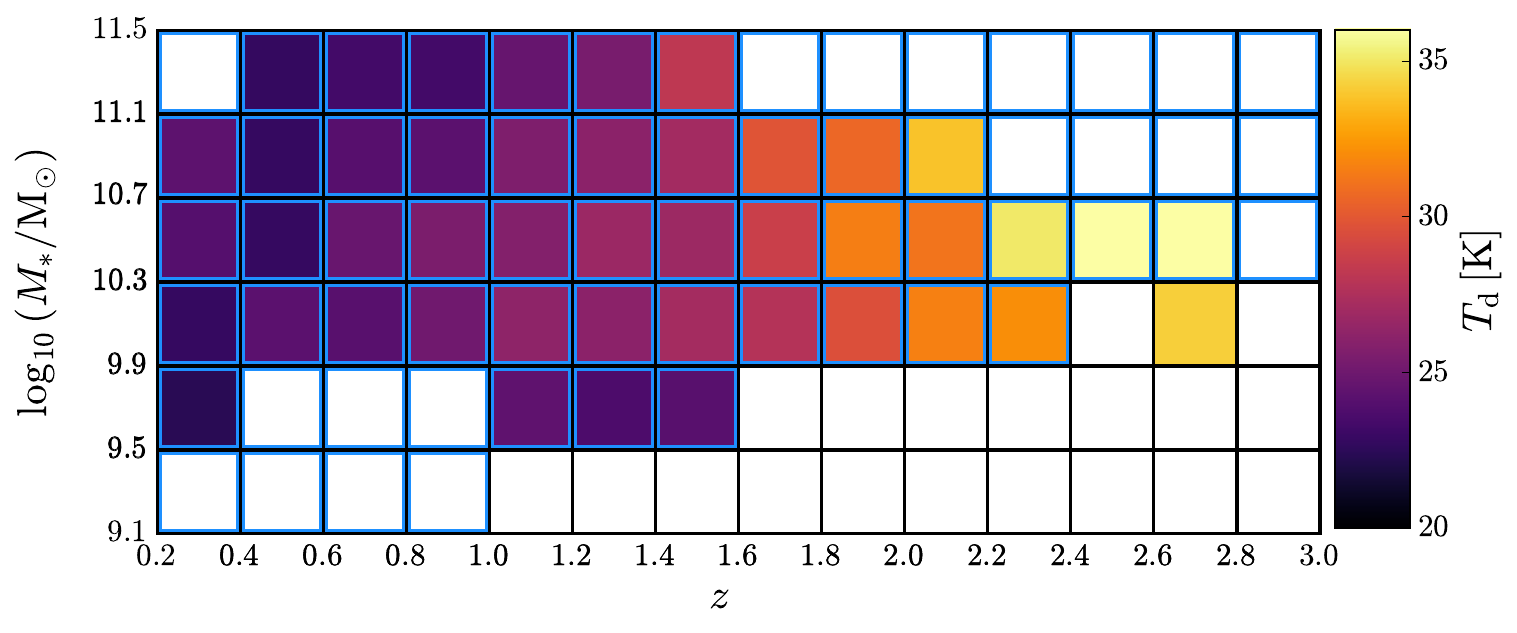}
    \includegraphics[width=0.65\textwidth]{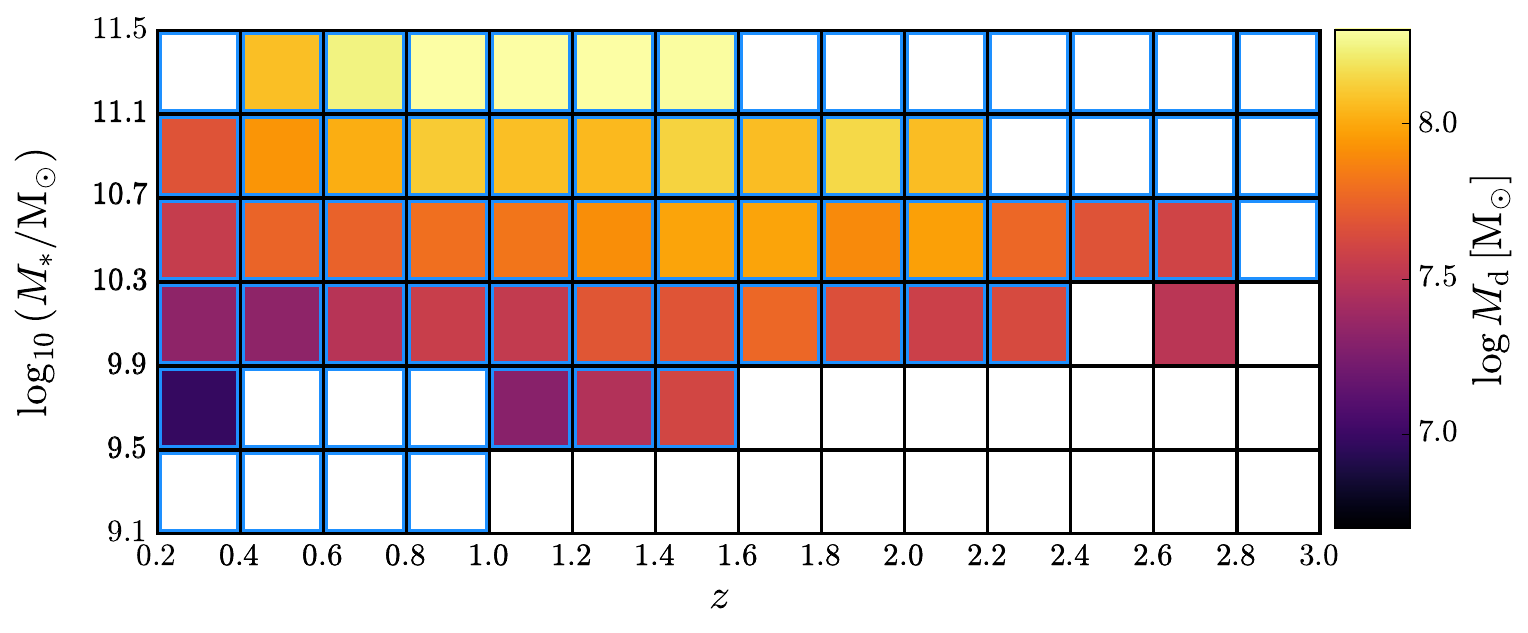}
    \includegraphics[width=0.65\textwidth]{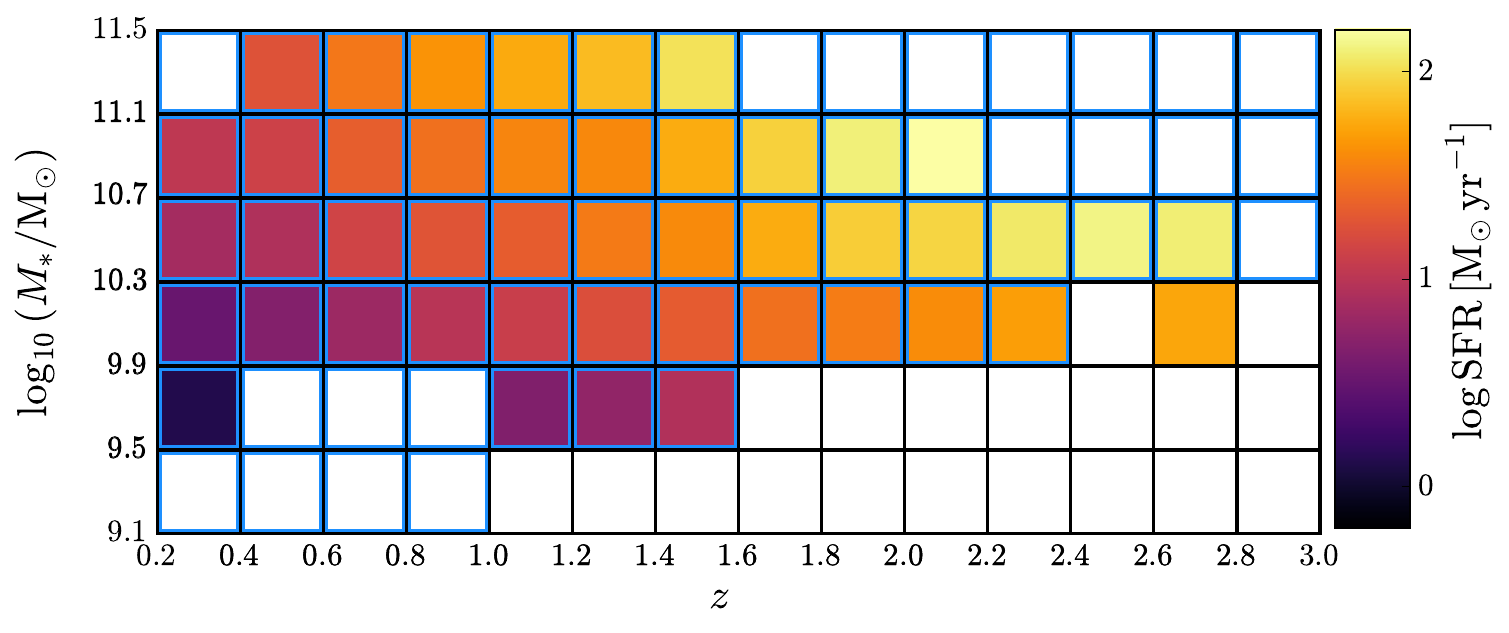}
    \caption{Physical parameters derived from the best-fit modified blackbody SEDs in \cref{fig:stack_sed_fits}. Bins that are ${>}\,95\%$ complete in stellar mass \citep{Q1-SP031} are highlighted in blue. {\it Top:} Best-fit dust temperatures, $T_{\rm d}$. {\it Middle:} Dust mass ($M_{\rm d}$), calculated by scaling the best-fit amplitude \citep[see e.g.][]{reuter2020,Eales2024,jolly2025}. {\it Bottom:} SFRs calculated from $L_{\rm IR}$ (the integral of the best-fit SED from 8 to 1000\,$\mu$m) multiplied by a factor of $1.49\,{\times}\,10^{-10}\,M_{\odot}\,$yr$^{-1}$\,L$_{\odot}^{-1}$.} 
    \label{fig:stack_params}
\end{figure}

Here we show the resulting physical parameters $T_{\rm d}$, $M_{\rm d}$, and SFR in each of the stellar mass and redshift bins where we have sufficient stacked photometry to derive these physical parameters, and we provide the best-fit SED parameters and derived physical quantities. \cref{fig:stack_params} shows the physical parameters for each redshift and stellar mass bin in our stacking analysis (where bins that are ${>}\,95\%$ complete in stellar mass are highlighted in blue -- see \citealt{Q1-SP031}.), and best-fit far-IR SED parameters and derived parameters are provided in \cref{table:bestfit_sed}. We omit showing the lowest stellar mass bins since they contain no data.

\clearpage
\onecolumn
\begin{center}
\setlength{\tabcolsep}{10pt}
\begin{longtable}{c c c c c c}
\caption{Best-fit far-IR SED parameters (see \cref{subsec:results_sed_core}).} \label{table:bestfit_sed} \\

\hline\noalign{\vskip 2pt}\hline\noalign{\vskip 2pt}
\multicolumn{1}{c}{$\logten$} & \multicolumn{1}{c}{$z$} & \multicolumn{1}{c}{$A$} & \multicolumn{1}{c}{$T_{\rm d}$} & \multicolumn{1}{c}{$M_{\rm d}$} & \multicolumn{1}{c}{SFR} \\ \multicolumn{1}{c}{$(M_\ast/M_\odot)$} & \multicolumn{1}{c}{} & \multicolumn{1}{c}{$[10^{-17}]$} & \multicolumn{1}{c}{$[\mathrm{K}]$} & \multicolumn{1}{c}{$[10^7\,\mathrm{M_{\odot}}]$} & \multicolumn{1}{c}{$[\mathrm{M_{\odot}\,yr^{-1}}]$} \\ \noalign{\vskip 1pt}\hline\noalign{\vskip 2pt} 
\endfirsthead

\multicolumn{6}{c}
{{\tablename\ \thetable{} -- continued from previous page.}} \\
\hline\noalign{\vskip 2pt}\hline\noalign{\vskip 2pt}
\multicolumn{1}{c}{$\logten$} & \multicolumn{1}{c}{$z$} & \multicolumn{1}{c}{$A$} & \multicolumn{1}{c}{$T_{\rm d}$} & \multicolumn{1}{c}{$M_{\rm d}$} & \multicolumn{1}{c}{SFR} \\ \multicolumn{1}{c}{$(M_\ast/M_\odot)$} & \multicolumn{1}{c}{} & \multicolumn{1}{c}{$[10^{-17}]$} & \multicolumn{1}{c}{$[\mathrm{K}]$} & \multicolumn{1}{c}{$[10^7\,\mathrm{M_{\odot}}]$} & \multicolumn{1}{c}{$[\mathrm{M_{\odot}\,yr^{-1}}]$} \\ \noalign{\vskip 1pt}\hline\noalign{\vskip 2pt}
\endhead

\noalign{\vskip 2pt}
\multicolumn{6}{r}{{Continued on next page.}} \\
\endfoot

\hline
\endlastfoot

8.5 & 0.3 & $\dots$ & $\dots$ & $\dots$ & $\dots$\\
8.5 & 0.5 & $\dots$ & $\dots$ & $\dots$ & $\dots$\\
8.5 & 0.7 & $\dots$ & $\dots$ & $\dots$ & $\dots$\\
8.5 & 0.9 & $\dots$ & $\dots$ & $\dots$ & $\dots$\\
8.5 & 1.1 & $\dots$ & $\dots$ & $\dots$ & $\dots$\\
8.5 & 1.3 & $\dots$ & $\dots$ & $\dots$ & $\dots$\\
8.5 & 1.5 & $\dots$ & $\dots$ & $\dots$ & $\dots$\\
8.5 & 1.7 & $\dots$ & $\dots$ & $\dots$ & $\dots$\\
8.5 & 1.9 & $\dots$ & $\dots$ & $\dots$ & $\dots$\\
8.5 & 2.1 & $\dots$ & $\dots$ & $\dots$ & $\dots$\\
8.5 & 2.3 & $\dots$ & $\dots$ & $\dots$ & $\dots$\\
8.5 & 2.5 & $\dots$ & $\dots$ & $\dots$ & $\dots$\\
8.5 & 2.7 & $\dots$ & $\dots$ & $\dots$ & $\dots$\\
8.5 & 2.9 & $\dots$ & $\dots$ & $\dots$ & $\dots$\\
8.9 & 0.3 & $\dots$ & $\dots$ & $\dots$ & $\dots$\\
8.9 & 0.5 & $\dots$ & $\dots$ & $\dots$ & $\dots$\\
8.9 & 0.7 & $\dots$ & $\dots$ & $\dots$ & $\dots$\\
8.9 & 0.9 & $\dots$ & $\dots$ & $\dots$ & $\dots$\\
8.9 & 1.1 & $\dots$ & $\dots$ & $\dots$ & $\dots$\\
8.9 & 1.3 & $\dots$ & $\dots$ & $\dots$ & $\dots$\\
8.9 & 1.5 & $\dots$ & $\dots$ & $\dots$ & $\dots$\\
8.9 & 1.7 & $\dots$ & $\dots$ & $\dots$ & $\dots$\\
8.9 & 1.9 & $\dots$ & $\dots$ & $\dots$ & $\dots$\\
8.9 & 2.1 & $\dots$ & $\dots$ & $\dots$ & $\dots$\\
8.9 & 2.3 & $\dots$ & $\dots$ & $\dots$ & $\dots$\\
8.9 & 2.5 & $\dots$ & $\dots$ & $\dots$ & $\dots$\\
8.9 & 2.7 & $\dots$ & $\dots$ & $\dots$ & $\dots$\\
8.9 & 2.9 & $\dots$ & $\dots$ & $\dots$ & $\dots$\\
9.3 & 0.3 & $\dots$ & $\dots$ & $\dots$ & $\dots$\\
9.3 & 0.5 & $\dots$ & $\dots$ & $\dots$ & $\dots$\\
9.3 & 0.7 & $\dots$ & $\dots$ & $\dots$ & $\dots$\\
9.3 & 0.9 & $\dots$ & $\dots$ & $\dots$ & $\dots$\\
9.3 & 1.1 & $\dots$ & $\dots$ & $\dots$ & $\dots$\\
9.3 & 1.3 & $\dots$ & $\dots$ & $\dots$ & $\dots$\\
9.3 & 1.5 & $\dots$ & $\dots$ & $\dots$ & $\dots$\\
9.3 & 1.7 & $\dots$ & $\dots$ & $\dots$ & $\dots$\\
9.3 & 1.9 & $\dots$ & $\dots$ & $\dots$ & $\dots$\\
9.3 & 2.1 & $\dots$ & $\dots$ & $\dots$ & $\dots$\\
9.3 & 2.3 & $\dots$ & $\dots$ & $\dots$ & $\dots$\\
9.3 & 2.5 & $\dots$ & $\dots$ & $\dots$ & $\dots$\\
9.3 & 2.7 & $\dots$ & $\dots$ & $\dots$ & $\dots$\\
9.3 & 2.9 & $\dots$ & $\dots$ & $\dots$ & $\dots$\\
9.7 & 0.3 & 59.35$\pm$4.82 & 22.3$\pm$0.4 & 1.0$\pm$0.3 & 1.3$\pm$0.2\\
9.7 & 0.5 & $\dots$ & $\dots$ & $\dots$ & $\dots$\\
9.7 & 0.7 & $\dots$ & $\dots$ & $\dots$ & $\dots$\\
9.7 & 0.9 & $\dots$ & $\dots$ & $\dots$ & $\dots$\\
9.7 & 1.1 & 5.43$\pm$0.89 & 24.3$\pm$0.9 & 2.0$\pm$0.6 & 4.5$\pm$1.3\\
9.7 & 1.3 & 5.34$\pm$0.84 & 23.6$\pm$0.7 & 2.9$\pm$0.9 & 5.7$\pm$1.3\\
9.7 & 1.5 & 5.14$\pm$0.74 & 24.1$\pm$0.7 & 4.0$\pm$1.2 & 8.6$\pm$2.0\\
9.7 & 1.7 & $\dots$ & $\dots$ & $\dots$ & $\dots$\\
9.7 & 1.9 & $\dots$ & $\dots$ & $\dots$ & $\dots$\\
9.7 & 2.1 & $\dots$ & $\dots$ & $\dots$ & $\dots$\\
9.7 & 2.3 & $\dots$ & $\dots$ & $\dots$ & $\dots$\\
9.7 & 2.5 & $\dots$ & $\dots$ & $\dots$ & $\dots$\\
9.7 & 2.7 & $\dots$ & $\dots$ & $\dots$ & $\dots$\\
9.7 & 2.9 & $\dots$ & $\dots$ & $\dots$ & $\dots$\\
10.1 & 0.3 & 131.04$\pm$5.78 & 22.8$\pm$0.2 & 2.1$\pm$0.6 & 3.2$\pm$0.3\\
10.1 & 0.5 & 39.39$\pm$2.58 & 24.2$\pm$0.4 & 2.1$\pm$0.6 & 4.7$\pm$0.5\\
10.1 & 0.7 & 25.35$\pm$1.74 & 24.1$\pm$0.4 & 3.1$\pm$0.8 & 6.5$\pm$0.8\\
10.1 & 0.9 & 16.30$\pm$1.21 & 25.0$\pm$0.4 & 3.6$\pm$1.0 & 9.8$\pm$1.3\\
10.1 & 1.1 & 9.42$\pm$0.70 & 26.2$\pm$0.4 & 3.4$\pm$0.9 & 12$\pm$2\\
10.1 & 1.3 & 8.78$\pm$0.70 & 26.1$\pm$0.5 & 4.8$\pm$1.3 & 17$\pm$2\\
10.1 & 1.5 & 6.09$\pm$0.54 & 27.1$\pm$0.5 & 4.8$\pm$1.3 & 20$\pm$3\\
10.1 & 1.7 & 5.38$\pm$0.55 & 27.7$\pm$0.6 & 5.7$\pm$1.6 & 28$\pm$5\\
10.1 & 1.9 & 3.22$\pm$0.48 & 29.5$\pm$0.9 & 4.5$\pm$1.3 & 32$\pm$8\\
10.1 & 2.1 & 2.08$\pm$0.32 & 31.5$\pm$1.0 & 3.7$\pm$1.1 & 39$\pm$10\\
10.1 & 2.3 & 1.92$\pm$0.25 & 32.0$\pm$0.9 & 4.3$\pm$1.2 & 49$\pm$11\\
10.1 & 2.5 & $\dots$ & $\dots$ & $\dots$ & $\dots$\\
10.1 & 2.7 & 0.96$\pm$0.20 & 34.2$\pm$1.5 & 3.1$\pm$1.0 & 54$\pm$18\\
10.1 & 2.9 & $\dots$ & $\dots$ & $\dots$ & $\dots$\\
10.5 & 0.3 & 217.31$\pm$10.24 & 24.0$\pm$0.2 & 3.5$\pm$0.9 & 7.3$\pm$0.6\\
10.5 & 0.5 & 104.44$\pm$4.20 & 22.8$\pm$0.2 & 5.6$\pm$1.5 & 8.6$\pm$0.6\\
10.5 & 0.7 & 45.94$\pm$1.93 & 24.7$\pm$0.2 & 5.5$\pm$1.5 & 14$\pm$1\\
10.5 & 0.9 & 27.55$\pm$1.47 & 25.5$\pm$0.3 & 6.1$\pm$1.6 & 18$\pm$2\\
10.5 & 1.1 & 17.88$\pm$1.04 & 25.8$\pm$0.3 & 6.5$\pm$1.7 & 21$\pm$2\\
10.5 & 1.3 & 14.51$\pm$1.03 & 26.7$\pm$0.4 & 8.0$\pm$2.2 & 32$\pm$4\\
10.5 & 1.5 & 12.21$\pm$0.93 & 26.8$\pm$0.5 & 9.6$\pm$2.6 & 38$\pm$5\\
10.5 & 1.7 & 8.93$\pm$0.88 & 28.7$\pm$0.6 & 9.5$\pm$2.6 & 57$\pm$9\\
10.5 & 1.9 & 5.59$\pm$0.77 & 31.5$\pm$1.0 & 7.8$\pm$2.3 & 82$\pm$18\\
10.5 & 2.1 & 5.17$\pm$0.74 & 31.1$\pm$1.0 & 9.2$\pm$2.7 & 90$\pm$19\\
10.5 & 2.3 & 2.59$\pm$0.42 & 35.0$\pm$1.3 & 5.8$\pm$1.8 & 113$\pm$33\\
10.5 & 2.5 & 1.72$\pm$0.30 & 37.2$\pm$1.4 & 4.7$\pm$1.5 & 132$\pm$36\\
10.5 & 2.7 & 1.19$\pm$0.27 & 37.9$\pm$1.8 & 3.9$\pm$1.3 & 122$\pm$47\\
10.5 & 2.9 & $\dots$ & $\dots$ & $\dots$ & $\dots$\\
10.9 & 0.3 & 293.26$\pm$28.24 & 24.3$\pm$0.5 & 4.7$\pm$1.3 & 11$\pm$2\\
10.9 & 0.5 & 158.19$\pm$7.67 & 22.8$\pm$0.3 & 8.5$\pm$2.2 & 13$\pm$1\\
10.9 & 0.7 & 84.67$\pm$3.13 & 24.0$\pm$0.2 & 10.2$\pm$2.7 & 22$\pm$1\\
10.9 & 0.9 & 57.28$\pm$2.52 & 24.2$\pm$0.3 & 12.8$\pm$3.4 & 28$\pm$2\\
10.9 & 1.1 & 31.88$\pm$1.59 & 25.6$\pm$0.3 & 11.6$\pm$3.1 & 36$\pm$3\\
10.9 & 1.3 & 20.18$\pm$1.86 & 26.1$\pm$0.5 & 11.1$\pm$3.1 & 38$\pm$6\\
10.9 & 1.5 & 17.10$\pm$1.70 & 27.0$\pm$0.6 & 13.4$\pm$3.7 & 57$\pm$9\\
10.9 & 1.7 & 10.81$\pm$1.41 & 29.8$\pm$0.9 & 11.5$\pm$3.3 & 86$\pm$20\\
10.9 & 1.9 & 10.03$\pm$1.50 & 30.6$\pm$1.0 & 14.0$\pm$4.2 & 124$\pm$31\\
10.9 & 2.1 & 6.43$\pm$1.34 & 33.8$\pm$1.6 & 11.5$\pm$3.8 & 183$\pm$67\\
10.9 & 2.3 & $\dots$ & $\dots$ & $\dots$ & $\dots$\\
10.9 & 2.5 & $\dots$ & $\dots$ & $\dots$ & $\dots$\\
10.9 & 2.7 & $\dots$ & $\dots$ & $\dots$ & $\dots$\\
10.9 & 2.9 & $\dots$ & $\dots$ & $\dots$ & $\dots$\\
11.3 & 0.3 & $\dots$ & $\dots$ & $\dots$ & $\dots$\\
11.3 & 0.5 & 216.49$\pm$35.47 & 22.7$\pm$0.8 & 11.6$\pm$3.6 & 18$\pm$5\\
11.3 & 0.7 & 145.44$\pm$10.79 & 23.3$\pm$0.4 & 17.5$\pm$4.7 & 31$\pm$4\\
11.3 & 0.9 & 109.95$\pm$7.87 & 23.3$\pm$0.4 & 24.5$\pm$6.6 & 43$\pm$5\\
11.3 & 1.1 & 61.73$\pm$5.11 & 24.6$\pm$0.5 & 22.5$\pm$6.1 & 55$\pm$8\\
11.3 & 1.3 & 41.94$\pm$6.66 & 25.3$\pm$0.9 & 23.1$\pm$7.0 & 67$\pm$18\\
11.3 & 1.5 & 24.91$\pm$3.86 & 28.1$\pm$1.0 & 19.5$\pm$5.9 & 105$\pm$27\\
11.3 & 1.7 & $\dots$ & $\dots$ & $\dots$ & $\dots$\\
11.3 & 1.9 & $\dots$ & $\dots$ & $\dots$ & $\dots$\\
11.3 & 2.1 & $\dots$ & $\dots$ & $\dots$ & $\dots$\\
11.3 & 2.3 & $\dots$ & $\dots$ & $\dots$ & $\dots$\\
11.3 & 2.5 & $\dots$ & $\dots$ & $\dots$ & $\dots$\\
11.3 & 2.7 & $\dots$ & $\dots$ & $\dots$ & $\dots$\\
11.3 & 2.9 & $\dots$ & $\dots$ & $\dots$ & $\dots$\\

\end{longtable}
\label{LastPage}

\footnotesize
 \tablefoot{The central stellar mass and redshift of each bin are given in the first two columns. The amplitude $A$ and dust temperature $T_{\rm d}$ come directly from the SED fit, while the dust mass $M_{\rm d}$ and SFR are calculated from the fit parameters.}

\end{center}
\clearpage
\twocolumn

\section{Mathematical models combining the star-forming main sequence with dust properties}
\label{app:4}

In this appendix we outline the mathematical details for combining the star-forming MS with the dust temperature, $T_{\rm d}$, and the dust mass, $M_{\rm d}$. We start by re-defining the MS, explicitly showing the dependence on time and stellar mass; here we focus on the parametrisation from \citet{popesso2023}, but it is easy to adjust our derivations for other parametrisations. The MS follows
\begin{equation}\label{eq:galaxy_ms_app}
    \mathrm{SFR}(t,M_{\ast}) = \frac{\mathrm{SFR}_{\mathrm{max}}(t)}{1+(M_0(t)/M_{\ast})^{\gamma}}\,,
\end{equation}
where we use $\gamma\,{=}\,1$, $\logten(\mathrm{SFR}_{\mathrm{max}})(t)\,{=}\,a_0\,{+}\,a_1 t$, $\logten(M_0)(t)\,{=}\,a_2\,{+}\,a_3 t$, $a_0\,{=}\,2.71$, $a_1\,{=}\,-0.186$, $a_2\,{=}\,10.86$, and $a_3\,{=}\,-0.0729$. Our equation for the dust temperature is only a function of time and not stellar mass and follows
\begin{equation}\label{eq:T_decay_app}
    T_{\rm d}(t) = T_2+(T_1 - T_2)\,{\rm e}^{-t/\tau}\,,
\end{equation}
where $T_1\,{=}\,79.7\,$K, $T_2\,{=}\,23.2\,$K, and $\tau\,{=}\,1.6\,$Gyr. The SFR is related to the time-evolving far-IR SED through
\begin{equation}\label{eq:lfir_app}
    \mathrm{SFR}(t,M_{\ast}) = (1.49 \times 10^{-10})\, 4 \pi D_{\rm L}^2(t) \int_{\nu_1}^{\nu_2} S_{\nu}(A(t,M_{\ast}),T_{\rm d}(t)) \,{\rm d}\nu,
\end{equation}
where $\nu_1\,{=}\,c/1000\,\mu$m, $\nu_2\,{=}\,c/8\,\mu$m, and $S_{\nu}$ is the rest-frame SED. The dust mass is calculated using similar quantities, with 
\begin{equation}\label{eq:dust_mass_app}
    M_{\rm d}(t,M_{\ast}) = \frac{D_{\rm L}^2(t)A(t,M_{\ast})}{\kappa_0}\,.
\end{equation}
We note that we are assuming only two free parameters in the time-evolving far-IR SED, the overall multiplicative amplitude $A$ (a function of both time and stellar mass) and the dust temperature $T_{\rm d}$ (a function of only time) -- see Sect.~\ref{subsec:results_sed_core}.

We start by solving for $M_{\rm d}(t)$. This effectively amounts to calculating the ratio between Eqs.~\ref{eq:galaxy_ms_app} and \ref{eq:lfir_app}:
\begin{equation}
    A(t,M_{\ast}) = \frac{\mathrm{SFR}_{\mathrm{max}}(t)/(1+(M_0(t)/M_{\ast})^{\gamma})}{(1.49 \times 10^{-10})\, 4 \pi D_{\rm L}^2(t) \int_{\nu_1}^{\nu_2} S_{\nu}(1,T_{\rm d}(t)) \,{\rm d}\nu}.
\end{equation}
This is then inserted into Eq.~\ref{eq:dust_mass_app} to find  $M_{\rm d}(t)$, and the dust-to-stellar mass ratio can be calculated by dividing by the stellar mass.

We can also calculate time-dependent correlations between quantities. As an example we show how to derive the equations relating SFR and $T_{\rm d}$, and we note that the same procedure can produce equations relating any other combination of far-IR properties. Eq.~\ref{eq:galaxy_ms_app} says that SFRs increase monotonically with time for galaxies of all stellar mass; therefore, given an SFR and a stellar mass we can calculate the appropriate age of the Universe. While in the current mathematical form this cannot be done analytically, a simple root-finding algorithm will provide a function $f$ that provides the age for a given SFR and stellar mass, $t\,{=}\,f(\mathrm{SFR},M_{\ast})$. The dust temperature can then be written as
\begin{equation}\label{eq:T_decay_sfr_app}
    T_{\rm d}(\mathrm{SFR},M_{\ast}) = T_2+(T_1 - T_2)\,{\rm e}^{-f(\mathrm{SFR},M_{\ast})/\tau}\,,
\end{equation}
where the SFR range being plotted is understood to correspond to a time range. This can be converted to $T_{\rm d}$ as a function of sSFR (SFR$/M_{\ast}$) by dividing Eq.~\ref{eq:galaxy_ms_app} by $M_{\ast}$ and writing a new root-finder to obtain a function $g$ that provides the age for a given sSFR and stellar mass, $t\,{=}\,g(\mathrm{sSFR},M_{\ast})$.

\end{document}